\documentclass[aps,pra,floatfix,longbibliography,superscriptaddress,amssymb,amsmath,twocolumn]{revtex4-2}
\pdfoutput=1
\usepackage{graphicx,braket,float,mathtools}
\usepackage{amsmath}
\usepackage{amsfonts}
\usepackage{amsbsy}
\usepackage{bbm}
\usepackage[table]{xcolor}
\usepackage{soul}
\usepackage{bm}
\usepackage[normalem]{ulem}
\usepackage[unicode]{hyperref}
\usepackage{braket}
\hypersetup{
    unicode=true, 
    plainpages=false,
    colorlinks=true,
    linkcolor=blue,
    citecolor=blue,
    filecolor=blue,
    urlcolor=blue
}
\begin{document}

\title{Reservoir-mediated spin entanglement in the mean-force Gibbs state}
\author{L. A. Williamson}
\affiliation{ARC Centre of Excellence for Engineered Quantum Systems, School of Mathematics and Physics, University of Queensland, St Lucia, Queensland 4072, Australia}
\author{W. McEniery}
\affiliation{ARC Centre of Excellence for Engineered Quantum Systems, School of Mathematics and Physics, University of Queensland, St Lucia, Queensland 4072, Australia}
\author{F. Cerisola}
\affiliation{Department of Physics and Astronomy, University of Exeter, Stocker Road, Exeter EX4 4QL, United Kingdom}
\author{J. Anders}
\affiliation{Department of Physics and Astronomy, University of Exeter, Stocker Road, Exeter EX4 4QL, United Kingdom}
\affiliation{Institut f\"{u}r Physik und Astronomie, University of Potsdam, 14476 Potsdam, Germany}
\date{\today}

\begin{abstract}
Two qubits strongly coupled to a common bosonic reservoir can become entangled with each other, despite having no direct interaction. In equilibrium, such coupling-induced coherences can be described by the mean-force Gibbs state. Here we derive approximate, analytic expressions for the two-qubit mean-force Gibbs state, and use these to characterize equilibrium qubit-qubit entanglement mediated by a thermal reservoir. Entanglement, which is highest at lowest temperatures, is a non-monotonic function of the system-reservoir coupling strength. Moreover, we find that broadening the reservoir spectral density beyond a single mode, as is realistic for typical baths, can enhance the qubit entanglement. Our results provide a comprehensive understanding of reservoir-mediated two-qubit entanglement in thermal equilibrium and provide a benchmark to compare with numerical methods, as well as demonstrating the utility of strong system-reservoir coupling as a resource.
\end{abstract}

\maketitle

\section{Introduction}

Coupling to a thermal reservoir is often regarded as detrimental to quantum coherence. For very weak system-reservoir coupling, this is typically the case, as the Born-Markov and secular approximations lead to decoherence in the energy eigenbasis and relaxation to a Gibbs state~\cite{breuer2002}. In contrast, strong system-reservoir coupling, which renders conventional master equation techniques inapplicable, can generate rather than destroy coherence, altering the dynamic and equilibrium properties of a system. The field of strong-coupling thermodynamics has become an active area of research~\cite{rivas2020,seifert2016,jarzynski2017,llobet2018,bruch2018,strasberg2019,esposito2015,esposito2015b,carrega2016,campisi2009,gallego2014,kato2016,uzdin2016,katz2016,newman2017,miller2018,talkner2020,campbell2026}, with applications in chemistry, biology, and engineering.

There is growing evidence that the equilibrium state of a system strongly coupled to a reservoir is the mean-force Gibbs (MFG) state~\cite{chiu2022,cerisola2024,kumar2025,becker2022,subasi2012,mori2008,latune2021,trushechkin2022b,brenes2024}. Since its original formulation describing chemical mixtures~\cite{kirkwood1935}, the MFG state has been used extensively in classical chemistry and biomolecular simulations~\cite{roux1995,darve2001,park2004,trzesniak2007,melo1997,jiang2002,hamelryck2010}. In a quantum system the MFG state is formally defined as~\cite{trushechkin2022}
\begin{equation}\label{MFG0}
\rho_\mathrm{MF}=\frac{1}{Z}\operatorname{Tr}_\mathrm{R}\left[e^{-\beta \hat{H}_\mathrm{tot}}\right],
\end{equation}
where $\beta=(k_\mathrm{B}T)^{-1}$ is the inverse temperature and $\hat{H}_\mathrm{tot}=\hat{H}_\mathrm{S}+\hat{H}_\mathrm{R}+\lambda\hat{V}$ is the global Hamiltonian, with system Hamiltonian $\hat{H}_\mathrm{S}$, reservoir Hamiltonian $\hat{H}_\mathrm{R}$, and system-reservoir coupling $\hat{V}$ with dimensionless strength $\lambda$. The trace $\operatorname{Tr}_\mathrm{R}$ is over reservoir states and $Z$ is the global partition function. Equation~\eqref{MFG0} reduces to the standard Gibbs state $\rho_\mathrm{G}=e^{-\beta \hat{H}_\mathrm{S}}/\operatorname{Tr}_\mathrm{S}e^{-\beta \hat{H}_\mathrm{S}}$ in the limit $\lambda\rightarrow 0$, with $\operatorname{Tr}_\mathrm{S}$ a trace over the system.

The spin-boson model consists of one or more qubits coupled to a common bosonic reservoir, and is one of the most important models within strong-coupling thermodynamics~\cite{leggett1987}. In addition to its theoretical significance, the spin-boson model describes a range of physical systems, including atom-light interactions~\cite{strack2011}, superconducting qubits in the presence of circuit noise~\cite{makhlin2001}, and damped electron transfer and decoherence in biomolecules~\cite{garg1985,xu1994,gilmore2005}. The mean-force Gibbs state for a single qubit coupled to a bosonic reservoir has recently been analysed, and can exhibit coherence in the $\hat{H}_\mathrm{S}$ eigenbasis~\cite{guarnieri2018,purkayastha2020,cresser2021,kumar2025}.

For two-qubits, coupling to a common reservoir can mediate qubit-qubit interactions, providing the possibility for reservoir-mediated entanglement. For $\hat{H}_\mathrm{S}=0$, the two-qubit steady-state is analytically solvable~\cite{braun2002} and entanglement is typically absent unless there is an occupied decoherence-free state~\cite{jakobczyk2002,maniscalco2008,francica2009,ng2009,mazzola2009,sahrapour2013}. For $\hat{H}_\mathrm{S}\ne 0$, numerical studies have shown dynamical convergence toward the MFG state~\cite{chiu2022}, and that the MFG state can exhibit entanglement~\cite{lee2019,chiu2022,hartmann2020}. A detailed understanding of this entanglement, however, such as its dependence on system parameters, temperature, and reservoir spectral density, is lacking. This is particularly relevant given advancements in analogue quantum simulation of the spin-boson model with engineered structured baths ~\cite{porras2008,lemmer2018,sun2025,mostame2012,leppakangas2018,magazzu2018}.

\begin{figure}
\includegraphics[trim=1cm 9cm 0cm 9cm,clip=true,width=0.5\textwidth]{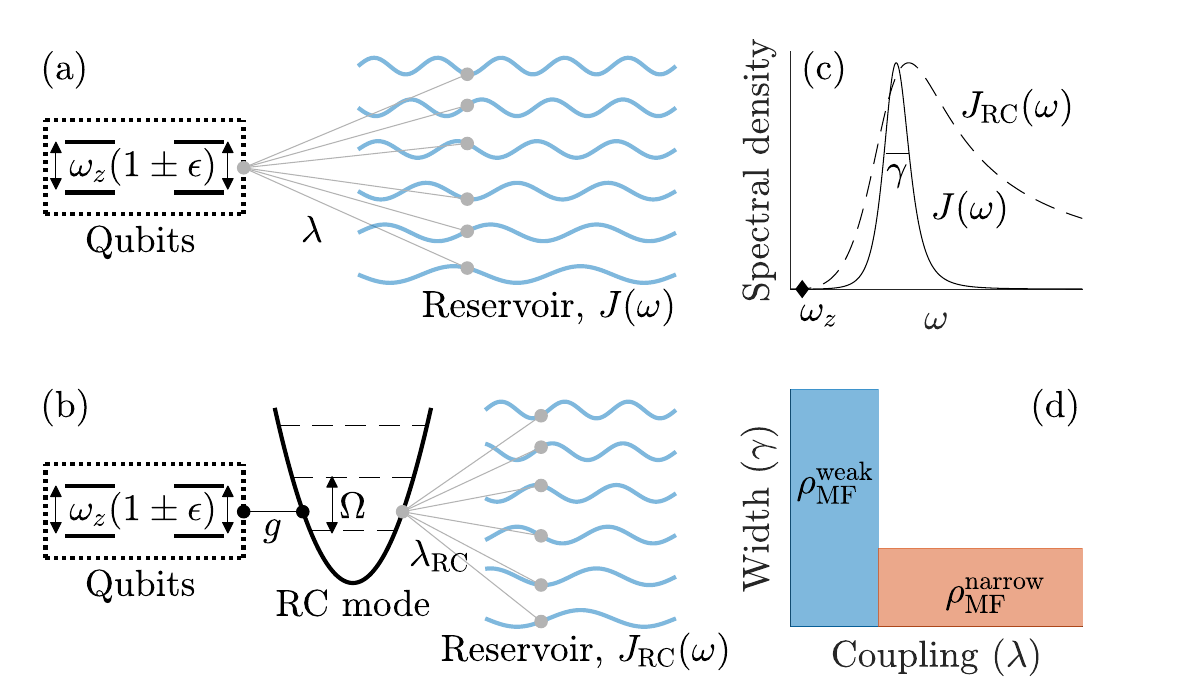}
\caption{\label{schem} (a) System setup: two qubits (level spacings $\omega_z(1\pm\epsilon)$) are coupled to a common bosonic reservoir with spectral density $J(\omega)$ and coupling strength $\lambda$. The mean-force Gibbs state is obtained by tracing out the bosonic reservoir. (b) After the reaction-coordinate mapping, the qubits couple to a single bosonic mode (the reaction coordinate, frequency $\Omega$) with coupling strength $g\propto \lambda$. The reaction coordinate couples to the remaining reservoir modes (spectral density $J_\mathrm{RC}(\omega)$) with coupling strength $\lambda_\mathrm{RC}$ characterised by the width $\gamma$ of $J(\omega)$. (c) Example reservoir spectral densities $J(\omega)$ (Eq.~\eqref{specden}, solid line) and $J_\mathrm{RC}(\omega)$ (Eq.~\eqref{JRC}, dashed line), scaled by their maximum value, with $\gamma$ (horizontal line) indicated. (d) We calculate the mean-force Gibbs state to lowest order in either the system-reservoir coupling $\lambda$ (Eq.~\eqref{MFG1}, blue region) or the spectral density width $\gamma$ (Eq.~\eqref{MFG2}, red region).}
\end{figure}

Here we analytically calculate the MFG state for two qubits coupled to a common bosonic reservoir to lowest order in either the system-reservoir coupling $\lambda$ or reservoir spectral density width $\gamma$ [see Fig.~\ref{schem}]. Focusing first on the case of qubits coupled to a single reservoir mode ($\gamma\rightarrow 0$), we confirm that the qubits exhibit entanglement at low temperatures, which decreases monotonically with increasing temperature and is absent above a critical temperature. We further show that entanglement peaks at a finite qubit-reservoir coupling strength, which at low temperatures is markedly different from the thermal-state entanglement of two qubits interacting directly. Next, we consider a broadened reservoir spectral density, which results in coupling to multiple reservoir modes. Remarkably, for moderate system-reservoir interactions, broadening of the reservoir spectral density increases qubit entanglement compared to that mediated by a single reservoir mode. For completeness, we also vary the magnitude and asymmetry of the qubit level spacing, finding that qubit entanglement is largest for identical qubits off-resonant from the reservoir spectral density peak.

\section{Two-qubit spin-boson model}

The global Hamiltonian for two qubits coupled to a bosonic reservoir [Fig.~\ref{schem}(a)] is~\cite{leggett1987}
\begin{equation}\label{H}
\begin{split}
\hat{H}_\mathrm{tot}&=\hat{H}_\mathrm{S}+\hat{H}_\mathrm{R}+\frac{1}{2}\lambda\left(\hat{\sigma}_x^{(1)}+\hat{\sigma}_x^{(2)}\right)\hat{R}.
\end{split}
\end{equation}
Here $\hat{H}_\mathrm{S}=\frac{1}{2}(1+\epsilon)\omega_z\hat{\sigma}_z^{(1)}+\frac{1}{2}(1-\epsilon)\omega_z\hat{\sigma}_z^{(2)}$ is the system Hamiltonian with qubit level spacings $\omega_z(1\pm \epsilon)$, $\hat{\sigma}_\mu^{(n)}$ are the spin-1/2 Pauli matrices for qubit $n=1,2$ ($[\hat{\sigma}_x^{(m)},\hat{\sigma}_y^{(n)}]=2i\delta_{mn}\hat{\sigma}_z^{(n)}$ etc.), and we set $\hbar\equiv 1$ throughout. The reservoir has Hamiltonian $\hat{H}_\mathrm{R}=\int_0^\infty \omega \hat{b}^\dagger(\omega)\hat{b}(\omega)\,d\omega$, where $\hat{b}(\omega)$ is a bosonic lowering operator at frequency $\omega$ ($[\hat{b}(\omega),\hat{b}^\dagger(\omega')]=\delta(\omega-\omega^\prime)$). The final term in Eq.~\eqref{H} describes a linear coupling between the collective spin $\hat{S}_x=\frac{1}{2}(\hat{\sigma}_x^{(1)}+\hat{\sigma}_x^{(2)})$ and the field amplitude $\hat{R}=\int_0^\infty \sqrt{J(\omega)}(\hat{b}(\omega)+\hat{b}^\dagger(\omega))\,d\omega$, with $J(\omega)$ the reservoir spectral density~\cite{leggett1987,weiss2012,hogg2024}. The strength of the system-reservoir coupling is characterized by the reorganization energy $Q=\lambda^2\int_0^\infty \omega^{-1}J(\omega)\,d\omega$~\cite{cheng2009,wu2010} (see also~\cite{ritschel2011,fruchtman2016}). We fix the scaling of $J(\omega)$ by setting $\int_0^\infty \omega^{-1}J(\omega)\,d\omega=\omega_z$, in which case $\lambda^2=Q/\omega_z$ measures the ratio of system-reservoir coupling to the energy scale of $\hat{H}_\mathrm{S}$ (the combination $\lambda^2 J(\omega)$ is independent of $\omega_z$).

Unequal level spacings ($\epsilon\ne 0$) in $\hat{H}_\mathrm{S}$ ensures the qubits have no decoherence-free state~\cite{zanardi1997,lidar1998,lidar1999,paz2008,paz2009}. This is important to ensure convergence to the MFG state when considering the system dynamics. In practice, a small asymmetry $\epsilon\rightarrow 0$ will allow the decoherence-free state to thermalize without affecting the eigendecomposition of $\hat{H}_\mathrm{tot}$ and hence the structure of the MFG state. We therefore mostly present results for $\epsilon=0$, reserving a discussion of the impact of $\epsilon$ to Sec.~\ref{sec:wz}.

\section{Approximate MFG states}

An exact calculation of the MFG state of Eq.~\eqref{H} for arbitrary reservoir spectral density and coupling strength is in general intractable. A perturbative approach~\cite{cresser2021} (summarized in Appendix~\ref{app:perturb}) allows the MFG state to be calculated to lowest order in $\lambda^2$, which gives [blue region in Fig.~\ref{schem}(d)]
\begin{equation}\label{MFG1}
\begin{split}
\rho_\mathrm{MF}&\approx\rho_\mathrm{MF}^\mathrm{weak}\\
&=\rho_\mathrm{G}-\frac{d\theta}{d\omega_z}\left(\hat{\sigma}_z^{(1)}\tau_2+\tau_1\hat{\sigma}_z^{(2)}\right)+\frac{(p_--p_+)\theta}{\omega_z}\hat{C}_-\\
&\phantom{=}+\left[2\beta\theta p_-p_+-(p_--p_+)\frac{d\theta}{d\omega_z}\right]\hat{C}_++O(\epsilon).
\end{split}
\end{equation}
Here $p_{\pm}=\frac{1}{2}\operatorname{sech}(\beta\omega_z/2)e^{\mp \frac{\beta\omega_z}{2}}$, $\tau_n=2\cosh(\beta\omega_z/2)e^{-\frac{\beta\omega_z}{2}\hat{\sigma}_z^{(n)}}$ and $\hat{C}_\pm=\frac{1}{2}(\hat{\sigma}_x^{(1)}\hat{\sigma}_x^{(2)}\pm\hat{\sigma}_y^{(1)}\hat{\sigma}_y^{(2)})$. Corrections to the Gibbs state $\rho_\mathrm{G}$ are characterized by the parameter
\begin{equation}
\theta=\frac{\lambda^2}{4}\mathcal{P}\int_0^\infty J(\omega)\frac{\omega+(p_+-p_-)\omega_z\coth \left(\frac{1}{2}\beta \omega\right)}{\omega^2-\omega_z^2}\,d\omega,
\end{equation}
which is related to an integral transform of imaginary-time reservoir correlation functions, with $\mathcal{P}\int$ a principal value integral. The first correction to $\rho_\mathrm{G}$ in Eq.~\eqref{MFG1} is a single-particle correction and is the same as that derived in~\cite{cresser2021}. The remaining two terms give rise to correlations between the qubits. The derivation of Eq.~\eqref{MFG1} is given in Appendix~\ref{app:MFG1}, explicitly including corrections due to $\epsilon$. The generalization of Eq.~\eqref{MFG1} to an arbitrary number of qubits is also provided.

To explore coupling strengths beyond order $O(\lambda^2)$ we utilise a recently developed effective Hamiltonian method~\cite{anto2023,brenes2024}. This enables a calculation of the MFG state that is perturbative in a new coupling strength $\lambda_\mathrm{eff}$ that is independent of $\lambda$ and instead depends on the spectral width of $J(\omega)$. The full details of this calculation are presented in Sec.~\ref{sec:derive} and Appendix~\ref{app:MFG2}. The final result is an expression for the MFG state to lowest order in $\lambda_\mathrm{eff}^2$ [red region in Fig.~\ref{schem}(d)],
\begin{equation}\label{MFG2}
    \rho_\mathrm{MF}\approx \rho_\mathrm{MF}^\mathrm{narrow}=\frac{L\left[e^{-\beta\hat{H}_\mathrm{S}^\mathrm{eff}}\right]}{\operatorname{Tr}\left[e^{-\beta\hat{H}_\mathrm{S}^\mathrm{eff}}\right]}+\lambda_\mathrm{eff}^2\delta\rho.
\end{equation}
Here~\cite{brenes2024}
\begin{equation}\label{HeffS}
    \hat{H}_\mathrm{S}^\mathrm{eff}=e^{-\frac{g^2}{2\Omega^2}}\hat{H}_\mathrm{S}-\frac{g^2}{\Omega}\hat{S}_x^2,
\end{equation}
with $\Omega=[\int_0^\infty \omega^3 J(\omega)\,d\omega/\int_0^\infty \omega J(\omega)\,d\omega]^{1/2}$ and $g=\lambda[\int_0^\infty \omega J(\omega)\,d\omega/\Omega]^{1/2}$, and
\begin{equation}\label{polaronTransform}
\begin{split}
L[\hat{O}]&=\hat{\mathcal{O}}+\frac{1}{2}\left(3+e^{-\frac{2g^2}{\Omega^2}}-4e^{-\frac{g^2}{2\Omega^2}}\right)\hat{S}_x^2\hat{\mathcal{O}}\hat{S}_x^2\\
&\phantom{=}+\frac{1}{2}\left(1-e^{-\frac{2g^2}{\Omega^2}}\right)\hat{S}_x\hat{\mathcal{O}}\hat{S}_x\\
&\phantom{=}-\left(1-e^{-\frac{g^2}{2\Omega^2}}\right)\left(\hat{S}_x^2\hat{\mathcal{O}}+\hat{\mathcal{O}}\hat{S}_x^2\right).
\end{split}
\end{equation}
The first term in Eq.~\eqref{MFG2} is the MFG state in the limit of vanishing reservoir spectral density width, in which case the qubits couple to a single reservoir mode. The term $\delta\rho$ describes corrections due to spectral density broadening. The full expression for $\delta\rho$ is complicated and not particularly illuminating; its evaluation, along with an explicit analytic expression for $\epsilon=0$, is discussed in Appendix~\ref{app:MFG2}.

Equation~\eqref{MFG1} and Eq.~\eqref{MFG2} enable characterization of the two-qubit MFG state in limits of either small $\lambda$ or small $\lambda_\mathrm{eff}$ (narrow reservoir spectral density), and constitute the main result of this work. Before discussing qubit entanglement in the MFG state, we present the derivation of Eq.~\eqref{MFG2}.

\section{Derivation of MFG state with narrow reservoir spectral density}\label{sec:derive}
The derivation of Eq.~\eqref{MFG2} starts with the reaction coordinate (RC) mapping~\cite{garg1985,martinazzo2011,smith2014,strasberg2016,correa2019,nazir2018} [Fig.~\ref{schem}(b)]. This is an exact reformulation of Eq.~\eqref{H} in which the system is enlarged to include the RC bosonic mode $\hat{a}$ defined by $\hat{a}+\hat{a}^\dagger=\hat{R}$. The RC then couples to the remaining reservoir modes $\hat{R}_\mathrm{rem}=\int_0^\infty \sqrt{J_\mathrm{RC}(\omega)}(\hat{c}(\omega)+\hat{c}^\dagger(\omega))\,d\omega$, where the bosonic lowering operators $\hat{c}(\omega)$ are obtained from a normal mode transformation of $\hat{b}(\omega)$, such that $[\hat{c}(\omega),\hat{a}^\dagger]=0$ and $[\hat{c}(\omega),\hat{c}^\dagger(\omega')]=\delta(\omega-\omega')$, and $J_\mathrm{RC}(\omega)$ is a modified spectral density.

\begin{figure*}
\includegraphics[trim=0cm 6.3cm 0cm 6.9cm,clip=true,width=\textwidth]{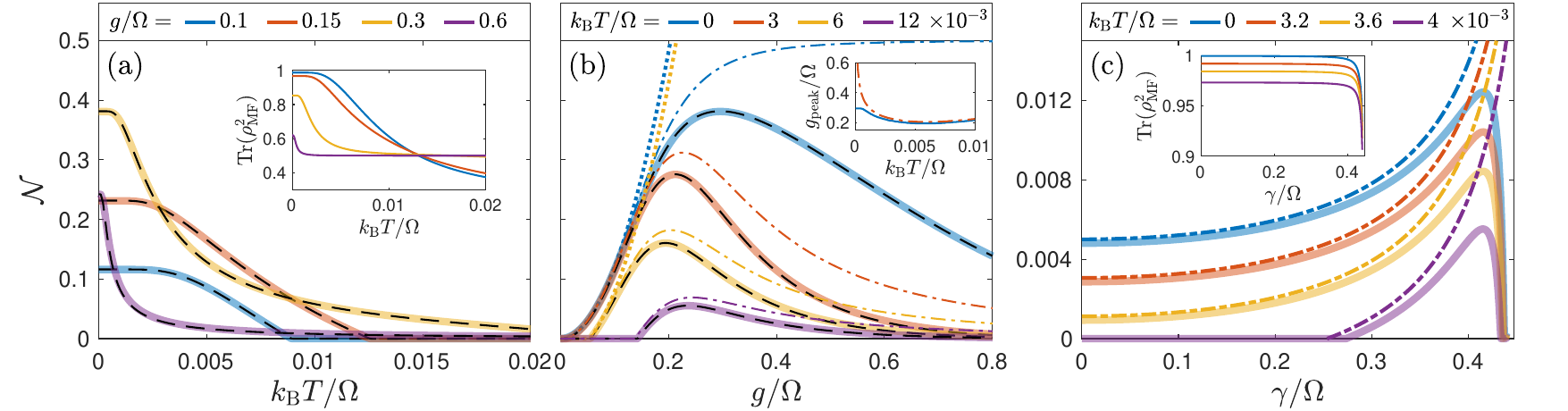}
\caption{\label{2spinA} Thermal entanglement, quantified by the negativity $\mathcal{N}$, for the mean-force Gibbs state of two qubits, which arises from the coupling to a common bosonic reservoir. (a,b) Entanglement mediated by a single-mode reservoir $(\gamma\rightarrow 0$) for varying $T$ and $g$. Solid lines are obtained from $\rho_\mathrm{MF}^\mathrm{narrow}$ [Eq.~\eqref{eq:rhogam0}], black dashed lines are obtained from numerical diagonalisation of $\hat{H}_\mathrm{S}^\mathrm{RC}$. (a) Entanglement decreases monotonically with increasing temperature and is absent for $T\ge T_\mathcal{N}$. Inset: purity of $\rho_\mathrm{MF}^\mathrm{narrow}$ decreases with increasing temperature and, for $k_\mathrm{B}T/\Omega\lesssim 0.013$, decreases with increasing $g$ due to entanglement with the RC. (b) For fixed $T\le T_\mathcal{N}$, entanglement of $\rho_\mathrm{MF}^\mathrm{narrow}$ peaks at $g=g_\mathrm{peak}$ and goes to zero for large $g$. The weak-coupling result $\rho_\mathrm{MF}^\mathrm{weak}$ [Eq.~\eqref{MFG1}] is shown by dotted lines for $k_\mathrm{B}T/\Omega=0$ (blue) and $k_\mathrm{B}T/\Omega=6\times 10^{-3}$ (yellow), and describes entanglement well for $g/\Omega\lesssim 0.1$. Dot-dashed lines are obtained from the state $\rho_\mathrm{G}^\mathrm{direct}$ (see text). Inset: $g_\mathrm{peak}$ from $\rho_\mathrm{MF}^\mathrm{weak}$ (solid line) and $\rho_\mathrm{G}^\mathrm{direct}$ (dot-dashed line). (c) Effect on entanglement of broadening the reservoir spectral density ($\gamma\ne 0)$ at weak coupling ($g=0.02\Omega$), evaluated from $\rho_\mathrm{MF}^\mathrm{weak}$ [Eq.~\eqref{MFG1}, solid lines]. Entanglement initially increases with $\gamma$, peaking at $\gamma/\Omega\approx 0.4$ and abruptly dropping to zero beyond this. Close to $T_\mathcal{N}$, non-zero $\gamma$ can give entanglement even when entanglement is absent for $\gamma=0$ (see results for $k_\mathrm{B}T/\Omega=4\times 10^{-3}$). Dash-dot lines are the approximation Eq.~\eqref{neganapprox}. Inset shows purity of $\rho_\mathrm{MF}^\mathrm{weak}$. All results are for $\omega_z/\Omega=0.02$ and $\epsilon=0$.}
\end{figure*}

Following the RC mapping, Eq.~\eqref{H} becomes 
\begin{equation}\label{HRC}
\hat{H}_\mathrm{tot}=\hat{H}_\mathrm{S}^\mathrm{RC}+\int_0^\infty \omega \hat{c}^\dagger(\omega)\hat{c}(\omega)\,d\omega+\lambda_\mathrm{RC}(\hat{a}+\hat{a}^\dagger)\hat{R}_\mathrm{rem}.
\end{equation}
The (enlarged) system Hamiltonian $\hat{H}_\mathrm{S}^\mathrm{RC}=\hat{H}_\mathrm{S}+\Omega\hat{a}^\dagger\hat{a}+g\hat{S}_x(\hat{a}+\hat{a}^\dagger)$ describes the coupling between the qubits and the RC and is the well-known Dicke Hamiltonian~\cite{hepp1973,hepp1973b,wang1973,hioe1973,carmichael1973,duncan1974,garraway2011}. The expressions for the RC mode frequency $\Omega$ and qubit-RC coupling strength $g$ are given below Eq.~\eqref{HeffS}~\cite{nazir2018}. The modified spectral density $J_\mathrm{RC}(\omega)$ [Fig.~\ref{schem}(c)] and coupling strength $\lambda_\mathrm{RC}$ satisfy~\cite{nazir2018}
\begin{equation}\label{JRCdef}
\lambda_\mathrm{RC}^2J_\mathrm{RC}(\omega)=\frac{g^2}{\lambda^2}\frac{2\pi J(\omega)}{\left[\mathcal{P}\int_{-\infty}^\infty \frac{J(\omega')}{\omega'-\omega}\,d\omega'\right]^2+\pi^2 J(\omega)^2}.
\end{equation}
Importantly, $\lambda_\mathrm{RC}^2J_\mathrm{RC}(\omega)$ is independent of $\lambda$ (noting that $g\propto \lambda$) and is instead characterised by the width $\gamma$ of the spectral density $J(\omega)$.

The global Hamiltonian following the RC mapping is exact but not yet analytically tractable. To obtain an analytically tractable description of the two-qubit MFG state, we derive an approximate Hamiltonian using the method from~\cite{anto2023}, which has been applied to the present system in~\cite{brenes2024}. We first move to the polaron frame via the unitary transformation $\hat{U}=e^{(g/\Omega)(\hat{a}^\dagger-\hat{a})\hat{S}_x}$, which gives $\hat{U}\hat{a}\hat{U}^\dagger=\hat{a}-(g/\Omega)\hat{S}_x$~\cite{lang1963} and
\begin{equation}\label{eq:HP}
\begin{split}    \hat{U}\hat{H}_\mathrm{tot}\hat{U}^\dagger&=\hat{U}\hat{H}_\mathrm{S}\hat{U}^\dagger+\Omega\hat{a}^\dagger\hat{a}+\frac{g^2}{\Omega}\hat{S}_x^2+\int_0^\infty \omega \hat{c}^\dagger(\omega)\hat{c}(\omega)\,d\omega\\
&\phantom{=}+\lambda_\mathrm{RC}\left(\hat{a}+\hat{a}^\dagger-\frac{2g}{\Omega}\hat{S}_x\right)\hat{R}_\mathrm{rem}.
\end{split}
\end{equation}
The polaron transform removes the explicit interaction between $\hat{S}_x$ and $\hat{a}$ and introduces a direct coupling between $\hat{S}_x$ and the remaining reservoir modes $\hat{R}_\mathrm{rem}$. Next, we assume the RC mode frequency $\Omega$ is the largest energy scale in the problem ($\Omega\gg k_\mathrm{B}T,g,\omega_z$), which allows us to project onto the vacuum $\ket{0}$ of $\hat{a}$ in Eq.~\eqref{eq:HP}. In the polaron frame, the global Hamiltonian following the projection is~\cite{anto2023}
\begin{equation}
\begin{split}\label{Heff1}
\hat{H}_\mathrm{tot}^\mathrm{P} &=\braket{0|\hat{U}\hat{H}_\mathrm{tot}\hat{U}^\dagger|0}\\
&=\hat{H}_\mathrm{S}^\mathrm{eff}+\int_0^\infty \omega \hat{c}^\dagger(\omega)\hat{c}(\omega)\,d\omega-\lambda_\mathrm{eff}\hat{S}_x\hat{R}_\mathrm{rem},
\end{split}
\end{equation}
with $\hat{H}_\mathrm{S}^\mathrm{eff}=\braket{0|\hat{U}\hat{H}_\mathrm{S}^\mathrm{RC}\hat{U}^\dagger|0}$ and $\lambda_\mathrm{eff}=2g\lambda_\mathrm{RC}/\Omega$. Expanding $\hat{U}$ in a power series and using $(\hat{a}^\dagger-\hat{a})^n\ket{0}=(\hat{a}^\dagger)^n\ket{0}$ gives~\cite{anto2023},
\begin{equation}\label{expand}
\braket{0|\hat{U}\hat{\mathcal{O}}\hat{U}^\dagger|0}=\sum_{n=0}^\infty \frac{g^{2n}}{\Omega^{2n}n!}e^{-\frac{g^2\hat{S}_x^2}{2\Omega^2}}\hat{S}_x^n\hat{\mathcal{O}}\hat{S}_x^n e^{-\frac{g^2\hat{S}_x^2}{2\Omega^2}}\equiv L[\hat{O}]
\end{equation}
for any qubit operator $\hat{\mathcal{O}}$. This also gives $\braket{0|\hat{U}^\dagger\hat{\mathcal{O}}\hat{U}|0}=\braket{0|\hat{U}\hat{\mathcal{O}}\hat{U}^\dagger|0}$ and therefore the transformation (with projection) to and from the polaron frame is the same.

The two-qubit MFG state is obtained from the Gibbs state $e^{-\beta\hat{H}_\mathrm{tot}^\mathrm{P}}/\operatorname{Tr}[^{-\beta\hat{H}_\mathrm{tot}^\mathrm{P}}]$ by first tracing out the reservoir modes $\hat{c}(\omega)$ and then transforming back from the polaron frame using Eq.~\eqref{expand}. For $\lambda_\mathrm{eff}\rightarrow 0$ (describing $J(\omega)\propto\delta(\omega-\Omega)$), this gives the first term in Eq.~\eqref{MFG2}, which contains interactions mediated by the RC. Corrections due to spectral-density broadening are computed perturbatively in $\lambda_\mathrm{eff}$ using the method from~\cite{cresser2021} (see Appendix~\ref{app:perturb} and~\ref{app:MFG2}), giving Eq.~\eqref{MFG2}. For two qubits we have $\hat{S}_x^3=\hat{S}_x$ and Eq.~\eqref{expand} can be simplified to give Eq.~\eqref{polaronTransform}.

\section{Entanglement mediated by a thermal reservoir}
Having established Eq.~\eqref{MFG1} and~\eqref{MFG2}, we are now ready to characterize qubit-qubit entanglement that arises in the MFG state due to coupling to a common reservoir. We quantify the entanglement of a two-qubit state $\rho$ by the negativity~\cite{vidal2002,kyczkowski1998,lee2000,eisert2001,plenio2005},
\begin{equation}\label{negativity}
\mathcal{N}(\rho)=\frac{||\rho^{T_1}||_1-1}{2},
\end{equation}
which is an entanglement monotone~\cite{vidal2000}. Here $\rho^{T_1}$ denotes the partial transpose of $\rho$ with respect to one of the two qubits and $||\rho^{T_1}||_1=\operatorname{Tr} \left[\sqrt{(\rho^{T_1})^\dagger\rho^{T_1}}\right]$ is its trace norm. For two qubits, $0\le \mathcal{N}\le 1/2$, with $\mathcal{N}>0$ being both a necessary and sufficient condition for entanglement~\cite{horodecki1996}.

\subsection{Single-mode reservoir}\label{sec:zerogam}
We first consider the limit of vanishing reservoir spectral density width, $\gamma\rightarrow 0$ ($\lambda_\mathrm{eff}\rightarrow 0$), so that the qubits couple to a single reservoir mode (the RC) and $\rho_\mathrm{MF}$ is described by the first term in Eq.~\eqref{MFG2},
\begin{equation}\label{eq:rhogam0}
    \rho_\mathrm{MF}^\mathrm{narrow}\overset{\gamma\rightarrow 0}= \frac{L\left[e^{-\beta\hat{H}_\mathrm{S}^\mathrm{eff}}\right]}{\operatorname{Tr}\left[e^{-\beta\hat{H}_\mathrm{S}^\mathrm{eff}}\right]}.
\end{equation}
The negativity of this state for varying $k_\mathrm{B}T$ and $g$ is shown in Fig.~\ref{2spinA}(a,b). As expected, entanglement is largest at $T=0$, decreasing monotonically with increasing $T$ and vanishing at sufficiently high temperatures $T\ge T_\mathcal{N}$. The negativity computed from Eq.~\eqref{eq:rhogam0} agrees almost perfectly with that obtained using exact diagonalisation of $\hat{H}_\mathrm{S}^\mathrm{RC}$ in a truncated bosonic basis, see Fig.~\ref{2spinA}(a,b), which justifies the projection in Eq.~\eqref{Heff1},\eqref{expand} (similar agreement holds as long as $\omega_z/\Omega\lesssim 0.1$). The small-$\lambda$ result $\rho_\mathrm{MF}^\mathrm{weak}$ [Eq.~\eqref{MFG1}] describes entanglement well for $g/\Omega\lesssim 0.1$ [see Fig.~\ref{2spinA}(b)]~\footnote{Note entanglement of $\rho_\mathrm{MF}^\mathrm{weak}$ requires the retention of co-rotating terms in Eq.~\eqref{H}. The ground state of $\hat{H}_\mathrm{S}^\mathrm{RC}$ with co-rotating terms removed exhibits entanglement only for $g>\sqrt{2\omega_z\Omega}$~\cite{buzek2005}.}. 

The origin of the two-qubit entanglement can be understood by considering $T=0$~\cite{lee2013,lambert2004,buzek2005}. Here the qubits are in the state $\rho_\mathrm{MF}\approx L[\ket{G}\bra{G}]$, with $\ket{G}$ the ground state of $\hat{H}_\mathrm{S}^\mathrm{eff}$. Explicitely, $\ket{G}\propto \alpha\ket{-}+(\sqrt{1+\alpha^2}-1)\ket{+}$, with $\ket{\pm}$ maximum-spin eigenstates of $\hat{S}_z=\frac{1}{2}(\hat{\sigma}_z^{(1)}+\hat{\sigma}_z^{(2)})$ in the polaron frame ($\hat{S}_z\ket{\pm}=\pm\ket{\pm}$) and $\alpha=g^2e^{g^2/(2\Omega^2)}/(2\omega_z\Omega)$ the ratio of the effective interaction strength $g^2/\Omega$ to the renormalized level spacing $\omega_ze^{-g^2/(2\Omega^2)}$. In the regime $g\ll\Omega$ the RC mediates qubit interactions without affecting the purity of the qubits~\cite{agarwal1997,zheng2000,zheng2001b,zheng2001}. This gives $\rho_\mathrm{MF}\approx\ket{G}\bra{G}$ and entanglement increases with increasing $\alpha$, with $\ket{G}$ approaching the Bell state $(\ket{-}+\ket{+})/\sqrt{2}$ for $\alpha\gg 1$. As $g/\Omega$ becomes appreciable, so too does entanglement with the RC, resulting in $L[\ket{G}\bra{G}]$ deviating from a pure state. The combination of these two effects --- $\ket{G}$ becoming more entangled while the purity of $\rho_\mathrm{MF}$ is reduced --- results in a peak in zero-temperature entanglement at $g=g_\mathrm{peak}\approx 0.3\Omega$.

The purity of $\rho_\mathrm{MF}$ and two-qubit entanglement is reduced by thermal fluctuations. For a fixed $T>0$, the susceptibility to thermal fluctuations increases with decreasing energy gap $\Delta$ between $\ket{G}$ and the first excited state of $\hat{H}_\mathrm{S}^\mathrm{eff}$. Explicitly,
\begin{equation}\label{eq:gap}
    \Delta=(1-\epsilon^2)\frac{\omega_z^2}{\Omega}\left(\frac{g^2}{\Omega^2} e^{g^2/\Omega^2}\right)^{-1}+O(\alpha^{-4}),
\end{equation}
which decreases with decreasing $\omega_z$, increasing $g$, or increasing $\epsilon$, assuming $\alpha\gg 1$ and fixed $\Omega$. The effect of varying $\omega_z$ and $\epsilon$ will be discussed later, and for now we suppose $\epsilon=0$ and $\omega_z\ll\Omega$. For $g\ll \Omega$, $\Delta\propto (g/\Omega)^{-2}$ decreases (i.e.\ the susceptibility to thermal fluctuations increases) moderately with increasing $g/\Omega$. This results in a small decrease in $g_\mathrm{peak}$ for $T>0$ [see inset to Fig.~\ref{2spinA}(b)]. As $g/\Omega$ becomes appreciable, so too does the renormalization of the qubit level-spacing due to the RC, in which case $\Delta$ decreases exponentially with $g/\Omega$. In this regime [$g/\Omega\gtrsim 0.3$ in Fig.~\ref{2spinA}(a)] the ground and first excited states are close to degenerate and the purity rapidly approaches $1/2$ for increasing temperature [see inset to Fig.~\ref{2spinA}(a)], coinciding with a steep reduction of entanglement. Generally, we find that qubit entanglement is both largest and most robust to thermal fluctuations for $g/\Omega\approx 0.2$, $\omega_z/\Omega\sim 10^{-2}$ and $\epsilon\ll 1$. Figure~\ref{2spinB}(b) suggests that entanglement vanishes for $g\rightarrow\infty$, however Eq.~\eqref{eq:rhogam0} is only valid for $g\lesssim \Omega$. An alternative calculation confirms that entanglement is absent in the MFG state in the ultrastrong-coupling limit $\lambda\rightarrow\infty$~\cite{cresser2021}.

We also compare reservoir-mediated two-qubit entanglement, computed from Eq.~\eqref{eq:rhogam0}, with the entanglement of two qubits interacting directly, for example via an exchange interaction. The two qubits with direct spin interactions are in a Gibbs state $\rho_\mathrm{G}^\mathrm{direct}=e^{-\beta\hat{H}_\mathrm{S}^\mathrm{direct}}/\operatorname{Tr}_\mathrm{S} [e^{-\beta\hat{H}_\mathrm{S}^\mathrm{direct}}]$, with $\hat{H}_\mathrm{S}^\mathrm{direct}=\hat{H}_\mathrm{S}-\frac{g^2}{\Omega}\hat{S}_x^2$. We parameterize the interaction strength as $g^2/\Omega$ to match the reservoir-mediated interaction in the small-$g$ limit. Thermal-state entanglement of qubits with similar Hamiltonians has been explored in~\cite{arnesen2001,wang2009}. At low temperatures, the reservoir-mediated entanglement is generally smaller than the entanglement of $\rho_\mathrm{G}^\mathrm{direct}$, due to the reduction of purity when the qubits are entangled with the RC, see Fig.~\ref{2spinA}(b). This is particularly pronounced at zero-temperature, where the entanglement of $\rho_\mathrm{G}^\mathrm{direct}$ increases monotonically with $g$, rather than peaking at a finite $g_\mathrm{peak}$ [the behavior of $g_\mathrm{peak}$ as a function of temperature is compared in the inset to Fig.~\ref{2spinA}(b)]. The reservoir-mediated entanglement approaches that of $\rho_\mathrm{G}^\mathrm{direct}$ for $g\ll\Omega$, since here entanglement with the RC is small ($L[\hat{O}]\approx\hat{O}$) and $\hat{H}_\mathrm{S}^\mathrm{eff}\approx\hat{H}_\mathrm{S}^\mathrm{direct}$~\cite{agarwal1997,zheng2000,zheng2001b,zheng2001}. The two also agree at higher temperatures, when the degradation of reservoir-mediated entanglement is primarily due to thermal fluctuations rather than entanglement with the RC.

\subsection{Effect of reservoir spectral density broadening}
We now consider the effect of broadening the reservoir spectral density, in which case interactions with modes additional to the RC become important. We consider the spectral density~\cite{nazir2018}
\begin{equation}\label{specden}
J(\omega)=\frac{1}{\pi}\frac{32\omega_z(\omega_0^2+\gamma^2)\gamma^3\omega^3}{[(\omega-\omega_0)^2+\gamma^2]^2[(\omega+\omega_0)^2+\gamma^2]^2},
\end{equation}
with peak position and width characterized by $\omega_0$ and $\gamma$ respectively. This gives~\cite{nazir2018}
\begin{equation}\label{JRC}
J_\mathrm{RC}(\omega)=\frac{1}{\pi}\frac{8\omega_z\gamma\omega^3}{\gamma^4+2\gamma^2(7\omega^2+\omega_0^2)+(\omega^2-\omega_0^2)^2}
\end{equation}
with RC mode frequency $\Omega=\sqrt{5\gamma^2+\omega_0^2}$ and system-RC coupling strength $g=\lambda\sqrt{\omega_z(\Omega^2-4\gamma^2)/\Omega}$. We have fixed the scaling of $J_\mathrm{RC}(\omega)$ by setting $\int_0^\infty \omega^{-1}J_\mathrm{RC}(\omega)\,d\omega\overset{\gamma\rightarrow 0}=\omega_z$, which gives the RC-reservoir coupling strength $\lambda_\mathrm{RC}^2=2\pi\gamma^2/(\omega_z\Omega)$. Note the spectral density Eq.~\eqref{specden} is more convenient than the commonly-used Lorentzian, since a Lorentzian gives $J_\mathrm{RC}(\omega)\sim\omega$ with a corresponding diverging reorganization energy. The spectral densities Eq.~\eqref{specden} and Eq.~\eqref{JRC} (scaled by their maximum value) are plotted in Fig.~\ref{schem}(c) for $\omega_z/\Omega=0.02$ and $\gamma/\Omega=0.2$.

We first consider the small-$\lambda$ limit $\rho_\mathrm{MF}^\mathrm{weak}$ [Eq.~\eqref{MFG1}], which allows for arbitrary spectral density broadening. The negativity computed from $\rho_\mathrm{MF}^\mathrm{weak}$ is plotted in Fig.~\ref{2spinA}(c) as a function of $\gamma$ for fixed $g$ and $\Omega$. Fixing $g,\Omega$ ensures changes in $\mathcal{N}$ are due to coupling to the remaining reservoir modes and not the dependence of $g$ and $\Omega$ on $\gamma$. For $T<T_\mathcal{N}$, entanglement initially increases with $\gamma$, peaking at $\gamma\approx 0.4\Omega$ before decreasing abruptly (fixing $\Omega$ gives $\omega_0=\sqrt{\Omega^2-5\gamma^2}$ and therefore $\omega_0\in\mathbb{R}$ requires $\gamma/\Omega\le 1/\sqrt{5}\approx 0.45$). Notably, entanglement can be present for $\gamma\ne 0$ even when absent for $\gamma=0$, as shown for the highest temperature in Fig.~\ref{2spinA}(c).

To understand this increase in entanglement with $\gamma$, we consider $k_BT\lesssim\omega_z\ll\omega_0$. We then have $\theta\approx Q/4$ and $d\theta/d\omega_z\approx 0$ in Eq.~\eqref{MFG1}, which gives
\begin{equation}\label{MFG1approx}
    \rho_\mathrm{MF}^\mathrm{weak}\approx \rho_\mathrm{G}+\frac{Q}{4\omega_z}\left[(p_--p_+)\hat{C}_-+2\beta\omega_zp_-p_+\hat{C}_+\right].
\end{equation}
(Recall that from our scaling convention for $J(\omega)$, the reorganization energy $Q$ is related to the coupling strength $\lambda$ via $Q/\omega_z=\lambda^2$. The form Eq.~\eqref{MFG1approx} holds irrespective of the scaling of $J(\omega)$.) With fixed $g$ and $\Omega$, $Q=g^2\Omega/(\Omega^2-4\gamma^2)$ increases with $\gamma$. Equation~\eqref{MFG1approx} gives
\begin{equation}\label{neganapprox}
    \mathcal{N}\approx \operatorname{max}\left[0,\frac{Q}{4\omega_z}(p_--p_+)-p_-p_+\right],
\end{equation}
and hence the increase in $Q$ translates directly into an increase in entanglement. Equation~\eqref{neganapprox} is plotted in Fig.~\ref{2spinA}(c) and approximates $\mathcal{N}$ well up to the peak in $\mathcal{N}$, beyond which $\omega_z\gtrsim\omega_0$. Note that broadening $J(\omega)$ increases $Q$ irrespective of the precise form of the spectral density: Defining $E[X]=\int_0^\infty X(\omega)J(\omega)\,d\omega$, the Cauchy-Schwarz inequality $E[X^2]E[Y^2]\geq E[XY]^2$ gives
\begin{equation}
    Q=\frac{g^2}{\Omega}\frac{\int_0^\infty\omega^3J(\omega)\,d\omega\int_0^\infty\omega^{-1}J(\omega)\,d\omega}{\left(\int_0^\infty \omega J(\omega)\,d\omega\right)^2}\geq \frac{g^2}{\Omega},
\end{equation}
with equality if and only if $J(\omega)\propto \delta(\omega-\Omega)$. Therefore broadening of $J(\omega)$ (with $g$ and $\Omega$ fixed) increases $Q$ compared to coupling to just the RC.

The purity of $\rho_\mathrm{MF}^\mathrm{weak}$ as a function of $\gamma/\Omega$ is shown in the inset to Fig.~\ref{2spinA}(c). For $\gamma\ll\Omega$, the purity is unaffected by broadening. Here the approximation Eq.~\eqref{MFG1approx} is valid, which gives $\operatorname{Tr}(\rho_\mathrm{MF}^2)\approx\operatorname{Tr}(\rho_\mathrm{G}^2)$, since the dressing terms in Eq.~\eqref{MFG1approx} are orthogonal to $\rho_\mathrm{G}$ ($\operatorname{Tr}[\rho_\mathrm{G}\hat{C}_\pm]=0$). As $\gamma/\Omega$ becomes appreciable, so too does $\omega_z/\omega_0=\omega_z/(\Omega^2-5\gamma^2)$ and the additional terms in Eq.~\eqref{MFG1} become important. In particular, the single-particle term reduces the purity of $\rho_\mathrm{MF}$, since $\operatorname{Tr}[\rho_\mathrm{G}(\hat{\sigma}_z^{(1)}\tau_2+\tau_1\hat{\sigma}_z^{(2)})]\ne 0$, resulting in the abrupt drop in entanglement in Fig.~\ref{2spinA}(c).

\begin{figure}
\includegraphics[trim=3cm 9.8cm 3cm 10cm,clip=true,width=0.48\textwidth]{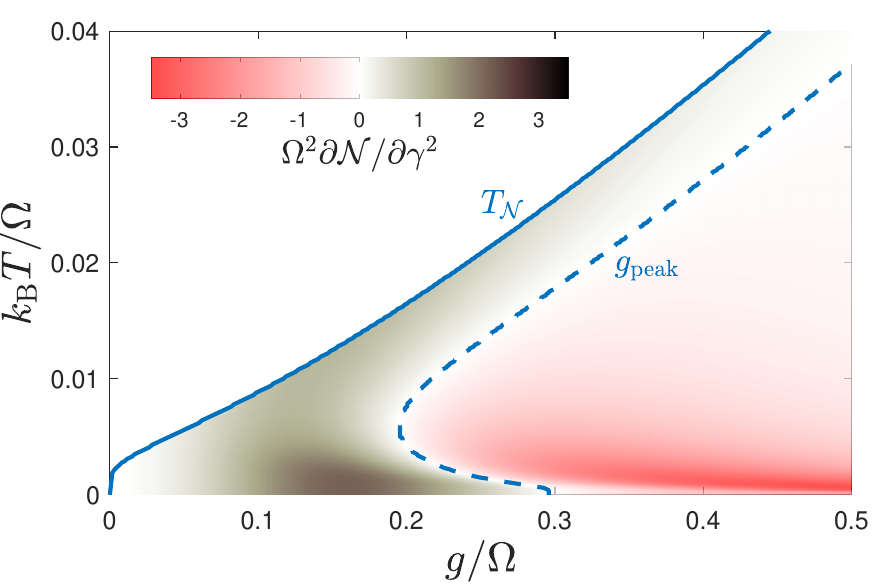}
\caption{\label{2spinB} Lowest order correction to entanglement due to reservoir spectral density broadening at strong coupling, quantified via $\partial\mathcal{N}/\partial\gamma^2|_{\gamma=0}$ evaluated from Eq.~\eqref{MFG2}. Entanglement increases with $\gamma$ ($\partial\mathcal{N}/\gamma^2>0$) in the gray region and decreases with $\gamma$ ($\partial\mathcal{N}/\partial\gamma^2<0$) in the red region. The boundary between these two regions follows $g_\mathrm{peak}$ (blue dashed line). Blue solid line indicates $T_\mathcal{N}$. Results are for $\omega_z/\Omega=0.02$ and $\epsilon=0$.}
\end{figure}

We can evaluate the effect of spectral density broadening perturbatively in the strong coupling regime using Eq.~\eqref{MFG2}, where $\lambda_\mathrm{eff}^2=8\pi g^2\gamma^2/(\omega_z\Omega^3)$. We quantify the effect of broadening on entanglement via $\partial \mathcal{N}/\partial \gamma^2|_{\gamma=0}$, see Fig.~\ref{2spinB}, with the partial derivative evaluated with $g$ and $\Omega$ fixed. For small to moderate $g/\Omega$, spectral density broadening increases entanglement ($\partial\mathcal{N}/\partial\gamma^2|_{\gamma=0}>0$), whereas for large $g/\Omega$, broadening decreases entanglement ($\partial\mathcal{N}/\partial\gamma^2|_{\gamma=0}<0$). The boundary between these two regimes coincides with $g_\mathrm{peak}$, suggesting an equivalence between increasing $g$ and increasing $\gamma$. For the result Eq.~\eqref{neganapprox} this equivalence can be seen directly, as here $\rho_\mathrm{MF}^\mathrm{weak}$ depends only on $g$ and $\gamma$ through the single parameter $Q$.

\begin{figure}
\includegraphics[trim=0cm 9.5cm 0cm 10cm,clip=true,width=0.48\textwidth]{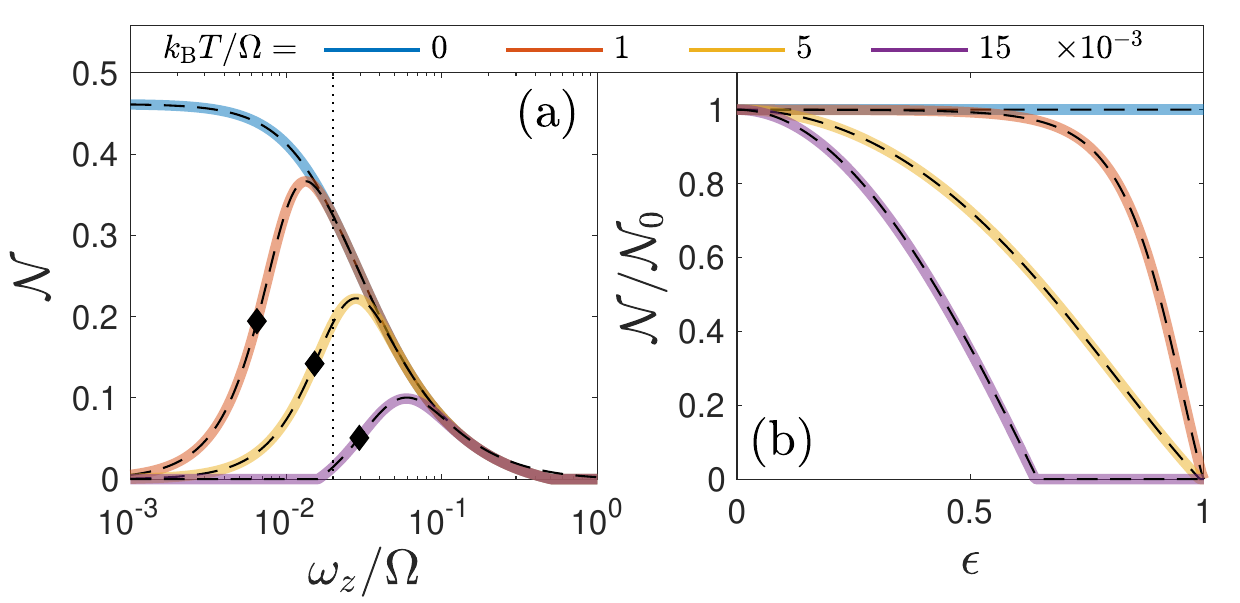}
\caption{\label{2spinC} (a) Negativity as a function of qubit level spacing $\omega_z$ with $\epsilon=0$. At $T=0$ the entanglement increases monotonically with decreasing $\omega_z/\Omega$. For a fixed $T>0$, thermal fluctuations increase with decreasing $\omega_z/\Omega$ (which decreases $\Delta$), resulting in a peak in entanglement. Diamonds mark the level spacing that gives $\Delta=k_\mathrm{B}T$, at which point $\mathcal{N}$ is approximately half its peak value. Vertical dotted line indicates $\omega_z/\Omega=0.02$. (b) Negativity as a function of asymmetry $\epsilon$ with $\omega_z/\Omega=0.02$, which is scaled by its $\epsilon=0$ value $\mathcal{N}_0$ for ease of comparison. At $T=0$ the entanglement is independent of $\epsilon$. For a fixed $T>0$, thermal fluctuations increase with increasing $\epsilon$, and hence $\mathcal{N}$ decreases. Solid lines in (a) and (b) are obtained from $\rho_\mathrm{MF}^\mathrm{narrow}$ [Eq.~\eqref{eq:rhogam0}], black dashed lines are obtained from numerical diagonalisation of $\hat{H}_\mathrm{S}^\mathrm{RC}$. All results are for $\gamma=0$ and $g/\Omega=0.2$.}
\end{figure}

\subsection{Effect of qubit level-spacing and asymmetry}\label{sec:wz}
Finally, for completeness we show in Fig.~\ref{2spinC} the effect of varying the qubit level-spacing $\omega_z$ and the asymmetry $\epsilon$ in the limit $\gamma\rightarrow 0$ [Eq.~\eqref{eq:rhogam0}]. At zero temperature, entanglement increases with decreasing $\omega_z/\Omega$ as the ground state of $\hat{H}_\mathrm{S}^\mathrm{eff}$ approaches a Bell state [see Sec.~\ref{sec:zerogam}]. This is insensitive to the asymmetry $\epsilon$. Almost identical behavior occurs at finite temperature for sufficiently large $\omega_z/\Omega$, as thermal fluctuations are frozen out. Reducing $\omega_z/\Omega$ reduces the energy gap $\Delta$ between the ground and first excited state of $\hat{H}_\mathrm{S}^\mathrm{eff}$ [Eq.~\eqref{eq:gap}], and hence the sensitivity to thermal fluctuations increases. For $T>0$, this results in a peak in entanglement at a small but non-zero value of $\omega_z/\Omega$. The energy gap $\Delta$ also decreases with increasing $\epsilon$, and therefore so too does $T>0$ entanglement, see Fig.~\ref{2spinC}(b). In particular, when $\epsilon=1$ (which gives $\Delta=0$), entanglement is zero for all $T>0$.

\section{Discussion}

We have derived analytic expressions for the mean-force Gibbs state of two qubits strongly coupled to a bosonic reservoir, valid to first order in either the system-reservoir coupling strength [Eq.~\eqref{MFG1}] or the reservoir spectral density width [Eq.~\eqref{MFG2}]. Using these results, we provide a comprehensive characterization of reservoir-mediated two-qubit entanglement in thermal equilibrium. Considering first coupling to a single reservoir mode, we show that entanglement is largest at zero temperature and intermediate system-reservoir coupling strengths. Thermal fluctuations and entanglement with the reservoir reduce the purity of the mean-force Gibbs state and the two-qubit entanglement. We show that broadening of the reservoir spectral density enhances entanglement compared to that mediated by a single reservoir mode as long as the system-reservoir coupling strength is not too large. We have focused our discussion on systems of two qubits, although generalizing the weak-coupling result Eq.~\eqref{MFG1} to an arbitrary number of qubits is straightforward (see Appendix~\ref{app:MFG1}).

 Exploring stronger couplings with broad spectral densities could be done numerically by extending the reaction coordinate mapping~\cite{hughes2009,hughes2009b}, or using hierarchical equations of motion~\cite{tanimura1989,tanimura1990,xu2007,hartmann2020}, density-matrix renormalization group~\cite{prior2010,chin2010,woods2015,chiu2022} or path integral~\cite{lee2019} methods. Our results provide a useful benchmark to compare with these and other numerical methods. An interesting area for future work is to consider spatially separated qubits, which results in interactions with multiple spatial modes of the reservoir~\cite{lehmberg1970,agarwal1970,javanainen1995,ruostekoski1997}. This can significantly alter entanglement~\cite{zell2009} and has relevance to non-local entanglement harvesting in quantum field theory~\cite{valentini1991,reznik2003,reznik2005,salton2015,pozas2015}. Our results also pave the way for exploring reservoir-mediated interactions in other systems, for example impurities in bosonic reservoirs~\cite{anhtai2024,gomez2024}.

\begin{acknowledgements}
This research was supported by the Australian Research Council Centre of Excellence for Engineered Quantum Systems (EQUS, CE170100009), the Australian Research Council Discovery Project DP260103158, and the Australian government Department of Industry, Science, and Resources via the Australia-India Strategic Research Fund (AIRXIV000025). FC and JA acknowledge funding from the Foundational Questions Institute Fund (FQXi-IAF19-01). FC also acknowledges support from EPSRC Standard Research grant EP/Z534250/1. JA gratefully acknowledges funding from EPSRC (EP/R045577/1) and from UKRI for projects ``QCS'' and ``Good enough''. JA thanks the Royal Society for support.
\end{acknowledgements}

\appendix

\begin{widetext}

\section{Perturbative MFG state}\label{app:perturb}
Here we present the framework for calculating the MFG state perturbatively in the system-reservoir coupling strength. Consider a system (Hamiltonian $\hat{H}_\mathrm{S}$) coupled to a bosonic reservoir via a system operator $\hat{X}$. The  Hamiltonian describing the system and reservoir is $\hat{H}_\mathrm{tot}=\hat{H}_\mathrm{S}+\hat{H}_\mathrm{R}+\lambda\hat{R}\hat{X}$ with $\hat{H}_\mathrm{R}=\int_0^\infty \omega \hat{b}^\dagger(\omega)\hat{b}(\omega)\,d\omega$ and $\hat{R}=\int_0^\infty \sqrt{J(\omega)}(\hat{b}(\omega)+\hat{b}^\dagger(\omega))\,d\omega$. Here $\hat{b}(\omega)$ are bosonic lowering operators ($[\hat{b}(\omega),\hat{b}^\dagger(\omega')]=\delta(\omega-\omega^\prime)$), $J(\omega)$ is the reservoir spectral density and $\lambda$ is the system-reservoir coupling strength. Although we adopt the same notation as Eq.~\eqref{H}, the system here is a generic system.

For small $\lambda$, Eq.~\eqref{MFG0} can be expanded to second order in $\lambda$ and the trace over the reservoir can be done analytically, assuming a thermal (i.e.\ Bose-Einstein) occupation of the bosonic reservoir modes $\langle \hat{b}^\dagger(\omega)\hat{b}(\omega)\rangle_\beta=1/(e^{\beta\omega}-1)$. This gives~\cite{cresser2021}
\begin{equation}\label{MFGweak}
\begin{split}
\rho_\mathrm{MF}=\rho_\mathrm{G}+&\lambda^2\beta\sum_n\rho_\mathrm{G}(\hat{X}_n\hat{X}_n^\dagger-\operatorname{Tr}[\rho_\mathrm{G}\hat{X}_n\hat{X}_n^\dagger]) D_\beta(\omega_n)+\lambda^2\sum_n[\hat{X}_n^\dagger,\rho_\mathrm{G}\hat{X}_n]\frac{dD_\beta(\omega_n)}{d\omega_n}\\
+&\lambda^2\sum_{\substack{m,n\\(m\ne n)}}\left([\hat{X}_m,\hat{X}_n^\dagger\rho_\mathrm{G}]-[\hat{X}_m^\dagger,\rho_\mathrm{G}\hat{X}_n]\right)\frac{D_\beta(\omega_n)}{\omega_m-\omega_n}.
\end{split}
\end{equation}
The operators $\hat{X}_n$ satisfy $[\hat{H}_\mathrm{S},\hat{X}_n]=\omega_n\hat{X}_n$ and $\sum_n\hat{X}_n=\hat{X}$, with $\omega_m\ne\omega_n$ for $m\ne n$. The coefficients
\begin{equation}\label{lambShift}
D_\beta(\omega)=\mathcal{P}\int_0^\infty\,J(\omega')\frac{\omega'+\omega\coth \left(\frac{1}{2}\beta \omega'\right)}{(\omega')^2-\omega^2}\,d\omega',
\end{equation}
with $\mathcal{P}\int$ a principal value integral, arise from tracing out the reservoir. These are related to an integral transform of imaginary-time temporal reservoir correlations,
\begin{equation}
D_\beta(\omega)+e^{\beta\omega_n}D_\beta(-\omega)=\int_0^\beta du\, e^{u\omega}\frac{\operatorname{Tr}[e^{-\beta\hat{H}_\mathrm{R}}\hat{R}(-iu)\hat{R}(0)]}{{\operatorname{Tr}e^{-\beta\hat{H}_\mathrm{R}}}}
\end{equation}
where
\begin{equation}
\frac{\operatorname{Tr}[e^{-\beta\hat{H}_\mathrm{R}}\hat{R}(t)\hat{R}(0)]}{\operatorname{Tr}e^{-\beta\hat{H}_\mathrm{R}}}=\int_0^\infty J(\omega)\frac{e^{i\omega t}+e^{\beta\omega}e^{-i\omega t}}{e^{\beta\omega}-1}\,d\omega
\end{equation}
are the temporal reservoir correlations in thermal equilibrium. Explicit calculations of Eq.~\eqref{lambShift} for the spectral densities used in this work [Eq.~\eqref{specden} and Eq.~\eqref{JRC}] are presented in Appendix~\ref{app:lamb}.

Equation~\eqref{MFGweak} is valid for~\cite{cresser2021}
\begin{equation}\label{validity}
\lambda^2\beta\left|\sum_n \operatorname{Tr}[\rho_\mathrm{G}\hat{X}_n\hat{X}_n^\dagger]D_\beta(\omega_n)\right|\ll 1.
\end{equation}
Equation~\eqref{MFGweak} is also valid for $\lambda\ll 1$ and $\beta=\infty$, in which case the term $\propto \beta$ in Eq.~\eqref{MFGweak} vanishes~\cite{cresser2021}. Equation~\eqref{MFGweak} can be used with Eq.~\eqref{H} for small $\lambda$ to give Eq.~\eqref{MFG1}. By modifying the system-reservoir demarcation using the reaction-coordinate mapping, Eq.~\eqref{MFGweak} can also be used for narrow reservoir spectral density to give Eq.~\eqref{MFG2}.

\section{Derivation of Eq.~\eqref{MFG1}}\label{app:MFG1}
We first evaluate Eq.~\eqref{MFGweak} for weak coupling between the spins and all reservoir modes. The system Hamiltonian is  $\hat{H}_\mathrm{S}=\frac{1}{2}(1+\epsilon)\omega_z\hat{\sigma}_z^{(1)}+\frac{1}{2}(1-\epsilon)\omega_z\hat{\sigma}_z^{(2)}$ and the coupling is $\hat{X}=\hat{S}_x$ (see Eq.~\eqref{H}). This gives $\hat{X}_{\pm n}=\frac{1}{4}(\hat{\sigma}_x^{(n)}\pm i\hat{\sigma}_y^{(n)})$ with $\omega_{\pm 1}=\pm \omega_z(1+\epsilon)$ and $\omega_{\pm 2}=\pm \omega_z(1-\epsilon)$. The terms in Eq.~\eqref{MFGweak} are

\begin{equation}\label{dressingCalc}
\begin{split}
\beta\left(\rho_\mathrm{G} X_{\pm n}X_{\pm n}^\dagger-\rho_\mathrm{G}\operatorname{Tr}[\rho_\mathrm{G} X_{\pm n}X_{\pm n}^\dagger]\right)D_\beta(\omega_{\pm n})&=\pm \frac{1}{4}\beta p_n p_{-n}\hat{\sigma}_z^{(n)}\tau_{m\ne n}D_\beta(\omega_{\pm n})\\
\left([X_{\pm n}^\dagger,\rho_\mathrm{G}X_{\pm n}]\right)\frac{dD_\beta(\omega_{\pm n})}{d\omega_{\pm n}}&=\mp \frac{1}{4}p_{\pm n}\hat{\sigma}_z^{(n)}\tau_{m\ne n}\frac{dD_\beta(\omega_{\pm n})}{d\omega_{\pm n}}\\
[X_{\mp n},X_{\pm n}^\dagger\rho_\mathrm{G}]-[X_{\mp n}^\dagger,\rho_\mathrm{G}X_{\pm n}]&=0\\
\left([X_2,X_1^\dagger\rho_\mathrm{G}]-[X_2^\dagger,\rho_\mathrm{G}X_ 1]\right)\frac{D_\beta(\omega_1)}{\omega_2-\omega_1}+1\leftrightarrow 2&=-\frac{1}{8\omega_z\epsilon}\left[(p_{-2}-p_2)p_1D_\beta(\omega_1)-(p_{-1}-p_1)p_2D_\beta(\omega_2)\right]\hat{C}_+\\
\left([X_{-2},X_1^\dagger\rho_\mathrm{G}]-[X_{-2}^\dagger,\rho_\mathrm{G}X_ 1]\right)\frac{D_\beta(\omega_1)}{\omega_{-2}-\omega_1}+1\leftrightarrow 2&=\frac{1}{8\omega_z}\left[(p_{-2}-p_2)p_1D_\beta(\omega_1)+(p_{-1}-p_1)p_2D_\beta(\omega_2)\right]\hat{C}_-\\
\left([X_2,X_{-1}^\dagger\rho_\mathrm{G}]-[X_2^\dagger,\rho_\mathrm{G}X_ {-1}]\right)\frac{D_\beta(\omega_{-1})}{\omega_2-\omega_{-1}}+1\leftrightarrow 2&=\frac{1}{8\omega_z}\left[(p_{-2}-p_2)p_{-1}D_\beta(\omega_{-1})+(p_{-1}-p_1)p_{-2}D_\beta(\omega_{-2})\right]\hat{C}_-\\
\left([X_{-2},X_{-1}^\dagger\rho_\mathrm{G}]-[X_{-2}^\dagger,\rho_\mathrm{G}X_ {-1}]\right)\frac{D_\beta(\omega_{-1})}{\omega_{-2}-\omega_{-1}}+1\leftrightarrow 2&=-\frac{1}{8\omega_z\epsilon}\left[(p_{-2}-p_2)p_{-1}D_\beta(\omega_{-1})-(p_{-1}-p_1)p_{-2}D_\beta(\omega_{-2})\right]\hat{C}_+
\end{split}
\end{equation}
Here $p_{\pm n}=\frac{1}{2}\operatorname{sech}(\beta\omega_n)e^{-\beta\omega_{\pm n}/2}$, $\tau_n=2\cosh(\beta\omega_n/2)e^{-\frac{\beta\omega_n}{2}\hat{\sigma}_z^{(n)}}$ and $\hat{C}_\pm=\frac{1}{2}(\hat{\sigma}_x^{(1)}\hat{\sigma}_x^{(2)}\pm\hat{\sigma}_y^{(1)}\hat{\sigma}_y^{(2)})$.

The first two lines in Eq.~\eqref{dressingCalc} are single-qubit terms while the remaining terms describe correlations. Assembling the terms in Eq.~\eqref{dressingCalc} gives
\begin{equation}\label{eq:rhoweakderive}
\begin{split}
    \rho_\mathrm{MF}^\mathrm{weak}&=\rho_\mathrm{G}-\left(\frac{\partial\theta_1}{\partial\omega_1}-\frac{\partial\theta_{-1}}{\partial\omega_{-1}}\right)\hat{\sigma}_z^{(1)}\tau_2-\left(\frac{\partial\theta_2}{\partial\omega_2}-\frac{\partial\theta_{-2}}{\partial\omega_{-2}}\right)\tau_1\hat{\sigma}_z^{(2)}\\
    &\phantom{=}-\frac{(p_{-2}-p_2)(\theta_1+\theta_{-1})-(p_{-1}-p_1)(\theta_2+\theta_{-2})}{2\omega_z\epsilon}\hat{C}_++\frac{(p_{-2}-p_2)(\theta_1+\theta_{-1})+(p_{-1}-p_1)(\theta_2+\theta_{-2})}{2\omega_z}\hat{C}_-
\end{split}
\end{equation}
where $\theta_n=\frac{\lambda^2}{4}p_nD_\beta(\omega_n)$. For $\epsilon\rightarrow 0$ this becomes
\begin{equation}
    \rho_\mathrm{MF}^\mathrm{weak}=\rho_\mathrm{G}-\frac{d\theta}{d\omega_z}\left(\hat{\sigma}_z^{(1)}\tau_2+\tau_1\hat{\sigma}_z^{(2)}\right)+\left[2\beta p_+p_-\theta-(p_--p_+)\frac{d\theta}{d\omega_z}\right]\hat{C}_++\frac{(p_--p_+)\theta}{\omega_z}\hat{C}_-
\end{equation}
with $p_\pm=\frac{1}{2}\operatorname{sech}(\beta\omega_z)e^{\mp\beta\omega_z/2}$ and $\theta=\frac{\lambda^2}{4}(p_+D_\beta(\omega_z)+p_-D_\beta(-\omega_z))$, which is Eq.~\eqref{MFG1}.

The derivation for $\rho_\mathrm{MF}^\mathrm{weak}$ is almost identical for $N>2$. The final result takes the form
\begin{equation}\label{MFGN}
    \rho_\mathrm{MF}^\mathrm{weak}=\rho_\mathrm{G}+\lambda^2\sum_{n=1}^N\rho^{(n)}+\lambda^2\sum_{m=n+1}^N\sum_{n=1}^{N-1}\rho^{(n,m)}.
\end{equation}
Here $\rho^{(n)}$ are single-qubit terms for qubit $n$ with the remaining qubits $m\ne n$ in a thermal state (first two lines of Eq.~\eqref{dressingCalc} with $\tau_m$ generalized to a thermal state of all qubits $m\ne n$). The $\rho^{(n,m)}$ are two-qubit terms for qubits $n$ and $m$ (remaining lines of Eq.~\eqref{dressingCalc} with qubits $\ell\ne n,m$ in a thermal state). For qubits with identical level spacings, Eq.~\eqref{MFGN} is
\begin{equation}\label{MFGN}
    \rho_\mathrm{MF}^\mathrm{weak}=\rho_\mathrm{G}-\frac{d\theta}{d\omega_z}\sum_{n=1}^N \hat{\sigma}_z^{(n)}\phi_n+\sum_{m=n+1}^N\sum_{n=1}^{N-1}\left\{\left[2\beta p_+p_-\theta-(p_--p_+)\frac{d\theta}{d\omega_z}\right]\hat{C}_+^{(m,n)}+\frac{(p_--p_+)\theta}{\omega_z}\hat{C}_-^{(m,n)}\right\}\phi_{m,n}
\end{equation}
with $\hat{C}_\pm^{(m,n)}=\frac{1}{2}(\hat{\sigma}_x^{(m)}\hat{\sigma}_x^{(n)}\pm\hat{\sigma}_y^{(m)}\hat{\sigma}_y^{(n)})$ and $\phi_n$($\phi_{m,n}$) the  Gibbs state $\rho_G$ with the density matrix for qubits $\ell\ne n$($\ell\ne m,n$) replaced by the identity. The two-qubit reduced density matrix computed from Eq.~\eqref{MFGN} is Eq.~\eqref{eq:rhoweakderive}. Identifying effects beyond two-body in an ensemble of $N>2$ qubits requires an expansion beyond order $O(\lambda^2)$.

For $N$ qubits with identical level spacings, the validity criterion~\eqref{validity} is
\begin{equation}\label{validity2}
    \frac{N\lambda^2\beta}{4}\left[p_+D_\beta(\omega_z)+p_-D_\beta(-\omega_z)\right]\ll 1.
\end{equation}
The left side of the inequality~\eqref{validity2} grows proportional to $N$, and therefore remaining in the weak-coupling regime as $N$ increases requires $\lambda$ to decrease as $\lambda\propto \frac{1}{\sqrt{N}}$. In the regime $\omega_z\ll\omega_0$, with $\omega_0$ characterizing the position of the peak of $J(\omega)$, we have $D_\beta(\pm\omega_z)\sim Q/\lambda^2$ and the criterion~\eqref{validity2} becomes
\begin{equation}
    \frac{N\beta Q}{4}\ll 1.
\end{equation}
We reiterate that Eq.~\eqref{MFGN} is also valid for $\beta=\infty$ as long as $N\lambda^2=NQ/\omega_z\ll 1$.

\section{Derivation of $\delta\rho$ in Eq.~\eqref{MFG2}}\label{app:MFG2}
We next evaluate Eq.~\eqref{MFGweak} for narrow reservoir spectral density. In the polaron frame, the total Hamiltonian is $\hat{H}_\mathrm{tot}^\mathrm{P}$ [Eq.~\eqref{Heff1}], and the MFG state is expanded in the effective coupling strength $\lambda_\mathrm{eff}$. The system has Hamiltonian
\begin{equation}
    \hat{H}_\mathrm{S}^\mathrm{eff}=e^{-\frac{g^2}{2\Omega^2}}\left[\frac{1}{2}(1+\epsilon)\omega_z\hat{\sigma}_z^{(1)}+\frac{1}{2}(1-\epsilon)\omega_z\hat{\sigma}_z^{(2)}\right]-\frac{g^2}{\Omega}\hat{S}_x^2
\end{equation}
and is coupled to the reservoir via $\hat{X}=\hat{S}_x$.

Focusing first on the case $\epsilon=0$, the eigenvectors of $\hat{H}_\mathrm{S}^\mathrm{eff}$ are
\begin{equation}\label{Heffev}
\begin{array}{ll}
\ket{1}=\frac{1}{\sqrt{2}}(\ket{eg}+\ket{ge},&e_1=-2\tilde{g}\\
\ket{2}=\frac{1}{\sqrt{2}}(\ket{eg}-\ket{ge}),&e_2=0\\
\ket{3}=\frac{1}{\sqrt{1+\mu_+^2}}(\ket{gg}+\mu_+\ket{ee}),&e_3=-\tilde{g}-\sqrt{\tilde{g}^2+\tilde{\omega}_z^2}\\
\ket{4}=\frac{1}{\sqrt{1+\mu_-^2}}(\ket{gg}+\mu_-\ket{ee}),&e_4=-\tilde{g}+\sqrt{\tilde{g}^2+\tilde{\omega}_z^2}\\
\end{array}
\end{equation}
with $\tilde{\omega}_z=e^{-g^2/(2\Omega^2)}\omega_z$, $\tilde{g}=g^2/(2\Omega)$, and $\mu_\pm=-\tilde{\omega}_z/\tilde{g}\pm\sqrt{1+\tilde{\omega}_z^2/\tilde{g}^2}$. This gives
\begin{equation}
    \hat{S}_x=A_+(\ket{1}\bra{3}+\ket{3}\bra{1})+A_-(\ket{1}\bra{4}+\ket{4}\bra{1})
\end{equation}
with $A_\pm=\sqrt{\frac{1+\mu_\pm^2}{2}}\frac{1-\mu_\mp}{\mu_\pm-\mu_\mp}$. The operators $\hat{X}_n$ and frequencies $\omega_n$ are $\hat{X}_1=A_+\ket{1}\bra{3}$, $\hat{X}_{-1}=A_+\ket{3}\bra{1}$, $\hat{X}_2=A_-\ket{1}\bra{4}$ and $\hat{X}_{-2}=A_-\ket{4}\bra{1}$, with $\omega_{\pm 1}=\pm(e_1-e_3)$ and $\omega_{\pm 2}=\pm(e_1-e_4)$. This gives,
\begin{equation}\label{MFGterms}
\begin{split}
    \rho_\mathrm{G}^\mathrm{eff}(\hat{X}_1\hat{X}_1^\dagger-\operatorname{Tr}[\rho_\mathrm{G}^\mathrm{eff}\hat{X}_1\hat{X}_1^\dagger])&= A_+^2(p_1\ket{1}\bra{1}-p_1\rho_\mathrm{G}^\mathrm{eff})\\
    \rho_\mathrm{G}^\mathrm{eff}(\hat{X}_{-1}\hat{X}_{-1}^\dagger-\operatorname{Tr}[\rho_\mathrm{G}^\mathrm{eff}\hat{X}_{-1}\hat{X}_{-1}^\dagger])&=A_+^2(p_3\ket{3}\bra{3}-p_3\rho_\mathrm{G}^\mathrm{eff})\\
    \rho_\mathrm{G}^\mathrm{eff}(\hat{X}_2\hat{X}_2^\dagger-\operatorname{Tr}[\rho_\mathrm{G}^\mathrm{eff}\hat{X}_2\hat{X}_2^\dagger])&= A_-^2(p_1\ket{1}\bra{1}-p_1\rho_\mathrm{G}^\mathrm{eff})\\
    \rho_\mathrm{G}^\mathrm{eff}(\hat{X}_{-2}\hat{X}_{-2}^\dagger-\operatorname{Tr}[\rho_\mathrm{G}^\mathrm{eff}\hat{X}_{-2}\hat{X}_{-2}^\dagger])&=A_-^2(p_4\ket{4}\bra{4}-p_4\rho_\mathrm{G}^\mathrm{eff})\\
    [\hat{X}_1^\dagger,\rho_\mathrm{G}^\mathrm{eff}\hat{X}_1]&=A_+^2p_1(\ket{3}\bra{3}-\ket{1}\bra{1})\\
    [\hat{X}_{-1}^\dagger,\rho_\mathrm{G}^\mathrm{eff}\hat{X}_{-1}]&=-A_+^2p_3(\ket{3}\bra{3}-\ket{1}\bra{1})\\
    [\hat{X}_2^\dagger,\rho_\mathrm{G}^\mathrm{eff}\hat{X}_2]&=A_-^2p_1(\ket{4}\bra{4}-\ket{1}\bra{1})\\
    [\hat{X}_{-2}^\dagger,\rho_\mathrm{G}^\mathrm{eff}\hat{X}_{-2}]&=-A_-^2p_4(\ket{4}\bra{4}-\ket{1}\bra{1})\\    [\hat{X}_2,\hat{X}_1^\dagger\rho_\mathrm{G}^\mathrm{eff}]&=-p_1A_+A_-\ket{3}\bra{4}\\
    [\hat{X}_1,\hat{X}_2^\dagger\rho_\mathrm{G}^\mathrm{eff}]&=-p_1A_+A_-\ket{4}\bra{3}\\
    [\hat{X}_{-2},\hat{X}_{-1}^\dagger\rho_\mathrm{G}^\mathrm{eff}]&=p_3A_+A_-\ket{4}\bra{3}\\
    [\hat{X}_{-1},\hat{X}_{-2}^\dagger\rho_\mathrm{G}^\mathrm{eff}]&=p_4A_+A_-\ket{3}\bra{4}
\end{split}
\end{equation}
with $\rho_\mathrm{G}^\mathrm{eff}=e^{-\beta\hat{H}_\mathrm{S}^\mathrm{eff}}/\operatorname{Tr}e^{-\beta\hat{H}_\mathrm{S}^\mathrm{eff}}\equiv\sum_{k=1}^4 p_k\ket{k}\bra{k}$ and the remaining $[\hat{X}_m,\hat{X}_n^\dagger\rho_\mathrm{G}^\mathrm{eff}]$ terms are zero.

Using Eq.~\eqref{Heffev} and Eq.~\eqref{MFGterms}, the MFG state Eq.~\eqref{MFGweak} can be evaluated in the polaron frame. To return to the original frame, we need to transform the MFG state using Eq.~\eqref{polaronTransform},
\begin{equation}\label{LtransformApp}
L[\hat{O}]=\hat{\mathcal{O}}+\frac{1}{2}\left(3+e^{-\frac{2g^2}{\Omega^2}}-4e^{-\frac{g^2}{2\Omega^2}}\right)\hat{S}_x^2\hat{\mathcal{O}}\hat{S}_x^2+\frac{1}{2}\left(1-e^{-\frac{2g^2}{\Omega^2}}\right)\hat{S}_x\hat{\mathcal{O}}\hat{S}_x-\left(1-e^{-\frac{g^2}{2\Omega^2}}\right)\left(\hat{S}_x^2\hat{\mathcal{O}}+\hat{\mathcal{O}}\hat{S}_x^2\right).
\end{equation}
To implement the transformation Eq.~\eqref{LtransformApp}, it is useful to change basis to $\{\ket{1},\ket{2},\ket{v},\ket{\tilde{v}}\}$, with $\ket{v}=\hat{S}_x\ket{1}=A_+\ket{3}+A_-\ket{4}=(\ket{ee}+\ket{gg})/\sqrt{2}$ and $\tilde{v}=A_-\ket{3}-A_+\ket{4}=(\ket{ee}-\ket{gg})/\sqrt{2}$. This gives $\hat{S}_x^2=\ket{1}\bra{1}+\ket{v}\bra{v}$ and,
\begin{equation}
    \begin{split}
        L[\ket{1}\bra{1}]&=\chi_+\ket{1}\bra{1}+\chi_-\ket{v}\bra{v}\\
        L[\ket{v}\bra{v}]&=\chi_+\ket{v}\bra{v}+\chi_-\ket{1}\bra{1}\\
        L[\ket{2}\bra{2}]&=\ket{2}\bra{2}\\
        L[\ket{\tilde{v}}\bra{\tilde{v}}]&=\ket{\tilde{v}}\bra{\tilde{v}}\\
        L[\ket{v}\bra{\tilde{v}}+\ket{\tilde{v}}\bra{v}]&=e^{-\frac{g^2}{2\Omega^2}}\left(\ket{v}\bra{\tilde{v}}+\ket{\tilde{v}}\bra{v}\right).
    \end{split}
\end{equation}
with $\chi_\pm=\frac{1}{2}(1\pm e^{-\frac{2g^2}{\Omega^2}})$. The final MFG state is therefore,
\begin{equation}\label{MFGeffFull}
\begin{split}
\rho_\mathrm{MF}^\mathrm{narrow}&=\left[\tilde{p}_1\chi_++\left(A_+^2\tilde{p}_3+A_-^2\tilde{p}_4+2A_+A_-C_{34}
        \right)\chi_-\right]\ket{1}\bra{1}+\left[\left(A_+^2\tilde{p}_3+A_-^2\tilde{p}_4+2A_+A_-C_{34}\right)\chi_++\tilde{p}_1\chi_-\right]\ket{v}\bra{v}\\
        &\phantom{=}+\tilde{p}_2\ket{2}\bra{2}+\left(A_-^2 \tilde{p}_3+A_+^2\tilde{p}_4-2 A_+A_-C_{34}\right)\ket{\tilde{v}}\bra{\tilde{v}}+e^{-\frac{g^2}{2\Omega^2}}\left[A_+A_-(\tilde{p}_3-\tilde{p}_4)-(A_+^2-A_-^2) C_{34}\right](\ket{v}\bra{\tilde{v}}+\ket{\tilde{v}}\bra{v})
\end{split}
\end{equation}
with $\tilde{p}_k=(1-C_1-C_3-C_4)p_k+C_k$ and
\begin{equation}
    \begin{split}
        C_1&=\lambda_\mathrm{eff}^2\left\{\beta A_+^2 p_1 D_\beta(\omega_1)+\beta A_-^2 p_1D_\beta(\omega_2)-A_+^2\left[p_1\frac{d D_\beta(\omega_1)}{d\omega_1}-p_3\frac{dD_\beta(\omega_{-1})}{d\omega_{-1}}\right]-A_-^2\left[p_1\frac{d D_\beta(\omega_2)}{d\omega_2}-p_4\frac{dD_\beta(\omega_{-2})}{d\omega_{-2}}\right]\right\}\\
        C_2&=0\\
        C_3&=\lambda_\mathrm{eff}^2\left\{\beta A_+^2 p_3 D_\beta(\omega_{-1})+A_+^2\left[p_1\frac{d D_\beta(\omega_1)}{d\omega_1}-p_3\frac{dD_\beta(\omega_{-1})}{d\omega_{-1}}\right]\right\}\\
        C_4&=\lambda_\mathrm{eff}^2\left\{\beta A_-^2 p_4D_\beta(\omega_{-2})+A_-^2\left[p_1\frac{d D_\beta(\omega_2)}{d\omega_2}-p_4\frac{dD_\beta(\omega_{-2})}{d\omega_{-2}}\right]\right\}\\
        C_{34}&=\frac{\lambda_\mathrm{eff}^2A_+A_-}{\omega_1-\omega_2}\left[p_1D_\beta(\omega_1)+p_3 D_\beta(\omega_{-1})-p_1 D_\beta(\omega_2)-p_4 D_\beta(\omega_{-2})\right]
    \end{split}
\end{equation}

For $\epsilon\ne 0$, the eigenvectors of $\hat{H}_\mathrm{S}^\mathrm{eff}$ are
\begin{equation}
\begin{array}{ll}
\ket{1}=\frac{1}{\sqrt{1+\kappa_+^2}}(\ket{eg}+\kappa_+\ket{ge}),&e_1=-\tilde{g}-\sqrt{\tilde{g}^2+\epsilon^2\tilde{\omega}_z^2}\\
\ket{2}=\frac{1}{\sqrt{1+\kappa_-^2}}(\ket{eg}+\kappa_-\ket{ge}),&e_2=-\tilde{g}+\sqrt{\tilde{g}^2+\epsilon^2\tilde{\omega}_z^2}\\
\ket{3}=\frac{1}{\sqrt{1+\mu_+^2}}(\ket{gg}+\mu_+\ket{ee}),&e_3=-\tilde{g}-\sqrt{\tilde{g}^2+\tilde{\omega}_z^2}\\
\ket{4}=\frac{1}{\sqrt{1+\mu_-^2}}(\ket{gg}+\mu_-\ket{ee}),&e_4=-\tilde{g}+\sqrt{\tilde{g}^2+\tilde{\omega}_z^2}\\
\end{array}
\end{equation}
with $\kappa_\pm=-\epsilon\tilde{\omega}_z/\tilde{g}\pm\sqrt{1+\epsilon^2\tilde{\omega}_z^2/\tilde{g}^2}$. This gives
\begin{equation}
    \hat{S}_x=A_+B_+(\ket{1}\bra{3}+\ket{3}\bra{1})+A_-B_+(\ket{1}\bra{4}+\ket{4}\bra{1})+A_+B_-(\ket{2}\bra{3}+\ket{3}\bra{2})+A_-B_-(\ket{2}\bra{4}+\ket{4}\bra{2})
\end{equation}
with $B_\pm=\sqrt{\frac{1+\kappa_\pm^2}{2}}\frac{1-\kappa_\mp}{\kappa_\pm-\kappa_\mp}$. An analytic expression for $\rho_\mathrm{MF}^\mathrm{narrow}$ can be obtained by following the same procedure that leads to Eq.~\eqref{MFGeffFull}. This requires calculating a large number of terms as in Eq.~\eqref{MFGterms}. In practice, a systematic numerical calculation of these terms and their transformation under Eq.~\eqref{polaronTransform} is straightforward.

\section{Calculation of $D_\beta(\omega)$}\label{app:lamb}

Here we evaluate the integral in Eq.~\eqref{lambShift} for the spectral densities considered in this work [Eq.~\eqref{specden} and Eq.~\eqref{JRC}]. The spectral densities are odd functions of $\omega$ and hence the integrand in Eq.~\eqref{lambShift} is even. Equation~\eqref{lambShift} can therefore be written as
\begin{equation}\label{lambShiftC}
\begin{split}
D_\beta(\omega_n)&=\frac{1}{2}\mathcal{P}\int_{-\infty}^\infty J(\omega)\frac{\omega+\omega_n\coth\left(\frac{1}{2}\beta\omega\right)}{\omega^2-\omega_n^2}\,d\omega\\
&=\frac{1}{2}\oint_\mathcal{C} J(z)\frac{z+\omega_n\coth\left(\frac{1}{2}\beta z\right)}{z^2-\omega_n^2}\,dz-\frac{1}{2}\int_{c_+} J(z)\frac{z+\omega_n\coth\left(\frac{1}{2}\beta z\right)}{z^2-\omega_n^2}\,dz-\frac{1}{2}\int_{c_-} J(z)\frac{z+\omega_n\coth\left(\frac{1}{2}\beta z\right)}{z^2-\omega_n^2}\,dz
\end{split}
\end{equation}
with $\mathcal{C}$ a semicircle contour with arc at $z\rightarrow\infty$ and diagonal along the real axis, excluding the poles at $\pm \omega_n$, and $c_\pm$ are infinitesimal semicircle arcs around the $\pm\omega_n$ poles. Using the residue theorem, Eq.~\eqref{lambShiftC} gives
\begin{equation}\label{lambShiftR}
    D_\beta(\omega_n)=i\pi\sum_k \mathcal{R}_k-\frac{i\pi}{2}(r_1+r_{-1}),
\end{equation}
with $\mathcal{R}_k$ the residues of $J(z)[z+\omega_n\coth\left(\frac{1}{2}\beta z\right)]/(z^2-\omega_n^2)$ contained in $\mathcal{C}$ (i.e.\ with positive imaginary part) and $r_{\pm 1}$ the residues from the circumvented poles at $\pm \omega_n$. We have $r_1=-r_{-1}$, since $J(z)[z+\omega_n\coth\left(\frac{1}{2}\beta z\right)]$ is even, and hence the $r_{\pm 1}$ residues cancel. Furthermore, for the results presented in this paper, the temperature is sufficiently low that $D_\beta(\omega_n)\approx D_\infty(\omega_n)$. Therefore the evaluation of Eq.~\eqref{lambShiftR} reduces to evaluating the residues of $J(z)[z+\omega_n\operatorname{sgn}(\operatorname{Re}z)]/(z^2-\omega_n^2)$, where we have used that $\lim_{\beta\rightarrow\infty}\coth(\frac{1}{2}\beta z)=\operatorname{sgn}(\operatorname{Re}z)$, with $\operatorname{Re}z$ the real part of $z$ and $\operatorname{sgn}()$ the sign function.

For the spectral density Eq.~\eqref{specden},
\begin{equation}
J(\omega)=\frac{1}{\pi}\frac{32\omega_z(\omega_0^2+\gamma^2)\gamma^3\omega^3}{[(\omega-\omega_0)^2+\gamma^2]^2[(\omega+\omega_0)^2+\gamma^2]^2},
\end{equation}
the poles inside $\mathcal{C}$ are $z_\pm=\pm\omega_0+i\gamma$ and are second order. The corresponding residues for $\beta=\infty$ are
\begin{equation}\label{R1}
\begin{split}
\mathcal{R}_\pm&=\lim_{z\rightarrow z_{\pm}}\frac{d}{dz}\left[(z-z_\pm)^2\frac{J(z)}{z\mp\omega_n}\right]\\
&=\frac{\omega_z(\omega_0^2+\gamma^2)\gamma}{\pi\omega_0^2}\left[\frac{z_\pm}{2(z_\pm\mp\omega_n)^2}\pm\frac{z_+^2+z_-^2}{(z_+^2-z_-^2)(z_\pm\mp\omega_n)}\right].
\end{split}
\end{equation}
The corresponding $D_\beta(\omega_n)$ are then computed using Eq.~\eqref{lambShiftR}, which gives
\begin{equation}\label{lambShift1}
    D_\beta(\omega_n)\approx D_\infty(\omega_n)=i\pi (\mathcal{R}_+-\mathcal{R}_+^*)=-2\pi \operatorname{Im} R_+
\end{equation}
with $\operatorname{Im}\mathcal{R}_+$ the imaginary part of $\mathcal{R}_+$ and we have used that $R_-=-R_+^*$.

For the spectral density Eq.~\eqref{JRC},
\begin{equation}
J_\mathrm{RC}(\omega)=\frac{1}{\pi}\frac{8\omega_z\gamma\omega^3}{\gamma^4+2\gamma^2(7\omega^2+\omega_0^2)+(\omega^2-\omega_0^2)^2},
\end{equation}
the poles inside $\mathcal{C}$ are $z_\pm=\pm\sqrt{\omega_0^2-7\gamma^2\pm 4\gamma\sqrt{-\omega_0^2+3\gamma^2}}$, assuming $7\gamma^2<\omega_0^2$. The corresponding residues for $\beta=\infty$ are
\begin{equation}
\begin{split}
\mathcal{R}_\pm&=\lim_{z\rightarrow z_{\pm}}(z-z_\pm)\frac{J_\mathrm{RC}(z)}{z\mp\omega_n}\\
&=\pm\frac{4\omega_z\gamma z_\pm^2}{\pi(z_+^2-z_-^2)(z_\pm\mp\omega_n)}.
\end{split}
\end{equation}
and
\begin{equation}
    D_\beta(\omega_n)\approx D_\infty(\omega_n)=-2\pi \operatorname{Im} R_+
\end{equation}

\end{widetext}


\begin{thebibliography}{123}%
\makeatletter
\providecommand \@ifxundefined [1]{%
 \@ifx{#1\undefined}
}%
\providecommand \@ifnum [1]{%
 \ifnum #1\expandafter \@firstoftwo
 \else \expandafter \@secondoftwo
 \fi
}%
\providecommand \@ifx [1]{%
 \ifx #1\expandafter \@firstoftwo
 \else \expandafter \@secondoftwo
 \fi
}%
\providecommand \natexlab [1]{#1}%
\providecommand \enquote  [1]{``#1''}%
\providecommand \bibnamefont  [1]{#1}%
\providecommand \bibfnamefont [1]{#1}%
\providecommand \citenamefont [1]{#1}%
\providecommand \href@noop [0]{\@secondoftwo}%
\providecommand \href [0]{\begingroup \@sanitize@url \@href}%
\providecommand \@href[1]{\@@startlink{#1}\@@href}%
\providecommand \@@href[1]{\endgroup#1\@@endlink}%
\providecommand \@sanitize@url [0]{\catcode `\\12\catcode `\$12\catcode
  `\&12\catcode `\#12\catcode `\^12\catcode `\_12\catcode `\%12\relax}%
\providecommand \@@startlink[1]{}%
\providecommand \@@endlink[0]{}%
\providecommand \url  [0]{\begingroup\@sanitize@url \@url }%
\providecommand \@url [1]{\endgroup\@href {#1}{\urlprefix }}%
\providecommand \urlprefix  [0]{URL }%
\providecommand \Eprint [0]{\href }%
\providecommand \doibase [0]{https://doi.org/}%
\providecommand \selectlanguage [0]{\@gobble}%
\providecommand \bibinfo  [0]{\@secondoftwo}%
\providecommand \bibfield  [0]{\@secondoftwo}%
\providecommand \translation [1]{[#1]}%
\providecommand \BibitemOpen [0]{}%
\providecommand \bibitemStop [0]{}%
\providecommand \bibitemNoStop [0]{.\EOS\space}%
\providecommand \EOS [0]{\spacefactor3000\relax}%
\providecommand \BibitemShut  [1]{\csname bibitem#1\endcsname}%
\let\auto@bib@innerbib\@empty
\bibitem [{\citenamefont {Breuer}\ and\ \citenamefont
  {Petruccione}(2002)}]{breuer2002}%
  \BibitemOpen
  \bibfield  {author} {\bibinfo {author} {\bibfnamefont {H.-P.}\ \bibnamefont
  {Breuer}}\ and\ \bibinfo {author} {\bibfnamefont {F.}~\bibnamefont
  {Petruccione}},\ }\href@noop {} {\emph {\bibinfo {title} {The theory of open
  quantum systems}}}\ (\bibinfo  {publisher} {Oxford University Press},\
  \bibinfo {address} {New York},\ \bibinfo {year} {2002})\BibitemShut {NoStop}%
\bibitem [{\citenamefont {Rivas}(2020)}]{rivas2020}%
  \BibitemOpen
  \bibfield  {author} {\bibinfo {author} {\bibfnamefont {A.}~\bibnamefont
  {Rivas}},\ }\bibfield  {title} {\bibinfo {title} {Strong coupling
  thermodynamics of open quantum systems},\ }\href
  {https://doi.org/10.1103/PhysRevLett.124.160601} {\bibfield  {journal}
  {\bibinfo  {journal} {Phys. Rev. Lett.}\ }\textbf {\bibinfo {volume} {124}},\
  \bibinfo {pages} {160601} (\bibinfo {year} {2020})}\BibitemShut {NoStop}%
\bibitem [{\citenamefont {Seifert}(2016)}]{seifert2016}%
  \BibitemOpen
  \bibfield  {author} {\bibinfo {author} {\bibfnamefont {U.}~\bibnamefont
  {Seifert}},\ }\bibfield  {title} {\bibinfo {title} {First and second law of
  thermodynamics at strong coupling},\ }\href
  {https://doi.org/10.1103/PhysRevLett.116.020601} {\bibfield  {journal}
  {\bibinfo  {journal} {Phys. Rev. Lett.}\ }\textbf {\bibinfo {volume} {116}},\
  \bibinfo {pages} {020601} (\bibinfo {year} {2016})}\BibitemShut {NoStop}%
\bibitem [{\citenamefont {Jarzynski}(2017)}]{jarzynski2017}%
  \BibitemOpen
  \bibfield  {author} {\bibinfo {author} {\bibfnamefont {C.}~\bibnamefont
  {Jarzynski}},\ }\bibfield  {title} {\bibinfo {title} {Stochastic and
  macroscopic thermodynamics of strongly coupled systems},\ }\href
  {https://doi.org/10.1103/PhysRevX.7.011008} {\bibfield  {journal} {\bibinfo
  {journal} {Phys. Rev. X}\ }\textbf {\bibinfo {volume} {7}},\ \bibinfo {pages}
  {011008} (\bibinfo {year} {2017})}\BibitemShut {NoStop}%
\bibitem [{\citenamefont {Perarnau-Llobet}\ \emph {et~al.}(2018)\citenamefont
  {Perarnau-Llobet}, \citenamefont {Wilming}, \citenamefont {Riera},
  \citenamefont {Gallego},\ and\ \citenamefont {Eisert}}]{llobet2018}%
  \BibitemOpen
  \bibfield  {author} {\bibinfo {author} {\bibfnamefont {M.}~\bibnamefont
  {Perarnau-Llobet}}, \bibinfo {author} {\bibfnamefont {H.}~\bibnamefont
  {Wilming}}, \bibinfo {author} {\bibfnamefont {A.}~\bibnamefont {Riera}},
  \bibinfo {author} {\bibfnamefont {R.}~\bibnamefont {Gallego}},\ and\ \bibinfo
  {author} {\bibfnamefont {J.}~\bibnamefont {Eisert}},\ }\bibfield  {title}
  {\bibinfo {title} {Strong coupling corrections in quantum thermodynamics},\
  }\href {https://doi.org/10.1103/PhysRevLett.120.120602} {\bibfield  {journal}
  {\bibinfo  {journal} {Phys. Rev. Lett.}\ }\textbf {\bibinfo {volume} {120}},\
  \bibinfo {pages} {120602} (\bibinfo {year} {2018})}\BibitemShut {NoStop}%
\bibitem [{\citenamefont {Bruch}\ \emph {et~al.}(2018)\citenamefont {Bruch},
  \citenamefont {Lewenkopf},\ and\ \citenamefont {von Oppen}}]{bruch2018}%
  \BibitemOpen
  \bibfield  {author} {\bibinfo {author} {\bibfnamefont {A.}~\bibnamefont
  {Bruch}}, \bibinfo {author} {\bibfnamefont {C.}~\bibnamefont {Lewenkopf}},\
  and\ \bibinfo {author} {\bibfnamefont {F.}~\bibnamefont {von Oppen}},\
  }\bibfield  {title} {\bibinfo {title} {Landauer-b\"uttiker approach to
  strongly coupled quantum thermodynamics: Inside-outside duality of entropy
  evolution},\ }\href {https://doi.org/10.1103/PhysRevLett.120.107701}
  {\bibfield  {journal} {\bibinfo  {journal} {Phys. Rev. Lett.}\ }\textbf
  {\bibinfo {volume} {120}},\ \bibinfo {pages} {107701} (\bibinfo {year}
  {2018})}\BibitemShut {NoStop}%
\bibitem [{\citenamefont {Strasberg}(2019)}]{strasberg2019}%
  \BibitemOpen
  \bibfield  {author} {\bibinfo {author} {\bibfnamefont {P.}~\bibnamefont
  {Strasberg}},\ }\bibfield  {title} {\bibinfo {title} {Repeated interactions
  and quantum stochastic thermodynamics at strong coupling},\ }\href
  {https://doi.org/10.1103/PhysRevLett.123.180604} {\bibfield  {journal}
  {\bibinfo  {journal} {Phys. Rev. Lett.}\ }\textbf {\bibinfo {volume} {123}},\
  \bibinfo {pages} {180604} (\bibinfo {year} {2019})}\BibitemShut {NoStop}%
\bibitem [{\citenamefont {Esposito}\ \emph
  {et~al.}(2015{\natexlab{a}})\citenamefont {Esposito}, \citenamefont {Ochoa},\
  and\ \citenamefont {Galperin}}]{esposito2015}%
  \BibitemOpen
  \bibfield  {author} {\bibinfo {author} {\bibfnamefont {M.}~\bibnamefont
  {Esposito}}, \bibinfo {author} {\bibfnamefont {M.~A.}\ \bibnamefont
  {Ochoa}},\ and\ \bibinfo {author} {\bibfnamefont {M.}~\bibnamefont
  {Galperin}},\ }\bibfield  {title} {\bibinfo {title} {Quantum thermodynamics:
  A nonequilibrium {G}reen's function approach},\ }\href
  {https://doi.org/10.1103/PhysRevLett.114.080602} {\bibfield  {journal}
  {\bibinfo  {journal} {Phys. Rev. Lett.}\ }\textbf {\bibinfo {volume} {114}},\
  \bibinfo {pages} {080602} (\bibinfo {year} {2015}{\natexlab{a}})}\BibitemShut
  {NoStop}%
\bibitem [{\citenamefont {Esposito}\ \emph
  {et~al.}(2015{\natexlab{b}})\citenamefont {Esposito}, \citenamefont {Ochoa},\
  and\ \citenamefont {Galperin}}]{esposito2015b}%
  \BibitemOpen
  \bibfield  {author} {\bibinfo {author} {\bibfnamefont {M.}~\bibnamefont
  {Esposito}}, \bibinfo {author} {\bibfnamefont {M.~A.}\ \bibnamefont
  {Ochoa}},\ and\ \bibinfo {author} {\bibfnamefont {M.}~\bibnamefont
  {Galperin}},\ }\bibfield  {title} {\bibinfo {title} {Nature of heat in
  strongly coupled open quantum systems},\ }\href
  {https://doi.org/10.1103/PhysRevB.92.235440} {\bibfield  {journal} {\bibinfo
  {journal} {Phys. Rev. B}\ }\textbf {\bibinfo {volume} {92}},\ \bibinfo
  {pages} {235440} (\bibinfo {year} {2015}{\natexlab{b}})}\BibitemShut
  {NoStop}%
\bibitem [{\citenamefont {Carrega}\ \emph {et~al.}(2016)\citenamefont
  {Carrega}, \citenamefont {Solinas}, \citenamefont {Sassetti},\ and\
  \citenamefont {Weiss}}]{carrega2016}%
  \BibitemOpen
  \bibfield  {author} {\bibinfo {author} {\bibfnamefont {M.}~\bibnamefont
  {Carrega}}, \bibinfo {author} {\bibfnamefont {P.}~\bibnamefont {Solinas}},
  \bibinfo {author} {\bibfnamefont {M.}~\bibnamefont {Sassetti}},\ and\
  \bibinfo {author} {\bibfnamefont {U.}~\bibnamefont {Weiss}},\ }\bibfield
  {title} {\bibinfo {title} {Energy exchange in driven open quantum systems at
  strong coupling},\ }\href {https://doi.org/10.1103/PhysRevLett.116.240403}
  {\bibfield  {journal} {\bibinfo  {journal} {Phys. Rev. Lett.}\ }\textbf
  {\bibinfo {volume} {116}},\ \bibinfo {pages} {240403} (\bibinfo {year}
  {2016})}\BibitemShut {NoStop}%
\bibitem [{\citenamefont {Campisi}\ \emph {et~al.}(2009)\citenamefont
  {Campisi}, \citenamefont {Talkner},\ and\ \citenamefont
  {H\"anggi}}]{campisi2009}%
  \BibitemOpen
  \bibfield  {author} {\bibinfo {author} {\bibfnamefont {M.}~\bibnamefont
  {Campisi}}, \bibinfo {author} {\bibfnamefont {P.}~\bibnamefont {Talkner}},\
  and\ \bibinfo {author} {\bibfnamefont {P.}~\bibnamefont {H\"anggi}},\
  }\bibfield  {title} {\bibinfo {title} {Fluctuation theorem for arbitrary open
  quantum systems},\ }\href {https://doi.org/10.1103/PhysRevLett.102.210401}
  {\bibfield  {journal} {\bibinfo  {journal} {Phys. Rev. Lett.}\ }\textbf
  {\bibinfo {volume} {102}},\ \bibinfo {pages} {210401} (\bibinfo {year}
  {2009})}\BibitemShut {NoStop}%
\bibitem [{\citenamefont {Gallego}\ \emph {et~al.}(2014)\citenamefont
  {Gallego}, \citenamefont {Riera},\ and\ \citenamefont
  {Eisert}}]{gallego2014}%
  \BibitemOpen
  \bibfield  {author} {\bibinfo {author} {\bibfnamefont {R.}~\bibnamefont
  {Gallego}}, \bibinfo {author} {\bibfnamefont {A.}~\bibnamefont {Riera}},\
  and\ \bibinfo {author} {\bibfnamefont {J.}~\bibnamefont {Eisert}},\
  }\bibfield  {title} {\bibinfo {title} {Thermal machines beyond the weak
  coupling regime},\ }\href {https://doi.org/10.1088/1367-2630/16/12/125009}
  {\bibfield  {journal} {\bibinfo  {journal} {New J. Phys.}\ }\textbf {\bibinfo
  {volume} {16}},\ \bibinfo {pages} {125009} (\bibinfo {year}
  {2014})}\BibitemShut {NoStop}%
\bibitem [{\citenamefont {Kato}\ and\ \citenamefont
  {Tanimura}(2016)}]{kato2016}%
  \BibitemOpen
  \bibfield  {author} {\bibinfo {author} {\bibfnamefont {A.}~\bibnamefont
  {Kato}}\ and\ \bibinfo {author} {\bibfnamefont {Y.}~\bibnamefont
  {Tanimura}},\ }\bibfield  {title} {\bibinfo {title} {Quantum heat current
  under non-perturbative and non-{Markovian} conditions: Applications to heat
  machines},\ }\href {https://doi.org/https://doi.org/10.1063/1.4971370}
  {\bibfield  {journal} {\bibinfo  {journal} {J. Chem. Phys.}\ }\textbf
  {\bibinfo {volume} {145}},\ \bibinfo {pages} {224105} (\bibinfo {year}
  {2016})}\BibitemShut {NoStop}%
\bibitem [{\citenamefont {Uzdin}\ \emph {et~al.}(2016)\citenamefont {Uzdin},
  \citenamefont {Levy},\ and\ \citenamefont {Kosloff}}]{uzdin2016}%
  \BibitemOpen
  \bibfield  {author} {\bibinfo {author} {\bibfnamefont {R.}~\bibnamefont
  {Uzdin}}, \bibinfo {author} {\bibfnamefont {A.}~\bibnamefont {Levy}},\ and\
  \bibinfo {author} {\bibfnamefont {R.}~\bibnamefont {Kosloff}},\ }\bibfield
  {title} {\bibinfo {title} {Quantum heat machines equivalence, work extraction
  beyond {Markovianity}, and strong coupling via heat exchangers},\ }\href
  {https://doi.org/https://doi.org/10.3390/e18040124} {\bibfield  {journal}
  {\bibinfo  {journal} {Entropy}\ }\textbf {\bibinfo {volume} {18}},\ \bibinfo
  {pages} {124} (\bibinfo {year} {2016})}\BibitemShut {NoStop}%
\bibitem [{\citenamefont {Katz}\ and\ \citenamefont
  {Kosloff}(2016)}]{katz2016}%
  \BibitemOpen
  \bibfield  {author} {\bibinfo {author} {\bibfnamefont {G.}~\bibnamefont
  {Katz}}\ and\ \bibinfo {author} {\bibfnamefont {R.}~\bibnamefont {Kosloff}},\
  }\bibfield  {title} {\bibinfo {title} {Quantum thermodynamics in strong
  coupling: Heat transport and refrigeration},\ }\href
  {https://doi.org/https://doi.org/10.3390/e18050186} {\bibfield  {journal}
  {\bibinfo  {journal} {Entropy}\ }\textbf {\bibinfo {volume} {18}},\ \bibinfo
  {pages} {186} (\bibinfo {year} {2016})}\BibitemShut {NoStop}%
\bibitem [{\citenamefont {Newman}\ \emph {et~al.}(2017)\citenamefont {Newman},
  \citenamefont {Mintert},\ and\ \citenamefont {Nazir}}]{newman2017}%
  \BibitemOpen
  \bibfield  {author} {\bibinfo {author} {\bibfnamefont {D.}~\bibnamefont
  {Newman}}, \bibinfo {author} {\bibfnamefont {F.}~\bibnamefont {Mintert}},\
  and\ \bibinfo {author} {\bibfnamefont {A.}~\bibnamefont {Nazir}},\ }\bibfield
   {title} {\bibinfo {title} {Performance of a quantum heat engine at strong
  reservoir coupling},\ }\href {https://doi.org/10.1103/PhysRevE.95.032139}
  {\bibfield  {journal} {\bibinfo  {journal} {Phys. Rev. E}\ }\textbf {\bibinfo
  {volume} {95}},\ \bibinfo {pages} {032139} (\bibinfo {year}
  {2017})}\BibitemShut {NoStop}%
\bibitem [{\citenamefont {Miller}(2018)}]{miller2018}%
  \BibitemOpen
  \bibfield  {author} {\bibinfo {author} {\bibfnamefont {H.~J.~D.}\
  \bibnamefont {Miller}},\ }\bibfield  {title} {\bibinfo {title} {Hamiltonian
  of mean force for strongly-coupled systems},\ }in\ \href@noop {} {\emph
  {\bibinfo {booktitle} {Thermodynamics in the Quantum Regime: Fundamental
  Aspects and New Directions}}},\ \bibinfo {editor} {edited by\ \bibinfo
  {editor} {\bibfnamefont {F.}~\bibnamefont {Binder}}, \bibinfo {editor}
  {\bibfnamefont {L.~A.}\ \bibnamefont {Correa}}, \bibinfo {editor}
  {\bibfnamefont {C.}~\bibnamefont {Gogolin}}, \bibinfo {editor} {\bibfnamefont
  {J.}~\bibnamefont {Anders}},\ and\ \bibinfo {editor} {\bibfnamefont
  {G.}~\bibnamefont {Adesso}}}\ (\bibinfo  {publisher} {Springer},\ \bibinfo
  {address} {Cham, Switzerland},\ \bibinfo {year} {2018})\ pp.\ \bibinfo
  {pages} {531--549}\BibitemShut {NoStop}%
\bibitem [{\citenamefont {Talkner}\ and\ \citenamefont
  {H\"anggi}(2020)}]{talkner2020}%
  \BibitemOpen
  \bibfield  {author} {\bibinfo {author} {\bibfnamefont {P.}~\bibnamefont
  {Talkner}}\ and\ \bibinfo {author} {\bibfnamefont {P.}~\bibnamefont
  {H\"anggi}},\ }\bibfield  {title} {\bibinfo {title} {Colloquium: Statistical
  mechanics and thermodynamics at strong coupling: Quantum and classical},\
  }\href {https://doi.org/10.1103/RevModPhys.92.041002} {\bibfield  {journal}
  {\bibinfo  {journal} {Rev. Mod. Phys.}\ }\textbf {\bibinfo {volume} {92}},\
  \bibinfo {pages} {041002} (\bibinfo {year} {2020})}\BibitemShut {NoStop}%
\bibitem [{\citenamefont {Campbell}\ \emph {et~al.}(2026)\citenamefont
  {Campbell} \emph {et~al.}}]{campbell2026}%
  \BibitemOpen
  \bibfield  {author} {\bibinfo {author} {\bibfnamefont {S.}~\bibnamefont
  {Campbell}} \emph {et~al.},\ }\bibfield  {title} {\bibinfo {title} {Roadmap
  on quantum thermodynamics},\ }\href
  {https://doi.org/10.1088/2058-9565/ae1e27} {\bibfield  {journal} {\bibinfo
  {journal} {Quantum Sci. Technol.}\ }\textbf {\bibinfo {volume} {11}},\
  \bibinfo {pages} {012501} (\bibinfo {year} {2026})}\BibitemShut {NoStop}%
\bibitem [{\citenamefont {Chiu}\ \emph {et~al.}(2022)\citenamefont {Chiu},
  \citenamefont {Strathearn},\ and\ \citenamefont {Keeling}}]{chiu2022}%
  \BibitemOpen
  \bibfield  {author} {\bibinfo {author} {\bibfnamefont {Y.-F.}\ \bibnamefont
  {Chiu}}, \bibinfo {author} {\bibfnamefont {A.}~\bibnamefont {Strathearn}},\
  and\ \bibinfo {author} {\bibfnamefont {J.}~\bibnamefont {Keeling}},\
  }\bibfield  {title} {\bibinfo {title} {Numerical evaluation and robustness of
  the quantum mean-force {Gibbs} state},\ }\href
  {https://doi.org/10.1103/PhysRevA.106.012204} {\bibfield  {journal} {\bibinfo
   {journal} {Phys. Rev. A}\ }\textbf {\bibinfo {volume} {106}},\ \bibinfo
  {pages} {012204} (\bibinfo {year} {2022})}\BibitemShut {NoStop}%
\bibitem [{\citenamefont {Cerisola}\ \emph {et~al.}(2024)\citenamefont
  {Cerisola}, \citenamefont {Berritta}, \citenamefont {Scali}, \citenamefont
  {Horsley}, \citenamefont {Cresser},\ and\ \citenamefont
  {Anders}}]{cerisola2024}%
  \BibitemOpen
  \bibfield  {author} {\bibinfo {author} {\bibfnamefont {F.}~\bibnamefont
  {Cerisola}}, \bibinfo {author} {\bibfnamefont {M.}~\bibnamefont {Berritta}},
  \bibinfo {author} {\bibfnamefont {S.}~\bibnamefont {Scali}}, \bibinfo
  {author} {\bibfnamefont {S.~A.~R.}\ \bibnamefont {Horsley}}, \bibinfo
  {author} {\bibfnamefont {J.~D.}\ \bibnamefont {Cresser}},\ and\ \bibinfo
  {author} {\bibfnamefont {J.}~\bibnamefont {Anders}},\ }\bibfield  {title}
  {\bibinfo {title} {Quantum--classical correspondence in spin--boson
  equilibrium states at arbitrary coupling},\ }\href
  {https://doi.org/10.1088/1367-2630/ad4818} {\bibfield  {journal} {\bibinfo
  {journal} {New J. Phys.}\ }\textbf {\bibinfo {volume} {26}},\ \bibinfo
  {pages} {053032} (\bibinfo {year} {2024})}\BibitemShut {NoStop}%
\bibitem [{\citenamefont {Kumar}\ \emph {et~al.}(2025)\citenamefont {Kumar},
  \citenamefont {Athulya},\ and\ \citenamefont {Ghosh}}]{kumar2025}%
  \BibitemOpen
  \bibfield  {author} {\bibinfo {author} {\bibfnamefont {P.}~\bibnamefont
  {Kumar}}, \bibinfo {author} {\bibfnamefont {K.~P.}\ \bibnamefont {Athulya}},\
  and\ \bibinfo {author} {\bibfnamefont {S.}~\bibnamefont {Ghosh}},\ }\bibfield
   {title} {\bibinfo {title} {Equivalence between the second order steady state
  for the spin-boson model and its quantum mean force {Gibbs} state},\ }\href
  {https://doi.org/10.1103/PhysRevB.111.115423} {\bibfield  {journal} {\bibinfo
   {journal} {Phys. Rev. B}\ }\textbf {\bibinfo {volume} {111}},\ \bibinfo
  {pages} {115423} (\bibinfo {year} {2025})}\BibitemShut {NoStop}%
\bibitem [{\citenamefont {Becker}\ \emph {et~al.}(2022)\citenamefont {Becker},
  \citenamefont {Schnell},\ and\ \citenamefont {Thingna}}]{becker2022}%
  \BibitemOpen
  \bibfield  {author} {\bibinfo {author} {\bibfnamefont {T.}~\bibnamefont
  {Becker}}, \bibinfo {author} {\bibfnamefont {A.}~\bibnamefont {Schnell}},\
  and\ \bibinfo {author} {\bibfnamefont {J.}~\bibnamefont {Thingna}},\
  }\bibfield  {title} {\bibinfo {title} {Canonically consistent quantum master
  equation},\ }\href {https://doi.org/10.1103/PhysRevLett.129.200403}
  {\bibfield  {journal} {\bibinfo  {journal} {Phys. Rev. Lett.}\ }\textbf
  {\bibinfo {volume} {129}},\ \bibinfo {pages} {200403} (\bibinfo {year}
  {2022})}\BibitemShut {NoStop}%
\bibitem [{\citenamefont
  {Suba\ifmmode\mbox{\c{s}}\else\c{s}\fi{}\ifmmode\imath\else\i\fi{}}\ \emph
  {et~al.}(2012)\citenamefont
  {Suba\ifmmode\mbox{\c{s}}\else\c{s}\fi{}\ifmmode\imath\else\i\fi{}},
  \citenamefont {Fleming}, \citenamefont {Taylor},\ and\ \citenamefont
  {Hu}}]{subasi2012}%
  \BibitemOpen
  \bibfield  {author} {\bibinfo {author} {\bibfnamefont {Y.}~\bibnamefont
  {Suba\ifmmode\mbox{\c{s}}\else\c{s}\fi{}\ifmmode\imath\else\i\fi{}}},
  \bibinfo {author} {\bibfnamefont {C.~H.}\ \bibnamefont {Fleming}}, \bibinfo
  {author} {\bibfnamefont {J.~M.}\ \bibnamefont {Taylor}},\ and\ \bibinfo
  {author} {\bibfnamefont {B.~L.}\ \bibnamefont {Hu}},\ }\bibfield  {title}
  {\bibinfo {title} {Equilibrium states of open quantum systems in the strong
  coupling regime},\ }\href {https://doi.org/10.1103/PhysRevE.86.061132}
  {\bibfield  {journal} {\bibinfo  {journal} {Phys. Rev. E}\ }\textbf {\bibinfo
  {volume} {86}},\ \bibinfo {pages} {061132} (\bibinfo {year}
  {2012})}\BibitemShut {NoStop}%
\bibitem [{\citenamefont {Mori}\ and\ \citenamefont
  {Miyashita}(2008)}]{mori2008}%
  \BibitemOpen
  \bibfield  {author} {\bibinfo {author} {\bibfnamefont {T.}~\bibnamefont
  {Mori}}\ and\ \bibinfo {author} {\bibfnamefont {S.}~\bibnamefont
  {Miyashita}},\ }\bibfield  {title} {\bibinfo {title} {Dynamics of the density
  matrix in contact with a thermal bath and the quantum master equation},\
  }\href {https://doi.org/10.1143/JPSJ.77.124005} {\bibfield  {journal}
  {\bibinfo  {journal} {J. Phys. Soc. Jpn.}\ }\textbf {\bibinfo {volume}
  {77}},\ \bibinfo {pages} {124005} (\bibinfo {year} {2008})}\BibitemShut
  {NoStop}%
\bibitem [{\citenamefont {Latune}(2022)}]{latune2021}%
  \BibitemOpen
  \bibfield  {author} {\bibinfo {author} {\bibfnamefont {C.~L.}\ \bibnamefont
  {Latune}},\ }\bibfield  {title} {\bibinfo {title} {Steady state in
  ultrastrong coupling regime: perturbative expansion and first orders},\
  }\href {https://doi.org/https://doi.org/10.12743/quanta.v11i1.167} {\bibfield
   {journal} {\bibinfo  {journal} {Quanta}\ }\textbf {\bibinfo {volume} {11}},\
  \bibinfo {pages} {53} (\bibinfo {year} {2022})}\BibitemShut {NoStop}%
\bibitem [{\citenamefont {Trushechkin}(2022)}]{trushechkin2022b}%
  \BibitemOpen
  \bibfield  {author} {\bibinfo {author} {\bibfnamefont {A.}~\bibnamefont
  {Trushechkin}},\ }\bibfield  {title} {\bibinfo {title} {Quantum master
  equations and steady states for the ultrastrong-coupling limit and the
  strong-decoherence limit},\ }\href
  {https://doi.org/10.1103/PhysRevA.106.042209} {\bibfield  {journal} {\bibinfo
   {journal} {Phys. Rev. A}\ }\textbf {\bibinfo {volume} {106}},\ \bibinfo
  {pages} {042209} (\bibinfo {year} {2022})}\BibitemShut {NoStop}%
\bibitem [{\citenamefont {Brenes}\ \emph {et~al.}(2024)\citenamefont {Brenes},
  \citenamefont {Min}, \citenamefont {Anto-Sztrikacs}, \citenamefont
  {Bar-Gill},\ and\ \citenamefont {Segal}}]{brenes2024}%
  \BibitemOpen
  \bibfield  {author} {\bibinfo {author} {\bibfnamefont {M.}~\bibnamefont
  {Brenes}}, \bibinfo {author} {\bibfnamefont {B.}~\bibnamefont {Min}},
  \bibinfo {author} {\bibfnamefont {N.}~\bibnamefont {Anto-Sztrikacs}},
  \bibinfo {author} {\bibfnamefont {N.}~\bibnamefont {Bar-Gill}},\ and\
  \bibinfo {author} {\bibfnamefont {D.}~\bibnamefont {Segal}},\ }\bibfield
  {title} {\bibinfo {title} {Bath-induced interactions and transient dynamics
  in open quantum systems at strong coupling: Effective {Hamiltonian}
  approach},\ }\href {https://doi.org/10.1063/5.0207028} {\bibfield  {journal}
  {\bibinfo  {journal} {J. Chem. Phys.}\ }\textbf {\bibinfo {volume} {160}},\
  \bibinfo {pages} {244106} (\bibinfo {year} {2024})}\BibitemShut {NoStop}%
\bibitem [{\citenamefont {Kirkwood}(1935)}]{kirkwood1935}%
  \BibitemOpen
  \bibfield  {author} {\bibinfo {author} {\bibfnamefont {J.~G.}\ \bibnamefont
  {Kirkwood}},\ }\bibfield  {title} {\bibinfo {title} {Statistical mechanics of
  fluid mixtures},\ }\href {https://doi.org/10.1063/1.1749657} {\bibfield
  {journal} {\bibinfo  {journal} {J. Chem. Phys.}\ }\textbf {\bibinfo {volume}
  {3}},\ \bibinfo {pages} {300} (\bibinfo {year} {1935})}\BibitemShut {NoStop}%
\bibitem [{\citenamefont {Roux}(1995)}]{roux1995}%
  \BibitemOpen
  \bibfield  {author} {\bibinfo {author} {\bibfnamefont {B.}~\bibnamefont
  {Roux}},\ }\bibfield  {title} {\bibinfo {title} {The calculation of the
  potential of mean force using computer simulations},\ }\href
  {https://doi.org/10.1016/0010-4655(95)00053-I} {\bibfield  {journal}
  {\bibinfo  {journal} {Comput. Phys. Commun.}\ }\textbf {\bibinfo {volume}
  {91}},\ \bibinfo {pages} {275} (\bibinfo {year} {1995})}\BibitemShut
  {NoStop}%
\bibitem [{\citenamefont {Darve}\ and\ \citenamefont
  {Pohorille}(2001)}]{darve2001}%
  \BibitemOpen
  \bibfield  {author} {\bibinfo {author} {\bibfnamefont {E.}~\bibnamefont
  {Darve}}\ and\ \bibinfo {author} {\bibfnamefont {A.}~\bibnamefont
  {Pohorille}},\ }\bibfield  {title} {\bibinfo {title} {Calculating free
  energies using average force},\ }\href {https://doi.org/10.1063/1.1410978}
  {\bibfield  {journal} {\bibinfo  {journal} {J. Chem. Phys.}\ }\textbf
  {\bibinfo {volume} {115}},\ \bibinfo {pages} {9169} (\bibinfo {year}
  {2001})}\BibitemShut {NoStop}%
\bibitem [{\citenamefont {Park}\ and\ \citenamefont
  {Schulten}(2004)}]{park2004}%
  \BibitemOpen
  \bibfield  {author} {\bibinfo {author} {\bibfnamefont {S.}~\bibnamefont
  {Park}}\ and\ \bibinfo {author} {\bibfnamefont {K.}~\bibnamefont
  {Schulten}},\ }\bibfield  {title} {\bibinfo {title} {Calculating potentials
  of mean force from steered molecular dynamics simulations},\ }\href
  {https://doi.org/10.1063/1.1651473} {\bibfield  {journal} {\bibinfo
  {journal} {J. Chem. Phys.}\ }\textbf {\bibinfo {volume} {120}},\ \bibinfo
  {pages} {5946} (\bibinfo {year} {2004})}\BibitemShut {NoStop}%
\bibitem [{\citenamefont {Trzesniak}\ \emph {et~al.}(2007)\citenamefont
  {Trzesniak}, \citenamefont {Kunz},\ and\ \citenamefont {van
  Gunsteren}}]{trzesniak2007}%
  \BibitemOpen
  \bibfield  {author} {\bibinfo {author} {\bibfnamefont {D.}~\bibnamefont
  {Trzesniak}}, \bibinfo {author} {\bibfnamefont {A.-P.~E.}\ \bibnamefont
  {Kunz}},\ and\ \bibinfo {author} {\bibfnamefont {W.~F.}\ \bibnamefont {van
  Gunsteren}},\ }\bibfield  {title} {\bibinfo {title} {A comparison of methods
  to compute the potential of mean force},\ }\href
  {https://doi.org/10.1002/cphc.200600527} {\bibfield  {journal} {\bibinfo
  {journal} {ChemPhysChem}\ }\textbf {\bibinfo {volume} {8}},\ \bibinfo {pages}
  {162} (\bibinfo {year} {2007})}\BibitemShut {NoStop}%
\bibitem [{\citenamefont {Melo}\ and\ \citenamefont
  {Feytmans}(1997)}]{melo1997}%
  \BibitemOpen
  \bibfield  {author} {\bibinfo {author} {\bibfnamefont {F.}~\bibnamefont
  {Melo}}\ and\ \bibinfo {author} {\bibfnamefont {E.}~\bibnamefont
  {Feytmans}},\ }\bibfield  {title} {\bibinfo {title} {Novel knowledge-based
  mean force potential at atomic level},\ }\href
  {https://doi.org/10.1006/jmbi.1996.0868} {\bibfield  {journal} {\bibinfo
  {journal} {J. Mol. Biol.}\ }\textbf {\bibinfo {volume} {267}},\ \bibinfo
  {pages} {207} (\bibinfo {year} {1997})}\BibitemShut {NoStop}%
\bibitem [{\citenamefont {Jiang}\ \emph {et~al.}(2002)\citenamefont {Jiang},
  \citenamefont {Gao}, \citenamefont {Mao}, \citenamefont {Liu},\ and\
  \citenamefont {Lai}}]{jiang2002}%
  \BibitemOpen
  \bibfield  {author} {\bibinfo {author} {\bibfnamefont {L.}~\bibnamefont
  {Jiang}}, \bibinfo {author} {\bibfnamefont {Y.}~\bibnamefont {Gao}}, \bibinfo
  {author} {\bibfnamefont {F.}~\bibnamefont {Mao}}, \bibinfo {author}
  {\bibfnamefont {Z.}~\bibnamefont {Liu}},\ and\ \bibinfo {author}
  {\bibfnamefont {L.}~\bibnamefont {Lai}},\ }\bibfield  {title} {\bibinfo
  {title} {Potential of mean force for protein--protein interaction studies},\
  }\href {https://doi.org/10.1002/prot.10031} {\bibfield  {journal} {\bibinfo
  {journal} {Proteins}\ }\textbf {\bibinfo {volume} {46}},\ \bibinfo {pages}
  {190} (\bibinfo {year} {2002})}\BibitemShut {NoStop}%
\bibitem [{\citenamefont {Hamelryck}\ \emph {et~al.}(2010)\citenamefont
  {Hamelryck}, \citenamefont {Borg}, \citenamefont {Paluszewski}, \citenamefont
  {Paulsen}, \citenamefont {Frellsen}, \citenamefont {Andreetta}, \citenamefont
  {Boomsma}, \citenamefont {Bottaro},\ and\ \citenamefont
  {Ferkinghoff-Borg}}]{hamelryck2010}%
  \BibitemOpen
  \bibfield  {author} {\bibinfo {author} {\bibfnamefont {T.}~\bibnamefont
  {Hamelryck}}, \bibinfo {author} {\bibfnamefont {M.}~\bibnamefont {Borg}},
  \bibinfo {author} {\bibfnamefont {M.}~\bibnamefont {Paluszewski}}, \bibinfo
  {author} {\bibfnamefont {J.}~\bibnamefont {Paulsen}}, \bibinfo {author}
  {\bibfnamefont {J.}~\bibnamefont {Frellsen}}, \bibinfo {author}
  {\bibfnamefont {C.}~\bibnamefont {Andreetta}}, \bibinfo {author}
  {\bibfnamefont {W.}~\bibnamefont {Boomsma}}, \bibinfo {author} {\bibfnamefont
  {S.}~\bibnamefont {Bottaro}},\ and\ \bibinfo {author} {\bibfnamefont
  {J.}~\bibnamefont {Ferkinghoff-Borg}},\ }\bibfield  {title} {\bibinfo {title}
  {Potentials of mean force for protein structure prediction vindicated,
  formalized and generalized},\ }\href
  {https://doi.org/10.1371/journal.pone.0013714} {\bibfield  {journal}
  {\bibinfo  {journal} {PLoS ONE}\ }\textbf {\bibinfo {volume} {5}},\ \bibinfo
  {pages} {e13714} (\bibinfo {year} {2010})}\BibitemShut {NoStop}%
\bibitem [{\citenamefont {Trushechkin}\ \emph {et~al.}(2022)\citenamefont
  {Trushechkin}, \citenamefont {Merkli}, \citenamefont {Cresser},\ and\
  \citenamefont {Anders}}]{trushechkin2022}%
  \BibitemOpen
  \bibfield  {author} {\bibinfo {author} {\bibfnamefont {A.~S.}\ \bibnamefont
  {Trushechkin}}, \bibinfo {author} {\bibfnamefont {M.}~\bibnamefont {Merkli}},
  \bibinfo {author} {\bibfnamefont {J.~D.}\ \bibnamefont {Cresser}},\ and\
  \bibinfo {author} {\bibfnamefont {J.}~\bibnamefont {Anders}},\ }\bibfield
  {title} {\bibinfo {title} {Open quantum system dynamics and the mean force
  {Gibbs} state},\ }\href {https://doi.org/10.1116/5.0073853} {\bibfield
  {journal} {\bibinfo  {journal} {AVS Quantum Sci.}\ }\textbf {\bibinfo
  {volume} {4}} (\bibinfo {year} {2022})}\BibitemShut {NoStop}%
\bibitem [{\citenamefont {Leggett}\ \emph {et~al.}(1987)\citenamefont
  {Leggett}, \citenamefont {Chakravarty}, \citenamefont {Dorsey}, \citenamefont
  {Fisher}, \citenamefont {Garg},\ and\ \citenamefont {Zwerger}}]{leggett1987}%
  \BibitemOpen
  \bibfield  {author} {\bibinfo {author} {\bibfnamefont {A.~J.}\ \bibnamefont
  {Leggett}}, \bibinfo {author} {\bibfnamefont {S.}~\bibnamefont
  {Chakravarty}}, \bibinfo {author} {\bibfnamefont {A.~T.}\ \bibnamefont
  {Dorsey}}, \bibinfo {author} {\bibfnamefont {M.~P.~A.}\ \bibnamefont
  {Fisher}}, \bibinfo {author} {\bibfnamefont {A.}~\bibnamefont {Garg}},\ and\
  \bibinfo {author} {\bibfnamefont {W.}~\bibnamefont {Zwerger}},\ }\bibfield
  {title} {\bibinfo {title} {Dynamics of the dissipative two-state system},\
  }\href {https://doi.org/10.1103/RevModPhys.59.1} {\bibfield  {journal}
  {\bibinfo  {journal} {Rev. Mod. Phys.}\ }\textbf {\bibinfo {volume} {59}},\
  \bibinfo {pages} {1} (\bibinfo {year} {1987})}\BibitemShut {NoStop}%
\bibitem [{\citenamefont {Strack}\ and\ \citenamefont
  {Sachdev}(2011)}]{strack2011}%
  \BibitemOpen
  \bibfield  {author} {\bibinfo {author} {\bibfnamefont {P.}~\bibnamefont
  {Strack}}\ and\ \bibinfo {author} {\bibfnamefont {S.}~\bibnamefont
  {Sachdev}},\ }\bibfield  {title} {\bibinfo {title} {Dicke quantum spin glass
  of atoms and photons},\ }\href
  {https://doi.org/10.1103/PhysRevLett.107.277202} {\bibfield  {journal}
  {\bibinfo  {journal} {Phys. Rev. Lett.}\ }\textbf {\bibinfo {volume} {107}},\
  \bibinfo {pages} {277202} (\bibinfo {year} {2011})}\BibitemShut {NoStop}%
\bibitem [{\citenamefont {Makhlin}\ \emph {et~al.}(2001)\citenamefont
  {Makhlin}, \citenamefont {Sch\"on},\ and\ \citenamefont
  {Shnirman}}]{makhlin2001}%
  \BibitemOpen
  \bibfield  {author} {\bibinfo {author} {\bibfnamefont {Y.}~\bibnamefont
  {Makhlin}}, \bibinfo {author} {\bibfnamefont {G.}~\bibnamefont {Sch\"on}},\
  and\ \bibinfo {author} {\bibfnamefont {A.}~\bibnamefont {Shnirman}},\
  }\bibfield  {title} {\bibinfo {title} {Quantum-state engineering with
  josephson-junction devices},\ }\href
  {https://doi.org/10.1103/RevModPhys.73.357} {\bibfield  {journal} {\bibinfo
  {journal} {Rev. Mod. Phys.}\ }\textbf {\bibinfo {volume} {73}},\ \bibinfo
  {pages} {357} (\bibinfo {year} {2001})}\BibitemShut {NoStop}%
\bibitem [{\citenamefont {Garg}\ \emph {et~al.}(1985)\citenamefont {Garg},
  \citenamefont {Onuchic},\ and\ \citenamefont {Ambegaokar}}]{garg1985}%
  \BibitemOpen
  \bibfield  {author} {\bibinfo {author} {\bibfnamefont {A.}~\bibnamefont
  {Garg}}, \bibinfo {author} {\bibfnamefont {J.~N.}\ \bibnamefont {Onuchic}},\
  and\ \bibinfo {author} {\bibfnamefont {V.}~\bibnamefont {Ambegaokar}},\
  }\bibfield  {title} {\bibinfo {title} {Effect of friction on electron
  transfer in biomolecules},\ }\href
  {https://doi.org/https://doi.org/10.1063/1.449017} {\bibfield  {journal}
  {\bibinfo  {journal} {J. Chem. Phys.}\ }\textbf {\bibinfo {volume} {83}},\
  \bibinfo {pages} {4491} (\bibinfo {year} {1985})}\BibitemShut {NoStop}%
\bibitem [{\citenamefont {Xu}\ and\ \citenamefont {Schulten}(1994)}]{xu1994}%
  \BibitemOpen
  \bibfield  {author} {\bibinfo {author} {\bibfnamefont {D.}~\bibnamefont
  {Xu}}\ and\ \bibinfo {author} {\bibfnamefont {K.}~\bibnamefont {Schulten}},\
  }\bibfield  {title} {\bibinfo {title} {Coupling of protein motion to electron
  transfer in a photosynthetic reaction center: investigating the low
  temperature behavior in the framework of the spin-boson model},\ }\href
  {https://doi.org/https://doi.org/10.1016/0301-0104(94)00016-6} {\bibfield
  {journal} {\bibinfo  {journal} {Chem. Phys.}\ }\textbf {\bibinfo {volume}
  {182}},\ \bibinfo {pages} {91} (\bibinfo {year} {1994})}\BibitemShut
  {NoStop}%
\bibitem [{\citenamefont {Gilmore}\ and\ \citenamefont
  {McKenzie}(2005)}]{gilmore2005}%
  \BibitemOpen
  \bibfield  {author} {\bibinfo {author} {\bibfnamefont {J.}~\bibnamefont
  {Gilmore}}\ and\ \bibinfo {author} {\bibfnamefont {R.~H.}\ \bibnamefont
  {McKenzie}},\ }\bibfield  {title} {\bibinfo {title} {Spin boson models for
  quantum decoherence of electronic excitations of biomolecules and quantum
  dots in a solvent},\ }\href {https://doi.org/10.1088/0953-8984/17/10/028}
  {\bibfield  {journal} {\bibinfo  {journal} {J. Phys.: Condens. Matter}\
  }\textbf {\bibinfo {volume} {17}},\ \bibinfo {pages} {1735} (\bibinfo {year}
  {2005})}\BibitemShut {NoStop}%
\bibitem [{\citenamefont {Guarnieri}\ \emph {et~al.}(2018)\citenamefont
  {Guarnieri}, \citenamefont {Kol\'a\ifmmode~\check{r}\else \v{r}\fi{}},\ and\
  \citenamefont {Filip}}]{guarnieri2018}%
  \BibitemOpen
  \bibfield  {author} {\bibinfo {author} {\bibfnamefont {G.}~\bibnamefont
  {Guarnieri}}, \bibinfo {author} {\bibfnamefont {M.}~\bibnamefont
  {Kol\'a\ifmmode~\check{r}\else \v{r}\fi{}}},\ and\ \bibinfo {author}
  {\bibfnamefont {R.}~\bibnamefont {Filip}},\ }\bibfield  {title} {\bibinfo
  {title} {Steady-state coherences by composite system-bath interactions},\
  }\href {https://doi.org/10.1103/PhysRevLett.121.070401} {\bibfield  {journal}
  {\bibinfo  {journal} {Phys. Rev. Lett.}\ }\textbf {\bibinfo {volume} {121}},\
  \bibinfo {pages} {070401} (\bibinfo {year} {2018})}\BibitemShut {NoStop}%
\bibitem [{\citenamefont {Purkayastha}\ \emph {et~al.}(2020)\citenamefont
  {Purkayastha}, \citenamefont {Guarnieri}, \citenamefont {Mitchison},
  \citenamefont {Filip},\ and\ \citenamefont {Goold}}]{purkayastha2020}%
  \BibitemOpen
  \bibfield  {author} {\bibinfo {author} {\bibfnamefont {A.}~\bibnamefont
  {Purkayastha}}, \bibinfo {author} {\bibfnamefont {G.}~\bibnamefont
  {Guarnieri}}, \bibinfo {author} {\bibfnamefont {M.~T.}\ \bibnamefont
  {Mitchison}}, \bibinfo {author} {\bibfnamefont {R.}~\bibnamefont {Filip}},\
  and\ \bibinfo {author} {\bibfnamefont {J.}~\bibnamefont {Goold}},\ }\bibfield
   {title} {\bibinfo {title} {Tunable phonon-induced steady-state coherence in
  a double-quantum-dot charge qubit},\ }\href
  {https://doi.org/10.1038/s41534-020-0256-6} {\bibfield  {journal} {\bibinfo
  {journal} {npj Quantum Inf.}\ }\textbf {\bibinfo {volume} {6}},\ \bibinfo
  {pages} {27} (\bibinfo {year} {2020})}\BibitemShut {NoStop}%
\bibitem [{\citenamefont {Cresser}\ and\ \citenamefont
  {Anders}(2021)}]{cresser2021}%
  \BibitemOpen
  \bibfield  {author} {\bibinfo {author} {\bibfnamefont {J.~D.}\ \bibnamefont
  {Cresser}}\ and\ \bibinfo {author} {\bibfnamefont {J.}~\bibnamefont
  {Anders}},\ }\bibfield  {title} {\bibinfo {title} {Weak and ultrastrong
  coupling limits of the quantum mean force {Gibbs} state},\ }\href
  {https://doi.org/10.1103/PhysRevLett.127.250601} {\bibfield  {journal}
  {\bibinfo  {journal} {Phys. Rev. Lett.}\ }\textbf {\bibinfo {volume} {127}},\
  \bibinfo {pages} {250601} (\bibinfo {year} {2021})}\BibitemShut {NoStop}%
\bibitem [{\citenamefont {Braun}(2002)}]{braun2002}%
  \BibitemOpen
  \bibfield  {author} {\bibinfo {author} {\bibfnamefont {D.}~\bibnamefont
  {Braun}},\ }\bibfield  {title} {\bibinfo {title} {Creation of entanglement by
  interaction with a common heat bath},\ }\href
  {https://doi.org/10.1103/PhysRevLett.89.277901} {\bibfield  {journal}
  {\bibinfo  {journal} {Phys. Rev. Lett.}\ }\textbf {\bibinfo {volume} {89}},\
  \bibinfo {pages} {277901} (\bibinfo {year} {2002})}\BibitemShut {NoStop}%
\bibitem [{\citenamefont {Jak{\'o}bczyk}(2002)}]{jakobczyk2002}%
  \BibitemOpen
  \bibfield  {author} {\bibinfo {author} {\bibfnamefont {L.}~\bibnamefont
  {Jak{\'o}bczyk}},\ }\bibfield  {title} {\bibinfo {title} {Entangling two
  qubits by dissipation},\ }\href {https://doi.org/10.1088/0305-4470/35/30/313}
  {\bibfield  {journal} {\bibinfo  {journal} {J. Phys. A: Math. Gen.}\ }\textbf
  {\bibinfo {volume} {35}},\ \bibinfo {pages} {6383} (\bibinfo {year}
  {2002})}\BibitemShut {NoStop}%
\bibitem [{\citenamefont {Maniscalco}\ \emph {et~al.}(2008)\citenamefont
  {Maniscalco}, \citenamefont {Francica}, \citenamefont {Zaffino},
  \citenamefont {Lo~Gullo},\ and\ \citenamefont {Plastina}}]{maniscalco2008}%
  \BibitemOpen
  \bibfield  {author} {\bibinfo {author} {\bibfnamefont {S.}~\bibnamefont
  {Maniscalco}}, \bibinfo {author} {\bibfnamefont {F.}~\bibnamefont
  {Francica}}, \bibinfo {author} {\bibfnamefont {R.~L.}\ \bibnamefont
  {Zaffino}}, \bibinfo {author} {\bibfnamefont {N.}~\bibnamefont {Lo~Gullo}},\
  and\ \bibinfo {author} {\bibfnamefont {F.}~\bibnamefont {Plastina}},\
  }\bibfield  {title} {\bibinfo {title} {Protecting entanglement via the
  quantum {Zeno} effect},\ }\href
  {https://doi.org/10.1103/PhysRevLett.100.090503} {\bibfield  {journal}
  {\bibinfo  {journal} {Phys. Rev. Lett.}\ }\textbf {\bibinfo {volume} {100}},\
  \bibinfo {pages} {090503} (\bibinfo {year} {2008})}\BibitemShut {NoStop}%
\bibitem [{\citenamefont {Francica}\ \emph {et~al.}(2009)\citenamefont
  {Francica}, \citenamefont {Maniscalco}, \citenamefont {Piilo}, \citenamefont
  {Plastina},\ and\ \citenamefont {Suominen}}]{francica2009}%
  \BibitemOpen
  \bibfield  {author} {\bibinfo {author} {\bibfnamefont {F.}~\bibnamefont
  {Francica}}, \bibinfo {author} {\bibfnamefont {S.}~\bibnamefont
  {Maniscalco}}, \bibinfo {author} {\bibfnamefont {J.}~\bibnamefont {Piilo}},
  \bibinfo {author} {\bibfnamefont {F.}~\bibnamefont {Plastina}},\ and\
  \bibinfo {author} {\bibfnamefont {K.-A.}\ \bibnamefont {Suominen}},\
  }\bibfield  {title} {\bibinfo {title} {Off-resonant entanglement generation
  in a lossy cavity},\ }\href {https://doi.org/10.1103/PhysRevA.79.032310}
  {\bibfield  {journal} {\bibinfo  {journal} {Phys. Rev. A}\ }\textbf {\bibinfo
  {volume} {79}},\ \bibinfo {pages} {032310} (\bibinfo {year}
  {2009})}\BibitemShut {NoStop}%
\bibitem [{\citenamefont {Ng}\ and\ \citenamefont {Bose}(2009)}]{ng2009}%
  \BibitemOpen
  \bibfield  {author} {\bibinfo {author} {\bibfnamefont {H.~T.}\ \bibnamefont
  {Ng}}\ and\ \bibinfo {author} {\bibfnamefont {S.}~\bibnamefont {Bose}},\
  }\bibfield  {title} {\bibinfo {title} {Quantum communication between trapped
  ions through a dissipative environment},\ }\href
  {https://doi.org/10.1088/1751-8113/42/18/182001} {\bibfield  {journal}
  {\bibinfo  {journal} {J. Phys. A: Math. Theor.}\ }\textbf {\bibinfo {volume}
  {42}},\ \bibinfo {pages} {182001} (\bibinfo {year} {2009})}\BibitemShut
  {NoStop}%
\bibitem [{\citenamefont {Mazzola}\ \emph {et~al.}(2009)\citenamefont
  {Mazzola}, \citenamefont {Maniscalco}, \citenamefont {Piilo}, \citenamefont
  {Suominen},\ and\ \citenamefont {Garraway}}]{mazzola2009}%
  \BibitemOpen
  \bibfield  {author} {\bibinfo {author} {\bibfnamefont {L.}~\bibnamefont
  {Mazzola}}, \bibinfo {author} {\bibfnamefont {S.}~\bibnamefont {Maniscalco}},
  \bibinfo {author} {\bibfnamefont {J.}~\bibnamefont {Piilo}}, \bibinfo
  {author} {\bibfnamefont {K.-A.}\ \bibnamefont {Suominen}},\ and\ \bibinfo
  {author} {\bibfnamefont {B.~M.}\ \bibnamefont {Garraway}},\ }\bibfield
  {title} {\bibinfo {title} {Sudden death and sudden birth of entanglement in
  common structured reservoirs},\ }\href
  {https://doi.org/10.1103/PhysRevA.79.042302} {\bibfield  {journal} {\bibinfo
  {journal} {Phys. Rev. A}\ }\textbf {\bibinfo {volume} {79}},\ \bibinfo
  {pages} {042302} (\bibinfo {year} {2009})}\BibitemShut {NoStop}%
\bibitem [{\citenamefont {Sahrapour}\ and\ \citenamefont
  {Makri}(2013)}]{sahrapour2013}%
  \BibitemOpen
  \bibfield  {author} {\bibinfo {author} {\bibfnamefont {M.~M.}\ \bibnamefont
  {Sahrapour}}\ and\ \bibinfo {author} {\bibfnamefont {N.}~\bibnamefont
  {Makri}},\ }\bibfield  {title} {\bibinfo {title} {Tunneling, decoherence, and
  entanglement of two spins interacting with a dissipative bath},\ }\href
  {https://doi.org/10.1063/1.4795159} {\bibfield  {journal} {\bibinfo
  {journal} {J. Chem. Phys.}\ }\textbf {\bibinfo {volume} {138}},\ \bibinfo
  {pages} {114109} (\bibinfo {year} {2013})}\BibitemShut {NoStop}%
\bibitem [{\citenamefont {Lee}\ \emph {et~al.}(2019)\citenamefont {Lee},
  \citenamefont {Najafabadi}, \citenamefont {Schumayer}, \citenamefont {Kwek},\
  and\ \citenamefont {Hutchinson}}]{lee2019}%
  \BibitemOpen
  \bibfield  {author} {\bibinfo {author} {\bibfnamefont {C.~K.}\ \bibnamefont
  {Lee}}, \bibinfo {author} {\bibfnamefont {M.~S.}\ \bibnamefont {Najafabadi}},
  \bibinfo {author} {\bibfnamefont {D.}~\bibnamefont {Schumayer}}, \bibinfo
  {author} {\bibfnamefont {L.~C.}\ \bibnamefont {Kwek}},\ and\ \bibinfo
  {author} {\bibfnamefont {D.~A.}\ \bibnamefont {Hutchinson}},\ }\bibfield
  {title} {\bibinfo {title} {Environment mediated multipartite and
  multidimensional entanglement},\ }\href
  {https://doi.org/10.1038/s41598-019-45496-2} {\bibfield  {journal} {\bibinfo
  {journal} {Sci. Rep.}\ }\textbf {\bibinfo {volume} {9}},\ \bibinfo {pages}
  {9147} (\bibinfo {year} {2019})}\BibitemShut {NoStop}%
\bibitem [{\citenamefont {Hartmann}\ and\ \citenamefont
  {Strunz}(2020)}]{hartmann2020}%
  \BibitemOpen
  \bibfield  {author} {\bibinfo {author} {\bibfnamefont {R.}~\bibnamefont
  {Hartmann}}\ and\ \bibinfo {author} {\bibfnamefont {W.~T.}\ \bibnamefont
  {Strunz}},\ }\bibfield  {title} {\bibinfo {title} {Environmentally induced
  entanglement --- anomalous behavior in the adiabatic regime},\ }\href
  {https://doi.org/10.22331/q-2020-10-22-347} {\bibfield  {journal} {\bibinfo
  {journal} {{Quantum}}\ }\textbf {\bibinfo {volume} {4}},\ \bibinfo {pages}
  {347} (\bibinfo {year} {2020})}\BibitemShut {NoStop}%
\bibitem [{\citenamefont {Porras}\ \emph {et~al.}(2008)\citenamefont {Porras},
  \citenamefont {Marquardt}, \citenamefont {von Delft},\ and\ \citenamefont
  {Cirac}}]{porras2008}%
  \BibitemOpen
  \bibfield  {author} {\bibinfo {author} {\bibfnamefont {D.}~\bibnamefont
  {Porras}}, \bibinfo {author} {\bibfnamefont {F.}~\bibnamefont {Marquardt}},
  \bibinfo {author} {\bibfnamefont {J.}~\bibnamefont {von Delft}},\ and\
  \bibinfo {author} {\bibfnamefont {J.~I.}\ \bibnamefont {Cirac}},\ }\bibfield
  {title} {\bibinfo {title} {Mesoscopic spin-boson models of trapped ions},\
  }\href {https://doi.org/10.1103/PhysRevA.78.010101} {\bibfield  {journal}
  {\bibinfo  {journal} {Phys. Rev. A}\ }\textbf {\bibinfo {volume} {78}},\
  \bibinfo {pages} {010101} (\bibinfo {year} {2008})}\BibitemShut {NoStop}%
\bibitem [{\citenamefont {Lemmer}\ \emph {et~al.}(2018)\citenamefont {Lemmer},
  \citenamefont {Cormick}, \citenamefont {Tamascelli}, \citenamefont {Schaetz},
  \citenamefont {Huelga},\ and\ \citenamefont {Plenio}}]{lemmer2018}%
  \BibitemOpen
  \bibfield  {author} {\bibinfo {author} {\bibfnamefont {A.}~\bibnamefont
  {Lemmer}}, \bibinfo {author} {\bibfnamefont {C.}~\bibnamefont {Cormick}},
  \bibinfo {author} {\bibfnamefont {D.}~\bibnamefont {Tamascelli}}, \bibinfo
  {author} {\bibfnamefont {T.}~\bibnamefont {Schaetz}}, \bibinfo {author}
  {\bibfnamefont {S.~F.}\ \bibnamefont {Huelga}},\ and\ \bibinfo {author}
  {\bibfnamefont {M.~B.}\ \bibnamefont {Plenio}},\ }\bibfield  {title}
  {\bibinfo {title} {A trapped-ion simulator for spin-boson models with
  structured environments},\ }\href {https://doi.org/10.1088/1367-2630/aac87d}
  {\bibfield  {journal} {\bibinfo  {journal} {New J. Phys.}\ }\textbf {\bibinfo
  {volume} {20}},\ \bibinfo {pages} {073002} (\bibinfo {year}
  {2018})}\BibitemShut {NoStop}%
\bibitem [{\citenamefont {Sun}\ \emph {et~al.}(2025)\citenamefont {Sun},
  \citenamefont {Kang}, \citenamefont {Nuomin}, \citenamefont {Schwartz},
  \citenamefont {Beratan}, \citenamefont {Brown},\ and\ \citenamefont
  {Kim}}]{sun2025}%
  \BibitemOpen
  \bibfield  {author} {\bibinfo {author} {\bibfnamefont {K.}~\bibnamefont
  {Sun}}, \bibinfo {author} {\bibfnamefont {M.}~\bibnamefont {Kang}}, \bibinfo
  {author} {\bibfnamefont {H.}~\bibnamefont {Nuomin}}, \bibinfo {author}
  {\bibfnamefont {G.}~\bibnamefont {Schwartz}}, \bibinfo {author}
  {\bibfnamefont {D.~N.}\ \bibnamefont {Beratan}}, \bibinfo {author}
  {\bibfnamefont {K.~R.}\ \bibnamefont {Brown}},\ and\ \bibinfo {author}
  {\bibfnamefont {J.}~\bibnamefont {Kim}},\ }\bibfield  {title} {\bibinfo
  {title} {Quantum simulation of spin-boson models with structured bath},\
  }\href {https://doi.org/https://doi.org/10.1038/s41467-025-59296-y}
  {\bibfield  {journal} {\bibinfo  {journal} {Nat. Commun.}\ }\textbf {\bibinfo
  {volume} {16}},\ \bibinfo {pages} {4042} (\bibinfo {year}
  {2025})}\BibitemShut {NoStop}%
\bibitem [{\citenamefont {Mostame}\ \emph {et~al.}(2012)\citenamefont
  {Mostame}, \citenamefont {Rebentrost}, \citenamefont {Eisfeld}, \citenamefont
  {Kerman}, \citenamefont {Tsomokos},\ and\ \citenamefont
  {Aspuru-Guzik}}]{mostame2012}%
  \BibitemOpen
  \bibfield  {author} {\bibinfo {author} {\bibfnamefont {S.}~\bibnamefont
  {Mostame}}, \bibinfo {author} {\bibfnamefont {P.}~\bibnamefont {Rebentrost}},
  \bibinfo {author} {\bibfnamefont {A.}~\bibnamefont {Eisfeld}}, \bibinfo
  {author} {\bibfnamefont {A.~J.}\ \bibnamefont {Kerman}}, \bibinfo {author}
  {\bibfnamefont {D.~I.}\ \bibnamefont {Tsomokos}},\ and\ \bibinfo {author}
  {\bibfnamefont {A.}~\bibnamefont {Aspuru-Guzik}},\ }\bibfield  {title}
  {\bibinfo {title} {Quantum simulator of an open quantum system using
  superconducting qubits: exciton transport in photosynthetic complexes},\
  }\href {https://doi.org/10.1088/1367-2630/14/10/105013} {\bibfield  {journal}
  {\bibinfo  {journal} {New J. Phys.}\ }\textbf {\bibinfo {volume} {14}},\
  \bibinfo {pages} {105013} (\bibinfo {year} {2012})}\BibitemShut {NoStop}%
\bibitem [{\citenamefont {Lepp\"akangas}\ \emph {et~al.}(2018)\citenamefont
  {Lepp\"akangas}, \citenamefont {Braum\"uller}, \citenamefont {Hauck},
  \citenamefont {Reiner}, \citenamefont {Schwenk}, \citenamefont {Zanker},
  \citenamefont {Fritz}, \citenamefont {Ustinov}, \citenamefont {Weides},\ and\
  \citenamefont {Marthaler}}]{leppakangas2018}%
  \BibitemOpen
  \bibfield  {author} {\bibinfo {author} {\bibfnamefont {J.}~\bibnamefont
  {Lepp\"akangas}}, \bibinfo {author} {\bibfnamefont {J.}~\bibnamefont
  {Braum\"uller}}, \bibinfo {author} {\bibfnamefont {M.}~\bibnamefont {Hauck}},
  \bibinfo {author} {\bibfnamefont {J.-M.}\ \bibnamefont {Reiner}}, \bibinfo
  {author} {\bibfnamefont {I.}~\bibnamefont {Schwenk}}, \bibinfo {author}
  {\bibfnamefont {S.}~\bibnamefont {Zanker}}, \bibinfo {author} {\bibfnamefont
  {L.}~\bibnamefont {Fritz}}, \bibinfo {author} {\bibfnamefont {A.~V.}\
  \bibnamefont {Ustinov}}, \bibinfo {author} {\bibfnamefont {M.}~\bibnamefont
  {Weides}},\ and\ \bibinfo {author} {\bibfnamefont {M.}~\bibnamefont
  {Marthaler}},\ }\bibfield  {title} {\bibinfo {title} {Quantum simulation of
  the spin-boson model with a microwave circuit},\ }\href
  {https://doi.org/10.1103/PhysRevA.97.052321} {\bibfield  {journal} {\bibinfo
  {journal} {Phys. Rev. A}\ }\textbf {\bibinfo {volume} {97}},\ \bibinfo
  {pages} {052321} (\bibinfo {year} {2018})}\BibitemShut {NoStop}%
\bibitem [{\citenamefont {Magazz{\`u}}\ \emph {et~al.}(2018)\citenamefont
  {Magazz{\`u}}, \citenamefont {Forn-D{\'\i}az}, \citenamefont {Belyansky},
  \citenamefont {Orgiazzi}, \citenamefont {Yurtalan}, \citenamefont {Otto},
  \citenamefont {Lupascu}, \citenamefont {Wilson},\ and\ \citenamefont
  {Grifoni}}]{magazzu2018}%
  \BibitemOpen
  \bibfield  {author} {\bibinfo {author} {\bibfnamefont {L.}~\bibnamefont
  {Magazz{\`u}}}, \bibinfo {author} {\bibfnamefont {P.}~\bibnamefont
  {Forn-D{\'\i}az}}, \bibinfo {author} {\bibfnamefont {R.}~\bibnamefont
  {Belyansky}}, \bibinfo {author} {\bibfnamefont {J.-L.}\ \bibnamefont
  {Orgiazzi}}, \bibinfo {author} {\bibfnamefont {M.}~\bibnamefont {Yurtalan}},
  \bibinfo {author} {\bibfnamefont {M.~R.}\ \bibnamefont {Otto}}, \bibinfo
  {author} {\bibfnamefont {A.}~\bibnamefont {Lupascu}}, \bibinfo {author}
  {\bibfnamefont {C.}~\bibnamefont {Wilson}},\ and\ \bibinfo {author}
  {\bibfnamefont {M.}~\bibnamefont {Grifoni}},\ }\bibfield  {title} {\bibinfo
  {title} {Probing the strongly driven spin-boson model in a superconducting
  quantum circuit},\ }\href
  {https://doi.org/https://doi.org/10.1038/s41467-018-03626-w} {\bibfield
  {journal} {\bibinfo  {journal} {Nat. Commun.}\ }\textbf {\bibinfo {volume}
  {9}},\ \bibinfo {pages} {1403} (\bibinfo {year} {2018})}\BibitemShut
  {NoStop}%
\bibitem [{\citenamefont {Weiss}(2012)}]{weiss2012}%
  \BibitemOpen
  \bibfield  {author} {\bibinfo {author} {\bibfnamefont {U.}~\bibnamefont
  {Weiss}},\ }\href@noop {} {\emph {\bibinfo {title} {Quantum Dissipative
  Systems}}},\ \bibinfo {edition} {4th}\ ed.\ (\bibinfo  {publisher} {World
  Scientific},\ \bibinfo {address} {Singapore, Hackensack, London},\ \bibinfo
  {year} {2012})\BibitemShut {NoStop}%
\bibitem [{\citenamefont {Hogg}\ \emph {et~al.}(2024)\citenamefont {Hogg},
  \citenamefont {Glatthard}, \citenamefont {Cerisola},\ and\ \citenamefont
  {Anders}}]{hogg2024}%
  \BibitemOpen
  \bibfield  {author} {\bibinfo {author} {\bibfnamefont {C.}~\bibnamefont
  {Hogg}}, \bibinfo {author} {\bibfnamefont {J.}~\bibnamefont {Glatthard}},
  \bibinfo {author} {\bibfnamefont {F.}~\bibnamefont {Cerisola}},\ and\
  \bibinfo {author} {\bibfnamefont {J.}~\bibnamefont {Anders}},\ }\bibfield
  {title} {\bibinfo {title} {Tutorial on the stochastic simulation of
  dissipative quantum oscillators},\ }\href {https://doi.org/10.1063/5.0222528}
  {\bibfield  {journal} {\bibinfo  {journal} {J. Chem. Phys.}\ }\textbf
  {\bibinfo {volume} {161}},\ \bibinfo {pages} {071501} (\bibinfo {year}
  {2024})}\BibitemShut {NoStop}%
\bibitem [{\citenamefont {Cheng}\ and\ \citenamefont
  {Fleming}(2009)}]{cheng2009}%
  \BibitemOpen
  \bibfield  {author} {\bibinfo {author} {\bibfnamefont {Y.-C.}\ \bibnamefont
  {Cheng}}\ and\ \bibinfo {author} {\bibfnamefont {G.~R.}\ \bibnamefont
  {Fleming}},\ }\bibfield  {title} {\bibinfo {title} {Dynamics of light
  harvesting in photosynthesis},\ }\href
  {https://doi.org/https://doi.org/10.1146/annurev.physchem.040808.090259}
  {\bibfield  {journal} {\bibinfo  {journal} {Annu. Rev. Phys. Chem.}\ }\textbf
  {\bibinfo {volume} {60}},\ \bibinfo {pages} {241} (\bibinfo {year}
  {2009})}\BibitemShut {NoStop}%
\bibitem [{\citenamefont {Wu}\ \emph {et~al.}(2010)\citenamefont {Wu},
  \citenamefont {Liu}, \citenamefont {Shen}, \citenamefont {Cao},\ and\
  \citenamefont {Silbey}}]{wu2010}%
  \BibitemOpen
  \bibfield  {author} {\bibinfo {author} {\bibfnamefont {J.}~\bibnamefont
  {Wu}}, \bibinfo {author} {\bibfnamefont {F.}~\bibnamefont {Liu}}, \bibinfo
  {author} {\bibfnamefont {Y.}~\bibnamefont {Shen}}, \bibinfo {author}
  {\bibfnamefont {J.}~\bibnamefont {Cao}},\ and\ \bibinfo {author}
  {\bibfnamefont {R.~J.}\ \bibnamefont {Silbey}},\ }\bibfield  {title}
  {\bibinfo {title} {Efficient energy transfer in light-harvesting systems, i:
  optimal temperature, reorganization energy and spatial--temporal
  correlations},\ }\href {https://doi.org/10.1088/1367-2630/12/10/105012}
  {\bibfield  {journal} {\bibinfo  {journal} {New J. Phys.}\ }\textbf {\bibinfo
  {volume} {12}},\ \bibinfo {pages} {105012} (\bibinfo {year}
  {2010})}\BibitemShut {NoStop}%
\bibitem [{\citenamefont {Ritschel}\ \emph {et~al.}(2011)\citenamefont
  {Ritschel}, \citenamefont {Roden}, \citenamefont {Strunz},\ and\
  \citenamefont {Eisfeld}}]{ritschel2011}%
  \BibitemOpen
  \bibfield  {author} {\bibinfo {author} {\bibfnamefont {G.}~\bibnamefont
  {Ritschel}}, \bibinfo {author} {\bibfnamefont {J.}~\bibnamefont {Roden}},
  \bibinfo {author} {\bibfnamefont {W.~T.}\ \bibnamefont {Strunz}},\ and\
  \bibinfo {author} {\bibfnamefont {A.}~\bibnamefont {Eisfeld}},\ }\bibfield
  {title} {\bibinfo {title} {An efficient method to calculate excitation energy
  transfer in light-harvesting systems: application to the
  {Fenna}--{Matthews}--{Olson} complex},\ }\href
  {https://doi.org/10.1088/1367-2630/13/11/113034} {\bibfield  {journal}
  {\bibinfo  {journal} {New J. Phys.}\ }\textbf {\bibinfo {volume} {13}},\
  \bibinfo {pages} {113034} (\bibinfo {year} {2011})}\BibitemShut {NoStop}%
\bibitem [{\citenamefont {Fruchtman}\ \emph {et~al.}(2016)\citenamefont
  {Fruchtman}, \citenamefont {Lambert},\ and\ \citenamefont
  {Gauger}}]{fruchtman2016}%
  \BibitemOpen
  \bibfield  {author} {\bibinfo {author} {\bibfnamefont {A.}~\bibnamefont
  {Fruchtman}}, \bibinfo {author} {\bibfnamefont {N.}~\bibnamefont {Lambert}},\
  and\ \bibinfo {author} {\bibfnamefont {E.~M.}\ \bibnamefont {Gauger}},\
  }\bibfield  {title} {\bibinfo {title} {When do perturbative approaches
  accurately capture the dynamics of complex quantum systems?},\ }\href
  {https://doi.org/https://doi.org/10.1038/srep28204} {\bibfield  {journal}
  {\bibinfo  {journal} {Sci. Rep.}\ }\textbf {\bibinfo {volume} {6}},\ \bibinfo
  {pages} {28204} (\bibinfo {year} {2016})}\BibitemShut {NoStop}%
\bibitem [{\citenamefont {Zanardi}\ and\ \citenamefont
  {Rasetti}(1997)}]{zanardi1997}%
  \BibitemOpen
  \bibfield  {author} {\bibinfo {author} {\bibfnamefont {P.}~\bibnamefont
  {Zanardi}}\ and\ \bibinfo {author} {\bibfnamefont {M.}~\bibnamefont
  {Rasetti}},\ }\bibfield  {title} {\bibinfo {title} {Noiseless quantum
  codes},\ }\href {https://doi.org/10.1103/PhysRevLett.79.3306} {\bibfield
  {journal} {\bibinfo  {journal} {Phys. Rev. Lett.}\ }\textbf {\bibinfo
  {volume} {79}},\ \bibinfo {pages} {3306} (\bibinfo {year}
  {1997})}\BibitemShut {NoStop}%
\bibitem [{\citenamefont {Lidar}\ \emph {et~al.}(1998)\citenamefont {Lidar},
  \citenamefont {Chuang},\ and\ \citenamefont {Whaley}}]{lidar1998}%
  \BibitemOpen
  \bibfield  {author} {\bibinfo {author} {\bibfnamefont {D.~A.}\ \bibnamefont
  {Lidar}}, \bibinfo {author} {\bibfnamefont {I.~L.}\ \bibnamefont {Chuang}},\
  and\ \bibinfo {author} {\bibfnamefont {K.~B.}\ \bibnamefont {Whaley}},\
  }\bibfield  {title} {\bibinfo {title} {Decoherence-free subspaces for quantum
  computation},\ }\href {https://doi.org/10.1103/PhysRevLett.81.2594}
  {\bibfield  {journal} {\bibinfo  {journal} {Phys. Rev. Lett.}\ }\textbf
  {\bibinfo {volume} {81}},\ \bibinfo {pages} {2594} (\bibinfo {year}
  {1998})}\BibitemShut {NoStop}%
\bibitem [{\citenamefont {Lidar}\ \emph {et~al.}(1999)\citenamefont {Lidar},
  \citenamefont {Bacon},\ and\ \citenamefont {Whaley}}]{lidar1999}%
  \BibitemOpen
  \bibfield  {author} {\bibinfo {author} {\bibfnamefont {D.~A.}\ \bibnamefont
  {Lidar}}, \bibinfo {author} {\bibfnamefont {D.}~\bibnamefont {Bacon}},\ and\
  \bibinfo {author} {\bibfnamefont {K.~B.}\ \bibnamefont {Whaley}},\ }\bibfield
   {title} {\bibinfo {title} {Concatenating decoherence-free subspaces with
  quantum error correcting codes},\ }\href
  {https://doi.org/10.1103/PhysRevLett.82.4556} {\bibfield  {journal} {\bibinfo
   {journal} {Phys. Rev. Lett.}\ }\textbf {\bibinfo {volume} {82}},\ \bibinfo
  {pages} {4556} (\bibinfo {year} {1999})}\BibitemShut {NoStop}%
\bibitem [{\citenamefont {Paz}\ and\ \citenamefont
  {Roncaglia}(2008)}]{paz2008}%
  \BibitemOpen
  \bibfield  {author} {\bibinfo {author} {\bibfnamefont {J.~P.}\ \bibnamefont
  {Paz}}\ and\ \bibinfo {author} {\bibfnamefont {A.~J.}\ \bibnamefont
  {Roncaglia}},\ }\bibfield  {title} {\bibinfo {title} {Dynamics of the
  entanglement between two oscillators in the same environment},\ }\href
  {https://doi.org/10.1103/PhysRevLett.100.220401} {\bibfield  {journal}
  {\bibinfo  {journal} {Phys. Rev. Lett.}\ }\textbf {\bibinfo {volume} {100}},\
  \bibinfo {pages} {220401} (\bibinfo {year} {2008})}\BibitemShut {NoStop}%
\bibitem [{\citenamefont {Paz}\ and\ \citenamefont
  {Roncaglia}(2009)}]{paz2009}%
  \BibitemOpen
  \bibfield  {author} {\bibinfo {author} {\bibfnamefont {J.~P.}\ \bibnamefont
  {Paz}}\ and\ \bibinfo {author} {\bibfnamefont {A.~J.}\ \bibnamefont
  {Roncaglia}},\ }\bibfield  {title} {\bibinfo {title} {Dynamical phases for
  the evolution of the entanglement between two oscillators coupled to the same
  environment},\ }\href {https://doi.org/10.1103/PhysRevA.79.032102} {\bibfield
   {journal} {\bibinfo  {journal} {Phys. Rev. A}\ }\textbf {\bibinfo {volume}
  {79}},\ \bibinfo {pages} {032102} (\bibinfo {year} {2009})}\BibitemShut
  {NoStop}%
\bibitem [{\citenamefont {Anto-Sztrikacs}\ \emph {et~al.}(2023)\citenamefont
  {Anto-Sztrikacs}, \citenamefont {Nazir},\ and\ \citenamefont
  {Segal}}]{anto2023}%
  \BibitemOpen
  \bibfield  {author} {\bibinfo {author} {\bibfnamefont {N.}~\bibnamefont
  {Anto-Sztrikacs}}, \bibinfo {author} {\bibfnamefont {A.}~\bibnamefont
  {Nazir}},\ and\ \bibinfo {author} {\bibfnamefont {D.}~\bibnamefont {Segal}},\
  }\bibfield  {title} {\bibinfo {title} {Effective-hamiltonian theory of open
  quantum systems at strong coupling},\ }\href
  {https://doi.org/10.1103/PRXQuantum.4.020307} {\bibfield  {journal} {\bibinfo
   {journal} {PRX Quantum}\ }\textbf {\bibinfo {volume} {4}},\ \bibinfo {pages}
  {020307} (\bibinfo {year} {2023})}\BibitemShut {NoStop}%
\bibitem [{\citenamefont {Martinazzo}\ \emph {et~al.}(2011)\citenamefont
  {Martinazzo}, \citenamefont {Vacchini}, \citenamefont {Hughes},\ and\
  \citenamefont {Burghardt}}]{martinazzo2011}%
  \BibitemOpen
  \bibfield  {author} {\bibinfo {author} {\bibfnamefont {R.}~\bibnamefont
  {Martinazzo}}, \bibinfo {author} {\bibfnamefont {B.}~\bibnamefont
  {Vacchini}}, \bibinfo {author} {\bibfnamefont {K.~H.}\ \bibnamefont
  {Hughes}},\ and\ \bibinfo {author} {\bibfnamefont {I.}~\bibnamefont
  {Burghardt}},\ }\bibfield  {title} {\bibinfo {title} {Communication:
  Universal {Markovian} reduction of {Brownian} particle dynamics},\ }\href
  {https://doi.org/10.1063/1.3532408} {\bibfield  {journal} {\bibinfo
  {journal} {J. Chem. Phys.}\ }\textbf {\bibinfo {volume} {134}},\ \bibinfo
  {pages} {011101} (\bibinfo {year} {2011})}\BibitemShut {NoStop}%
\bibitem [{\citenamefont {Iles-Smith}\ \emph {et~al.}(2014)\citenamefont
  {Iles-Smith}, \citenamefont {Lambert},\ and\ \citenamefont
  {Nazir}}]{smith2014}%
  \BibitemOpen
  \bibfield  {author} {\bibinfo {author} {\bibfnamefont {J.}~\bibnamefont
  {Iles-Smith}}, \bibinfo {author} {\bibfnamefont {N.}~\bibnamefont
  {Lambert}},\ and\ \bibinfo {author} {\bibfnamefont {A.}~\bibnamefont
  {Nazir}},\ }\bibfield  {title} {\bibinfo {title} {Environmental dynamics,
  correlations, and the emergence of noncanonical equilibrium states in open
  quantum systems},\ }\href {https://doi.org/10.1103/PhysRevA.90.032114}
  {\bibfield  {journal} {\bibinfo  {journal} {Phys. Rev. A}\ }\textbf {\bibinfo
  {volume} {90}},\ \bibinfo {pages} {032114} (\bibinfo {year}
  {2014})}\BibitemShut {NoStop}%
\bibitem [{\citenamefont {Strasberg}\ \emph {et~al.}(2016)\citenamefont
  {Strasberg}, \citenamefont {Schaller}, \citenamefont {Lambert},\ and\
  \citenamefont {Brandes}}]{strasberg2016}%
  \BibitemOpen
  \bibfield  {author} {\bibinfo {author} {\bibfnamefont {P.}~\bibnamefont
  {Strasberg}}, \bibinfo {author} {\bibfnamefont {G.}~\bibnamefont {Schaller}},
  \bibinfo {author} {\bibfnamefont {N.}~\bibnamefont {Lambert}},\ and\ \bibinfo
  {author} {\bibfnamefont {T.}~\bibnamefont {Brandes}},\ }\bibfield  {title}
  {\bibinfo {title} {Nonequilibrium thermodynamics in the strong coupling and
  non-{Markovian} regime based on a reaction coordinate mapping},\ }\href
  {https://doi.org/10.1088/1367-2630/18/7/073007} {\bibfield  {journal}
  {\bibinfo  {journal} {New J. Phys.}\ }\textbf {\bibinfo {volume} {18}},\
  \bibinfo {pages} {073007} (\bibinfo {year} {2016})}\BibitemShut {NoStop}%
\bibitem [{\citenamefont {Correa}\ \emph {et~al.}(2019)\citenamefont {Correa},
  \citenamefont {Xu}, \citenamefont {Morris},\ and\ \citenamefont
  {Adesso}}]{correa2019}%
  \BibitemOpen
  \bibfield  {author} {\bibinfo {author} {\bibfnamefont {L.~A.}\ \bibnamefont
  {Correa}}, \bibinfo {author} {\bibfnamefont {B.}~\bibnamefont {Xu}}, \bibinfo
  {author} {\bibfnamefont {B.}~\bibnamefont {Morris}},\ and\ \bibinfo {author}
  {\bibfnamefont {G.}~\bibnamefont {Adesso}},\ }\bibfield  {title} {\bibinfo
  {title} {Pushing the limits of the reaction-coordinate mapping},\ }\href
  {https://doi.org/10.1063/1.5114690} {\bibfield  {journal} {\bibinfo
  {journal} {J. Chem. Phys.}\ }\textbf {\bibinfo {volume} {151}},\ \bibinfo
  {pages} {094107} (\bibinfo {year} {2019})}\BibitemShut {NoStop}%
\bibitem [{\citenamefont {Nazir}\ and\ \citenamefont
  {Schaller}(2018)}]{nazir2018}%
  \BibitemOpen
  \bibfield  {author} {\bibinfo {author} {\bibfnamefont {A.}~\bibnamefont
  {Nazir}}\ and\ \bibinfo {author} {\bibfnamefont {G.}~\bibnamefont
  {Schaller}},\ }\bibfield  {title} {\bibinfo {title} {The reaction coordinate
  mapping in quantum thermodynamics},\ }in\ \href@noop {} {\emph {\bibinfo
  {booktitle} {Thermodynamics in the Quantum Regime: Fundamental Aspects and
  New Directions}}},\ \bibinfo {editor} {edited by\ \bibinfo {editor}
  {\bibfnamefont {F.}~\bibnamefont {Binder}}, \bibinfo {editor} {\bibfnamefont
  {L.~A.}\ \bibnamefont {Correa}}, \bibinfo {editor} {\bibfnamefont
  {C.}~\bibnamefont {Gogolin}}, \bibinfo {editor} {\bibfnamefont
  {J.}~\bibnamefont {Anders}},\ and\ \bibinfo {editor} {\bibfnamefont
  {G.}~\bibnamefont {Adesso}}}\ (\bibinfo  {publisher} {Springer},\ \bibinfo
  {address} {Cham, Switzerland},\ \bibinfo {year} {2018})\ pp.\ \bibinfo
  {pages} {551--577}\BibitemShut {NoStop}%
\bibitem [{\citenamefont {Hepp}\ and\ \citenamefont
  {Lieb}(1973{\natexlab{a}})}]{hepp1973}%
  \BibitemOpen
  \bibfield  {author} {\bibinfo {author} {\bibfnamefont {K.}~\bibnamefont
  {Hepp}}\ and\ \bibinfo {author} {\bibfnamefont {E.~H.}\ \bibnamefont
  {Lieb}},\ }\bibfield  {title} {\bibinfo {title} {On the superradiant phase
  transition for molecules in a quantized radiation field: the {Dicke} maser
  model},\ }\href {https://doi.org/10.1016/0003-4916(73)90039-0} {\bibfield
  {journal} {\bibinfo  {journal} {Ann. Phys.}\ }\textbf {\bibinfo {volume}
  {76}},\ \bibinfo {pages} {360} (\bibinfo {year}
  {1973}{\natexlab{a}})}\BibitemShut {NoStop}%
\bibitem [{\citenamefont {Hepp}\ and\ \citenamefont
  {Lieb}(1973{\natexlab{b}})}]{hepp1973b}%
  \BibitemOpen
  \bibfield  {author} {\bibinfo {author} {\bibfnamefont {K.}~\bibnamefont
  {Hepp}}\ and\ \bibinfo {author} {\bibfnamefont {E.~H.}\ \bibnamefont
  {Lieb}},\ }\bibfield  {title} {\bibinfo {title} {Equilibrium statistical
  mechanics of matter interacting with the quantized radiation field},\ }\href
  {https://doi.org/10.1103/PhysRevA.8.2517} {\bibfield  {journal} {\bibinfo
  {journal} {Phys. Rev. A}\ }\textbf {\bibinfo {volume} {8}},\ \bibinfo {pages}
  {2517} (\bibinfo {year} {1973}{\natexlab{b}})}\BibitemShut {NoStop}%
\bibitem [{\citenamefont {Wang}\ and\ \citenamefont {Hioe}(1973)}]{wang1973}%
  \BibitemOpen
  \bibfield  {author} {\bibinfo {author} {\bibfnamefont {Y.~K.}\ \bibnamefont
  {Wang}}\ and\ \bibinfo {author} {\bibfnamefont {F.~T.}\ \bibnamefont
  {Hioe}},\ }\bibfield  {title} {\bibinfo {title} {Phase transition in the
  {Dicke} model of superradiance},\ }\href
  {https://doi.org/10.1103/PhysRevA.7.831} {\bibfield  {journal} {\bibinfo
  {journal} {Phys. Rev. A}\ }\textbf {\bibinfo {volume} {7}},\ \bibinfo {pages}
  {831} (\bibinfo {year} {1973})}\BibitemShut {NoStop}%
\bibitem [{\citenamefont {Hioe}(1973)}]{hioe1973}%
  \BibitemOpen
  \bibfield  {author} {\bibinfo {author} {\bibfnamefont {F.~T.}\ \bibnamefont
  {Hioe}},\ }\bibfield  {title} {\bibinfo {title} {Phase transitions in some
  generalized {Dicke} models of superradiance},\ }\href
  {https://doi.org/10.1103/PhysRevA.8.1440} {\bibfield  {journal} {\bibinfo
  {journal} {Phys. Rev. A}\ }\textbf {\bibinfo {volume} {8}},\ \bibinfo {pages}
  {1440} (\bibinfo {year} {1973})}\BibitemShut {NoStop}%
\bibitem [{\citenamefont {Carmichael}\ \emph {et~al.}(1973)\citenamefont
  {Carmichael}, \citenamefont {Gardiner},\ and\ \citenamefont
  {Walls}}]{carmichael1973}%
  \BibitemOpen
  \bibfield  {author} {\bibinfo {author} {\bibfnamefont {H.}~\bibnamefont
  {Carmichael}}, \bibinfo {author} {\bibfnamefont {C.}~\bibnamefont
  {Gardiner}},\ and\ \bibinfo {author} {\bibfnamefont {D.}~\bibnamefont
  {Walls}},\ }\bibfield  {title} {\bibinfo {title} {Higher order corrections to
  the {Dicke} superradiant phase transition},\ }\href
  {https://doi.org/10.1016/0375-9601(73)90679-8} {\bibfield  {journal}
  {\bibinfo  {journal} {Phys. Lett. A}\ }\textbf {\bibinfo {volume} {46}},\
  \bibinfo {pages} {47} (\bibinfo {year} {1973})}\BibitemShut {NoStop}%
\bibitem [{\citenamefont {Duncan}(1974)}]{duncan1974}%
  \BibitemOpen
  \bibfield  {author} {\bibinfo {author} {\bibfnamefont {G.~C.}\ \bibnamefont
  {Duncan}},\ }\bibfield  {title} {\bibinfo {title} {Effect of antiresonant
  atom-field interactions on phase transitions in the {Dicke} model},\ }\href
  {https://doi.org/10.1103/PhysRevA.9.418} {\bibfield  {journal} {\bibinfo
  {journal} {Phys. Rev. A}\ }\textbf {\bibinfo {volume} {9}},\ \bibinfo {pages}
  {418} (\bibinfo {year} {1974})}\BibitemShut {NoStop}%
\bibitem [{\citenamefont {Garraway}(2011)}]{garraway2011}%
  \BibitemOpen
  \bibfield  {author} {\bibinfo {author} {\bibfnamefont {B.~M.}\ \bibnamefont
  {Garraway}},\ }\bibfield  {title} {\bibinfo {title} {The {Dicke} model in
  quantum optics: {Dicke} model revisited},\ }\href
  {https://doi.org/10.1098/rsta.2010.0333} {\bibfield  {journal} {\bibinfo
  {journal} {Phil. Trans. R. Soc. A}\ }\textbf {\bibinfo {volume} {369}},\
  \bibinfo {pages} {1137} (\bibinfo {year} {2011})}\BibitemShut {NoStop}%
\bibitem [{\citenamefont {Lang}\ and\ \citenamefont {Firsov}(1963)}]{lang1963}%
  \BibitemOpen
  \bibfield  {author} {\bibinfo {author} {\bibfnamefont {I.~G.}\ \bibnamefont
  {Lang}}\ and\ \bibinfo {author} {\bibfnamefont {Y.~A.}\ \bibnamefont
  {Firsov}},\ }\bibfield  {title} {\bibinfo {title} {Kinetic theory of
  semiconductors with low mobility},\ }\href@noop {} {\bibfield  {journal}
  {\bibinfo  {journal} {Sov. Phys. JETP}\ }\textbf {\bibinfo {volume} {16}},\
  \bibinfo {pages} {1301} (\bibinfo {year} {1963})}\BibitemShut {NoStop}%
\bibitem [{\citenamefont {Vidal}\ and\ \citenamefont
  {Werner}(2002)}]{vidal2002}%
  \BibitemOpen
  \bibfield  {author} {\bibinfo {author} {\bibfnamefont {G.}~\bibnamefont
  {Vidal}}\ and\ \bibinfo {author} {\bibfnamefont {R.~F.}\ \bibnamefont
  {Werner}},\ }\bibfield  {title} {\bibinfo {title} {Computable measure of
  entanglement},\ }\href {https://doi.org/10.1103/PhysRevA.65.032314}
  {\bibfield  {journal} {\bibinfo  {journal} {Phys. Rev. A}\ }\textbf {\bibinfo
  {volume} {65}},\ \bibinfo {pages} {032314} (\bibinfo {year}
  {2002})}\BibitemShut {NoStop}%
\bibitem [{\citenamefont {\ifmmode~\dot{Z}\else \.{Z}\fi{}yczkowski}\ \emph
  {et~al.}(1998)\citenamefont {\ifmmode~\dot{Z}\else \.{Z}\fi{}yczkowski},
  \citenamefont {Horodecki}, \citenamefont {Sanpera},\ and\ \citenamefont
  {Lewenstein}}]{kyczkowski1998}%
  \BibitemOpen
  \bibfield  {author} {\bibinfo {author} {\bibfnamefont {K.}~\bibnamefont
  {\ifmmode~\dot{Z}\else \.{Z}\fi{}yczkowski}}, \bibinfo {author}
  {\bibfnamefont {P.}~\bibnamefont {Horodecki}}, \bibinfo {author}
  {\bibfnamefont {A.}~\bibnamefont {Sanpera}},\ and\ \bibinfo {author}
  {\bibfnamefont {M.}~\bibnamefont {Lewenstein}},\ }\bibfield  {title}
  {\bibinfo {title} {Volume of the set of separable states},\ }\href
  {https://doi.org/10.1103/PhysRevA.58.883} {\bibfield  {journal} {\bibinfo
  {journal} {Phys. Rev. A}\ }\textbf {\bibinfo {volume} {58}},\ \bibinfo
  {pages} {883} (\bibinfo {year} {1998})}\BibitemShut {NoStop}%
\bibitem [{\citenamefont {Lee}\ \emph {et~al.}(2000)\citenamefont {Lee},
  \citenamefont {Kim}, \citenamefont {Park},\ and\ \citenamefont
  {Lee}}]{lee2000}%
  \BibitemOpen
  \bibfield  {author} {\bibinfo {author} {\bibfnamefont {J.}~\bibnamefont
  {Lee}}, \bibinfo {author} {\bibfnamefont {M.}~\bibnamefont {Kim}}, \bibinfo
  {author} {\bibfnamefont {Y.}~\bibnamefont {Park}},\ and\ \bibinfo {author}
  {\bibfnamefont {S.}~\bibnamefont {Lee}},\ }\bibfield  {title} {\bibinfo
  {title} {Partial teleportation of entanglement in a noisy environment},\
  }\href {https://doi.org/10.1080/09500340008235138} {\bibfield  {journal}
  {\bibinfo  {journal} {J. Mod. Opt.}\ }\textbf {\bibinfo {volume} {47}},\
  \bibinfo {pages} {2151} (\bibinfo {year} {2000})}\BibitemShut {NoStop}%
\bibitem [{\citenamefont {Eisert}(2001)}]{eisert2001}%
  \BibitemOpen
  \bibfield  {author} {\bibinfo {author} {\bibfnamefont {J.}~\bibnamefont
  {Eisert}},\ }\emph {\bibinfo {title} {Entanglement in Quantum Information
  Theory}},\ \href@noop {} {Ph.D. thesis},\ \bibinfo  {school} {University of
  Potsdam} (\bibinfo {year} {2001})\BibitemShut {NoStop}%
\bibitem [{\citenamefont {Plenio}(2005)}]{plenio2005}%
  \BibitemOpen
  \bibfield  {author} {\bibinfo {author} {\bibfnamefont {M.~B.}\ \bibnamefont
  {Plenio}},\ }\bibfield  {title} {\bibinfo {title} {Logarithmic negativity: A
  full entanglement monotone that is not convex},\ }\href
  {https://doi.org/10.1103/PhysRevLett.95.090503} {\bibfield  {journal}
  {\bibinfo  {journal} {Phys. Rev. Lett.}\ }\textbf {\bibinfo {volume} {95}},\
  \bibinfo {pages} {090503} (\bibinfo {year} {2005})}\BibitemShut {NoStop}%
\bibitem [{\citenamefont {Vidal}(2000)}]{vidal2000}%
  \BibitemOpen
  \bibfield  {author} {\bibinfo {author} {\bibfnamefont {G.}~\bibnamefont
  {Vidal}},\ }\bibfield  {title} {\bibinfo {title} {Entanglement monotones},\
  }\href {https://doi.org/10.1080/09500340008244048} {\bibfield  {journal}
  {\bibinfo  {journal} {J. Mod. Opt.}\ }\textbf {\bibinfo {volume} {47}},\
  \bibinfo {pages} {355} (\bibinfo {year} {2000})}\BibitemShut {NoStop}%
\bibitem [{\citenamefont {Horodecki}\ \emph {et~al.}(1996)\citenamefont
  {Horodecki}, \citenamefont {Horodecki},\ and\ \citenamefont
  {Horodecki}}]{horodecki1996}%
  \BibitemOpen
  \bibfield  {author} {\bibinfo {author} {\bibfnamefont {M.}~\bibnamefont
  {Horodecki}}, \bibinfo {author} {\bibfnamefont {P.}~\bibnamefont
  {Horodecki}},\ and\ \bibinfo {author} {\bibfnamefont {R.}~\bibnamefont
  {Horodecki}},\ }\bibfield  {title} {\bibinfo {title} {Separability of mixed
  states: necessary and sufficient conditions},\ }\href
  {https://doi.org/10.1016/S0375-9601(96)00706-2} {\bibfield  {journal}
  {\bibinfo  {journal} {Phys. Lett. A}\ }\textbf {\bibinfo {volume} {223}},\
  \bibinfo {pages} {1} (\bibinfo {year} {1996})}\BibitemShut {NoStop}%
\bibitem [{Note1()}]{Note1}%
  \BibitemOpen
  \bibinfo {note} {Note entanglement of $\rho _\protect \mathrm {MF}^\protect
  \mathrm {weak}$ requires the retention of co-rotating terms in Eq.~\protect
  \textup {\hbox {\mathsurround \z@ \protect \normalfont (\ignorespaces \ref
  {H}\unskip \@@italiccorr )}}. The ground state of $\protect \hat {H}_\protect
  \mathrm {S}^\protect \mathrm {RC}$ with co-rotating terms removed exhibits
  entanglement only for $g>\protect \sqrt {2\omega _z\Omega }$~\cite
  {buzek2005}.}\BibitemShut {Stop}%
\bibitem [{\citenamefont {Lee}\ and\ \citenamefont {Law}(2013)}]{lee2013}%
  \BibitemOpen
  \bibfield  {author} {\bibinfo {author} {\bibfnamefont {K.~M.~C.}\
  \bibnamefont {Lee}}\ and\ \bibinfo {author} {\bibfnamefont {C.~K.}\
  \bibnamefont {Law}},\ }\bibfield  {title} {\bibinfo {title} {Ground state of
  a resonant two-qubit cavity system in the ultrastrong-coupling regime},\
  }\href {https://doi.org/10.1103/PhysRevA.88.015802} {\bibfield  {journal}
  {\bibinfo  {journal} {Phys. Rev. A}\ }\textbf {\bibinfo {volume} {88}},\
  \bibinfo {pages} {015802} (\bibinfo {year} {2013})}\BibitemShut {NoStop}%
\bibitem [{\citenamefont {Lambert}\ \emph {et~al.}(2004)\citenamefont
  {Lambert}, \citenamefont {Emary},\ and\ \citenamefont
  {Brandes}}]{lambert2004}%
  \BibitemOpen
  \bibfield  {author} {\bibinfo {author} {\bibfnamefont {N.}~\bibnamefont
  {Lambert}}, \bibinfo {author} {\bibfnamefont {C.}~\bibnamefont {Emary}},\
  and\ \bibinfo {author} {\bibfnamefont {T.}~\bibnamefont {Brandes}},\
  }\bibfield  {title} {\bibinfo {title} {Entanglement and the phase transition
  in single-mode superradiance},\ }\href
  {https://doi.org/10.1103/PhysRevLett.92.073602} {\bibfield  {journal}
  {\bibinfo  {journal} {Phys. Rev. Lett.}\ }\textbf {\bibinfo {volume} {92}},\
  \bibinfo {pages} {073602} (\bibinfo {year} {2004})}\BibitemShut {NoStop}%
\bibitem [{\citenamefont {Bu\ifmmode~\check{z}\else \v{z}\fi{}ek}\ \emph
  {et~al.}(2005)\citenamefont {Bu\ifmmode~\check{z}\else \v{z}\fi{}ek},
  \citenamefont {Orszag},\ and\ \citenamefont {Ro\ifmmode~\check{s}\else
  \v{s}\fi{}ko}}]{buzek2005}%
  \BibitemOpen
  \bibfield  {author} {\bibinfo {author} {\bibfnamefont {V.}~\bibnamefont
  {Bu\ifmmode~\check{z}\else \v{z}\fi{}ek}}, \bibinfo {author} {\bibfnamefont
  {M.}~\bibnamefont {Orszag}},\ and\ \bibinfo {author} {\bibfnamefont
  {M.}~\bibnamefont {Ro\ifmmode~\check{s}\else \v{s}\fi{}ko}},\ }\bibfield
  {title} {\bibinfo {title} {Instability and entanglement of the ground state
  of the {Dicke} model},\ }\href
  {https://doi.org/10.1103/PhysRevLett.94.163601} {\bibfield  {journal}
  {\bibinfo  {journal} {Phys. Rev. Lett.}\ }\textbf {\bibinfo {volume} {94}},\
  \bibinfo {pages} {163601} (\bibinfo {year} {2005})}\BibitemShut {NoStop}%
\bibitem [{\citenamefont {Agarwal}\ \emph {et~al.}(1997)\citenamefont
  {Agarwal}, \citenamefont {Puri},\ and\ \citenamefont {Singh}}]{agarwal1997}%
  \BibitemOpen
  \bibfield  {author} {\bibinfo {author} {\bibfnamefont {G.~S.}\ \bibnamefont
  {Agarwal}}, \bibinfo {author} {\bibfnamefont {R.~R.}\ \bibnamefont {Puri}},\
  and\ \bibinfo {author} {\bibfnamefont {R.~P.}\ \bibnamefont {Singh}},\
  }\bibfield  {title} {\bibinfo {title} {Atomic {Schr\"odinger} cat states},\
  }\href {https://doi.org/10.1103/PhysRevA.56.2249} {\bibfield  {journal}
  {\bibinfo  {journal} {Phys. Rev. A}\ }\textbf {\bibinfo {volume} {56}},\
  \bibinfo {pages} {2249} (\bibinfo {year} {1997})}\BibitemShut {NoStop}%
\bibitem [{\citenamefont {Zheng}\ and\ \citenamefont {Guo}(2000)}]{zheng2000}%
  \BibitemOpen
  \bibfield  {author} {\bibinfo {author} {\bibfnamefont {S.-B.}\ \bibnamefont
  {Zheng}}\ and\ \bibinfo {author} {\bibfnamefont {G.-C.}\ \bibnamefont
  {Guo}},\ }\bibfield  {title} {\bibinfo {title} {Efficient scheme for two-atom
  entanglement and quantum information processing in cavity {QED}},\ }\href
  {https://doi.org/10.1103/PhysRevLett.85.2392} {\bibfield  {journal} {\bibinfo
   {journal} {Phys. Rev. Lett.}\ }\textbf {\bibinfo {volume} {85}},\ \bibinfo
  {pages} {2392} (\bibinfo {year} {2000})}\BibitemShut {NoStop}%
\bibitem [{\citenamefont {Zheng}(2001)}]{zheng2001b}%
  \BibitemOpen
  \bibfield  {author} {\bibinfo {author} {\bibfnamefont {S.-B.}\ \bibnamefont
  {Zheng}},\ }\bibfield  {title} {\bibinfo {title} {One-step synthesis of
  multiatom {Greenberger}-{Horne}-{Zeilinger} states},\ }\href
  {https://doi.org/10.1103/PhysRevLett.87.230404} {\bibfield  {journal}
  {\bibinfo  {journal} {Phys. Rev. Lett.}\ }\textbf {\bibinfo {volume} {87}},\
  \bibinfo {pages} {230404} (\bibinfo {year} {2001})}\BibitemShut {NoStop}%
\bibitem [{\citenamefont {Zheng}\ and\ \citenamefont {Guo}(2001)}]{zheng2001}%
  \BibitemOpen
  \bibfield  {author} {\bibinfo {author} {\bibfnamefont {S.-B.}\ \bibnamefont
  {Zheng}}\ and\ \bibinfo {author} {\bibfnamefont {G.-C.}\ \bibnamefont
  {Guo}},\ }\bibfield  {title} {\bibinfo {title} {Teleportation of atomic
  states within cavities in thermal states},\ }\href
  {https://doi.org/10.1103/PhysRevA.63.044302} {\bibfield  {journal} {\bibinfo
  {journal} {Phys. Rev. A}\ }\textbf {\bibinfo {volume} {63}},\ \bibinfo
  {pages} {044302} (\bibinfo {year} {2001})}\BibitemShut {NoStop}%
\bibitem [{\citenamefont {Arnesen}\ \emph {et~al.}(2001)\citenamefont
  {Arnesen}, \citenamefont {Bose},\ and\ \citenamefont {Vedral}}]{arnesen2001}%
  \BibitemOpen
  \bibfield  {author} {\bibinfo {author} {\bibfnamefont {M.~C.}\ \bibnamefont
  {Arnesen}}, \bibinfo {author} {\bibfnamefont {S.}~\bibnamefont {Bose}},\ and\
  \bibinfo {author} {\bibfnamefont {V.}~\bibnamefont {Vedral}},\ }\bibfield
  {title} {\bibinfo {title} {Natural thermal and magnetic entanglement in the
  {1D} {Heisenberg} model},\ }\href
  {https://doi.org/10.1103/PhysRevLett.87.017901} {\bibfield  {journal}
  {\bibinfo  {journal} {Phys. Rev. Lett.}\ }\textbf {\bibinfo {volume} {87}},\
  \bibinfo {pages} {017901} (\bibinfo {year} {2001})}\BibitemShut {NoStop}%
\bibitem [{\citenamefont {Wang}\ \emph {et~al.}(2009)\citenamefont {Wang},
  \citenamefont {Liu},\ and\ \citenamefont {He}}]{wang2009}%
  \BibitemOpen
  \bibfield  {author} {\bibinfo {author} {\bibfnamefont {H.}~\bibnamefont
  {Wang}}, \bibinfo {author} {\bibfnamefont {S.}~\bibnamefont {Liu}},\ and\
  \bibinfo {author} {\bibfnamefont {J.}~\bibnamefont {He}},\ }\bibfield
  {title} {\bibinfo {title} {Thermal entanglement in two-atom cavity {QED} and
  the entangled quantum {Otto} engine},\ }\href
  {https://doi.org/10.1103/PhysRevE.79.041113} {\bibfield  {journal} {\bibinfo
  {journal} {Phys. Rev. E}\ }\textbf {\bibinfo {volume} {79}},\ \bibinfo
  {pages} {041113} (\bibinfo {year} {2009})}\BibitemShut {NoStop}%
\bibitem [{\citenamefont {Hughes}\ \emph
  {et~al.}(2009{\natexlab{a}})\citenamefont {Hughes}, \citenamefont {Christ},\
  and\ \citenamefont {Burghardt}}]{hughes2009}%
  \BibitemOpen
  \bibfield  {author} {\bibinfo {author} {\bibfnamefont {K.~H.}\ \bibnamefont
  {Hughes}}, \bibinfo {author} {\bibfnamefont {C.~D.}\ \bibnamefont {Christ}},\
  and\ \bibinfo {author} {\bibfnamefont {I.}~\bibnamefont {Burghardt}},\
  }\bibfield  {title} {\bibinfo {title} {Effective-mode representation of
  non-{Markovian} dynamics: A hierarchical approximation of the spectral
  density. {I}. application to single surface dynamics},\ }\href
  {https://doi.org/10.1063/1.3159671} {\bibfield  {journal} {\bibinfo
  {journal} {J. Chem. Phys.}\ }\textbf {\bibinfo {volume} {131}},\ \bibinfo
  {pages} {024109} (\bibinfo {year} {2009}{\natexlab{a}})}\BibitemShut
  {NoStop}%
\bibitem [{\citenamefont {Hughes}\ \emph
  {et~al.}(2009{\natexlab{b}})\citenamefont {Hughes}, \citenamefont {Christ},\
  and\ \citenamefont {Burghardt}}]{hughes2009b}%
  \BibitemOpen
  \bibfield  {author} {\bibinfo {author} {\bibfnamefont {K.~H.}\ \bibnamefont
  {Hughes}}, \bibinfo {author} {\bibfnamefont {C.~D.}\ \bibnamefont {Christ}},\
  and\ \bibinfo {author} {\bibfnamefont {I.}~\bibnamefont {Burghardt}},\
  }\bibfield  {title} {\bibinfo {title} {Effective-mode representation of
  non-{Markovian} dynamics: A hierarchical approximation of the spectral
  density. {II}. application to environment-induced nonadiabatic dynamics},\
  }\href {https://doi.org/10.1063/1.3226343} {\bibfield  {journal} {\bibinfo
  {journal} {J. Chem. Phys.}\ }\textbf {\bibinfo {volume} {131}},\ \bibinfo
  {pages} {124108} (\bibinfo {year} {2009}{\natexlab{b}})}\BibitemShut
  {NoStop}%
\bibitem [{\citenamefont {Tanimura}\ and\ \citenamefont
  {Kubo}(1989)}]{tanimura1989}%
  \BibitemOpen
  \bibfield  {author} {\bibinfo {author} {\bibfnamefont {Y.}~\bibnamefont
  {Tanimura}}\ and\ \bibinfo {author} {\bibfnamefont {R.}~\bibnamefont
  {Kubo}},\ }\bibfield  {title} {\bibinfo {title} {Time evolution of a quantum
  system in contact with a nearly {Gaussian}-{Markoffian} noise bath},\ }\href
  {https://doi.org/https://doi.org/10.1143/JPSJ.58.101} {\bibfield  {journal}
  {\bibinfo  {journal} {J. Phys. Soc. Jpn.}\ }\textbf {\bibinfo {volume}
  {58}},\ \bibinfo {pages} {101} (\bibinfo {year} {1989})}\BibitemShut
  {NoStop}%
\bibitem [{\citenamefont {Tanimura}(1990)}]{tanimura1990}%
  \BibitemOpen
  \bibfield  {author} {\bibinfo {author} {\bibfnamefont {Y.}~\bibnamefont
  {Tanimura}},\ }\bibfield  {title} {\bibinfo {title} {Nonperturbative
  expansion method for a quantum system coupled to a harmonic-oscillator
  bath},\ }\href {https://doi.org/10.1103/PhysRevA.41.6676} {\bibfield
  {journal} {\bibinfo  {journal} {Phys. Rev. A}\ }\textbf {\bibinfo {volume}
  {41}},\ \bibinfo {pages} {6676} (\bibinfo {year} {1990})}\BibitemShut
  {NoStop}%
\bibitem [{\citenamefont {Xu}\ and\ \citenamefont {Yan}(2007)}]{xu2007}%
  \BibitemOpen
  \bibfield  {author} {\bibinfo {author} {\bibfnamefont {R.-X.}\ \bibnamefont
  {Xu}}\ and\ \bibinfo {author} {\bibfnamefont {Y.}~\bibnamefont {Yan}},\
  }\bibfield  {title} {\bibinfo {title} {Dynamics of quantum dissipation
  systems interacting with bosonic canonical bath: Hierarchical equations of
  motion approach},\ }\href {https://doi.org/10.1103/PhysRevE.75.031107}
  {\bibfield  {journal} {\bibinfo  {journal} {Phys. Rev. E}\ }\textbf {\bibinfo
  {volume} {75}},\ \bibinfo {pages} {031107} (\bibinfo {year}
  {2007})}\BibitemShut {NoStop}%
\bibitem [{\citenamefont {Prior}\ \emph {et~al.}(2010)\citenamefont {Prior},
  \citenamefont {Chin}, \citenamefont {Huelga},\ and\ \citenamefont
  {Plenio}}]{prior2010}%
  \BibitemOpen
  \bibfield  {author} {\bibinfo {author} {\bibfnamefont {J.}~\bibnamefont
  {Prior}}, \bibinfo {author} {\bibfnamefont {A.~W.}\ \bibnamefont {Chin}},
  \bibinfo {author} {\bibfnamefont {S.~F.}\ \bibnamefont {Huelga}},\ and\
  \bibinfo {author} {\bibfnamefont {M.~B.}\ \bibnamefont {Plenio}},\ }\bibfield
   {title} {\bibinfo {title} {Efficient simulation of strong system-environment
  interactions},\ }\href {https://doi.org/10.1103/PhysRevLett.105.050404}
  {\bibfield  {journal} {\bibinfo  {journal} {Phys. Rev. Lett.}\ }\textbf
  {\bibinfo {volume} {105}},\ \bibinfo {pages} {050404} (\bibinfo {year}
  {2010})}\BibitemShut {NoStop}%
\bibitem [{\citenamefont {Chin}\ \emph {et~al.}(2010)\citenamefont {Chin},
  \citenamefont {Rivas}, \citenamefont {Huelga},\ and\ \citenamefont
  {Plenio}}]{chin2010}%
  \BibitemOpen
  \bibfield  {author} {\bibinfo {author} {\bibfnamefont {A.~W.}\ \bibnamefont
  {Chin}}, \bibinfo {author} {\bibfnamefont {{\'A}.}~\bibnamefont {Rivas}},
  \bibinfo {author} {\bibfnamefont {S.~F.}\ \bibnamefont {Huelga}},\ and\
  \bibinfo {author} {\bibfnamefont {M.~B.}\ \bibnamefont {Plenio}},\ }\bibfield
   {title} {\bibinfo {title} {Exact mapping between system-reservoir quantum
  models and semi-infinite discrete chains using orthogonal polynomials},\
  }\href {https://doi.org/10.1063/1.3490188} {\bibfield  {journal} {\bibinfo
  {journal} {J. Math. Phys.}\ }\textbf {\bibinfo {volume} {51}},\ \bibinfo
  {pages} {092109} (\bibinfo {year} {2010})}\BibitemShut {NoStop}%
\bibitem [{\citenamefont {Woods}\ \emph {et~al.}(2015)\citenamefont {Woods},
  \citenamefont {Cramer},\ and\ \citenamefont {Plenio}}]{woods2015}%
  \BibitemOpen
  \bibfield  {author} {\bibinfo {author} {\bibfnamefont {M.~P.}\ \bibnamefont
  {Woods}}, \bibinfo {author} {\bibfnamefont {M.}~\bibnamefont {Cramer}},\ and\
  \bibinfo {author} {\bibfnamefont {M.~B.}\ \bibnamefont {Plenio}},\ }\bibfield
   {title} {\bibinfo {title} {Simulating bosonic baths with error bars},\
  }\href {https://doi.org/10.1103/PhysRevLett.115.130401} {\bibfield  {journal}
  {\bibinfo  {journal} {Phys. Rev. Lett.}\ }\textbf {\bibinfo {volume} {115}},\
  \bibinfo {pages} {130401} (\bibinfo {year} {2015})}\BibitemShut {NoStop}%
\bibitem [{\citenamefont {Lehmberg}(1970)}]{lehmberg1970}%
  \BibitemOpen
  \bibfield  {author} {\bibinfo {author} {\bibfnamefont {R.~H.}\ \bibnamefont
  {Lehmberg}},\ }\bibfield  {title} {\bibinfo {title} {Radiation from an
  {$N$}-atom system. {I}. general formalism},\ }\href
  {https://doi.org/10.1103/PhysRevA.2.883} {\bibfield  {journal} {\bibinfo
  {journal} {Phys. Rev. A}\ }\textbf {\bibinfo {volume} {2}},\ \bibinfo {pages}
  {883} (\bibinfo {year} {1970})}\BibitemShut {NoStop}%
\bibitem [{\citenamefont {Agarwal}(1970)}]{agarwal1970}%
  \BibitemOpen
  \bibfield  {author} {\bibinfo {author} {\bibfnamefont {G.~S.}\ \bibnamefont
  {Agarwal}},\ }\bibfield  {title} {\bibinfo {title} {Master-equation approach
  to spontaneous emission},\ }\href {https://doi.org/10.1103/PhysRevA.2.2038}
  {\bibfield  {journal} {\bibinfo  {journal} {Phys. Rev. A}\ }\textbf {\bibinfo
  {volume} {2}},\ \bibinfo {pages} {2038} (\bibinfo {year} {1970})}\BibitemShut
  {NoStop}%
\bibitem [{\citenamefont {Javanainen}\ and\ \citenamefont
  {Ruostekoski}(1995)}]{javanainen1995}%
  \BibitemOpen
  \bibfield  {author} {\bibinfo {author} {\bibfnamefont {J.}~\bibnamefont
  {Javanainen}}\ and\ \bibinfo {author} {\bibfnamefont {J.}~\bibnamefont
  {Ruostekoski}},\ }\bibfield  {title} {\bibinfo {title} {Off-resonance light
  scattering from low-temperature {Bose} and {Fermi} gases},\ }\href
  {https://doi.org/10.1103/PhysRevA.52.3033} {\bibfield  {journal} {\bibinfo
  {journal} {Phys. Rev. A}\ }\textbf {\bibinfo {volume} {52}},\ \bibinfo
  {pages} {3033} (\bibinfo {year} {1995})}\BibitemShut {NoStop}%
\bibitem [{\citenamefont {Ruostekoski}\ and\ \citenamefont
  {Javanainen}(1997)}]{ruostekoski1997}%
  \BibitemOpen
  \bibfield  {author} {\bibinfo {author} {\bibfnamefont {J.}~\bibnamefont
  {Ruostekoski}}\ and\ \bibinfo {author} {\bibfnamefont {J.}~\bibnamefont
  {Javanainen}},\ }\bibfield  {title} {\bibinfo {title} {Quantum field theory
  of cooperative atom response: Low light intensity},\ }\href
  {https://doi.org/10.1103/PhysRevA.55.513} {\bibfield  {journal} {\bibinfo
  {journal} {Phys. Rev. A}\ }\textbf {\bibinfo {volume} {55}},\ \bibinfo
  {pages} {513} (\bibinfo {year} {1997})}\BibitemShut {NoStop}%
\bibitem [{\citenamefont {Zell}\ \emph {et~al.}(2009)\citenamefont {Zell},
  \citenamefont {Queisser},\ and\ \citenamefont {Klesse}}]{zell2009}%
  \BibitemOpen
  \bibfield  {author} {\bibinfo {author} {\bibfnamefont {T.}~\bibnamefont
  {Zell}}, \bibinfo {author} {\bibfnamefont {F.}~\bibnamefont {Queisser}},\
  and\ \bibinfo {author} {\bibfnamefont {R.}~\bibnamefont {Klesse}},\
  }\bibfield  {title} {\bibinfo {title} {Distance dependence of entanglement
  generation via a bosonic heat bath},\ }\href
  {https://doi.org/10.1103/PhysRevLett.102.160501} {\bibfield  {journal}
  {\bibinfo  {journal} {Phys. Rev. Lett.}\ }\textbf {\bibinfo {volume} {102}},\
  \bibinfo {pages} {160501} (\bibinfo {year} {2009})}\BibitemShut {NoStop}%
\bibitem [{\citenamefont {Valentini}(1991)}]{valentini1991}%
  \BibitemOpen
  \bibfield  {author} {\bibinfo {author} {\bibfnamefont {A.}~\bibnamefont
  {Valentini}},\ }\bibfield  {title} {\bibinfo {title} {Non-local correlations
  in quantum electrodynamics},\ }\href
  {https://doi.org/10.1016/0375-9601(91)90952-5} {\bibfield  {journal}
  {\bibinfo  {journal} {Phys. Lett. A}\ }\textbf {\bibinfo {volume} {153}},\
  \bibinfo {pages} {321} (\bibinfo {year} {1991})}\BibitemShut {NoStop}%
\bibitem [{\citenamefont {Reznik}(2003)}]{reznik2003}%
  \BibitemOpen
  \bibfield  {author} {\bibinfo {author} {\bibfnamefont {B.}~\bibnamefont
  {Reznik}},\ }\bibfield  {title} {\bibinfo {title} {Entanglement from the
  vacuum},\ }\href {https://doi.org/10.1023/A:1022875910744} {\bibfield
  {journal} {\bibinfo  {journal} {Found. Phys.}\ }\textbf {\bibinfo {volume}
  {33}},\ \bibinfo {pages} {167} (\bibinfo {year} {2003})}\BibitemShut
  {NoStop}%
\bibitem [{\citenamefont {Reznik}\ \emph {et~al.}(2005)\citenamefont {Reznik},
  \citenamefont {Retzker},\ and\ \citenamefont {Silman}}]{reznik2005}%
  \BibitemOpen
  \bibfield  {author} {\bibinfo {author} {\bibfnamefont {B.}~\bibnamefont
  {Reznik}}, \bibinfo {author} {\bibfnamefont {A.}~\bibnamefont {Retzker}},\
  and\ \bibinfo {author} {\bibfnamefont {J.}~\bibnamefont {Silman}},\
  }\bibfield  {title} {\bibinfo {title} {Violating {Bell's} inequalities in
  vacuum},\ }\href {https://doi.org/10.1103/PhysRevA.71.042104} {\bibfield
  {journal} {\bibinfo  {journal} {Phys. Rev. A}\ }\textbf {\bibinfo {volume}
  {71}},\ \bibinfo {pages} {042104} (\bibinfo {year} {2005})}\BibitemShut
  {NoStop}%
\bibitem [{\citenamefont {Salton}\ \emph {et~al.}(2015)\citenamefont {Salton},
  \citenamefont {Mann},\ and\ \citenamefont {Menicucci}}]{salton2015}%
  \BibitemOpen
  \bibfield  {author} {\bibinfo {author} {\bibfnamefont {G.}~\bibnamefont
  {Salton}}, \bibinfo {author} {\bibfnamefont {R.~B.}\ \bibnamefont {Mann}},\
  and\ \bibinfo {author} {\bibfnamefont {N.~C.}\ \bibnamefont {Menicucci}},\
  }\bibfield  {title} {\bibinfo {title} {Acceleration-assisted entanglement
  harvesting and rangefinding},\ }\href
  {https://doi.org/10.1088/1367-2630/17/3/035001} {\bibfield  {journal}
  {\bibinfo  {journal} {New J. Phys.}\ }\textbf {\bibinfo {volume} {17}},\
  \bibinfo {pages} {035001} (\bibinfo {year} {2015})}\BibitemShut {NoStop}%
\bibitem [{\citenamefont {Pozas-Kerstjens}\ and\ \citenamefont
  {Mart\'{\i}n-Mart\'{\i}nez}(2015)}]{pozas2015}%
  \BibitemOpen
  \bibfield  {author} {\bibinfo {author} {\bibfnamefont {A.}~\bibnamefont
  {Pozas-Kerstjens}}\ and\ \bibinfo {author} {\bibfnamefont {E.}~\bibnamefont
  {Mart\'{\i}n-Mart\'{\i}nez}},\ }\bibfield  {title} {\bibinfo {title}
  {Harvesting correlations from the quantum vacuum},\ }\href
  {https://doi.org/10.1103/PhysRevD.92.064042} {\bibfield  {journal} {\bibinfo
  {journal} {Phys. Rev. D}\ }\textbf {\bibinfo {volume} {92}},\ \bibinfo
  {pages} {064042} (\bibinfo {year} {2015})}\BibitemShut {NoStop}%
\bibitem [{\citenamefont {Anh-Tai}\ \emph {et~al.}(2024)\citenamefont
  {Anh-Tai}, \citenamefont {Fogarty}, \citenamefont
  {de~Mar\'{\i}a-Garc\'{\i}a}, \citenamefont {Busch},\ and\ \citenamefont
  {Garc\'{\i}a-March}}]{anhtai2024}%
  \BibitemOpen
  \bibfield  {author} {\bibinfo {author} {\bibfnamefont {T.~D.}\ \bibnamefont
  {Anh-Tai}}, \bibinfo {author} {\bibfnamefont {T.}~\bibnamefont {Fogarty}},
  \bibinfo {author} {\bibfnamefont {S.}~\bibnamefont
  {de~Mar\'{\i}a-Garc\'{\i}a}}, \bibinfo {author} {\bibfnamefont
  {T.}~\bibnamefont {Busch}},\ and\ \bibinfo {author} {\bibfnamefont {M.~A.}\
  \bibnamefont {Garc\'{\i}a-March}},\ }\bibfield  {title} {\bibinfo {title}
  {Engineering impurity {Bell} states through coupling with a quantum bath},\
  }\href {https://doi.org/10.1103/PhysRevResearch.6.043042} {\bibfield
  {journal} {\bibinfo  {journal} {Phys. Rev. Res.}\ }\textbf {\bibinfo {volume}
  {6}},\ \bibinfo {pages} {043042} (\bibinfo {year} {2024})}\BibitemShut
  {NoStop}%
\bibitem [{\citenamefont {G\'omez-Lozada}\ \emph {et~al.}(2025)\citenamefont
  {G\'omez-Lozada}, \citenamefont {Hiyane}, \citenamefont {Busch},\ and\
  \citenamefont {Fogarty}}]{gomez2024}%
  \BibitemOpen
  \bibfield  {author} {\bibinfo {author} {\bibfnamefont {F.}~\bibnamefont
  {G\'omez-Lozada}}, \bibinfo {author} {\bibfnamefont {H.}~\bibnamefont
  {Hiyane}}, \bibinfo {author} {\bibfnamefont {T.}~\bibnamefont {Busch}},\ and\
  \bibinfo {author} {\bibfnamefont {T.}~\bibnamefont {Fogarty}},\ }\bibfield
  {title} {\bibinfo {title} {{Bose}--{Fermi} {$N$}-polaron state emergence from
  correlation-mediated blocking of phase separation},\ }\href
  {https://doi.org/10.1103/PhysRevResearch.7.023053} {\bibfield  {journal}
  {\bibinfo  {journal} {Phys. Rev. Res.}\ }\textbf {\bibinfo {volume} {7}},\
  \bibinfo {pages} {023053} (\bibinfo {year} {2025})}\BibitemShut {NoStop}%
\end{thebibliography}
\end{document}